# Development of polar nematic fluids with giant-κ dielectric properties

## Authors

J. Li[1]†, H. Nishikawa[2]†, J. Kougo[1]†, J. Zhou[1], S. Dai[1], W. Tang[1], X. Zhao[1], Y. Hisai[3], M. Huang[1]*, S. Aya[1]*

† J. X. L., H. N. and J. K. contributed equally to this work.

## Affiliations

[1] South China Advanced Institute for Soft Matter Science and Technology (AISMST), South China University of Technology, Guangzhou 510640, China.

[2] Physicochemical Soft Matter Research Team, RIKEN Center for Emergent Matter Science (CEMS), Japan.

[3] Money Forward, Inc. Shibaura, Minato-ku, Tokyo, Japan.

*Corresponding author. Email: huangmj25@scut.edu.cn (M.H.); satoshiaya@scut.edu.cn (S.A.)

## Abstract

Super-high-$\kappa$ materials that exhibit exceptionally high dielectric permittivity are recognized as potential candidates for a wide range of next-generation photonic and electronic devices. Generally, the high dielectricity for achieving a high-$\kappa$ state requires a low symmetry of materials so that most of the discovered high-$\kappa$ materials are symmetry-broken crystals. There are scarce reports on fluidic high-$\kappa$ dielectrics. Here we demonstrate a rational molecular design, supported by machine-learning analyses, that introduces high polarity to asymmetric molecules, successfully realizing super-high-$\kappa$ fluid materials (dielectric permittivity, $\varepsilon > 10^4$) and strong second harmonic generation with macroscopic spontaneous polar ordering. The polar structures are confirmed to be identical for all the synthesized materials. Our experiments and computational calculation reveal the unique orientational structures coupled with the emerging polarity. Furthermore, adopting this strategy to high-molecular-weight systems additionally extends the novel material category from monomer to polar polymer materials, creating polar soft matters with spontaneous symmetry breaking.

## MAIN TEXT

## Introduction

High-$\kappa$ dielectrics are classified as the materials exhibiting dielectric permittivity, $\kappa = \varepsilon + 1$, larger than 8 of silicon nitride. In recent decades, there emerges a new trend of high-$\kappa$ electrics by utilizing barium-titanate-type (1-5) and 2D crystalline materials (6-8), where $\varepsilon$ reaches up to about $10^2 - 10^3$, making it possible to realize high-quality electronics such as high-energy-density dielectric films (9), high-performance capacitor (10) and insulator (11-12) and high-density memory (13). While barium-titanate-type crystals serve as high-$\kappa$ dielectrics owing to the breaking of the cubic symmetry in bulk, 2D crystals do so primarily due to the huge specific surface area of ordered atoms on the surface. These paradigms give a common perspective that the incidence of the high-$\kappa$ state is only allowed when the symmetry of material is broken (thereby generally non-fluidic) and the thickness of the material is small, so that a net switchable polarization can remain at a high density and contribute to the high value of $\varepsilon$. Massive progress in this framework has been made in the last few decades, and it has been recognized that a majority of other materials, e.g. non-crystalline materials, are out of consideration for achieving a high-$\kappa$ state because the two characteristics of broken symmetry and fluidity barely can coexist in the same material. Therefore, the selection variety of materials is considerably limited, so many other unknown properties may not show up so far.

Very recently in 2017, two outliers appeared. One of our authors, Nishikawa et al. reported a new class of liquid crystalline phase state of matters, termed as polar nematic, that serves as the first example of fluidic super-high-$\kappa$ (or giant-$\kappa$) dielectric, exhibiting e.g. $\varepsilon \sim 10^4$ at 1 kHz characterized by undefined polar structures and inherent high fluidity (14). At almost the same period, Mandle et al. reported an unknown liquid crystalline phase state in RM734, which has been assigned to be splay nematic later (15-18). Most recently, Chen et al. demonstrated RM734 exhibits the local ferroelectric nature (19). The polar nematic liquid crystal phases with a fascinating combination of huge κ and high fluidity are attracting the increasing endeavor for deep study in underlying physics and material properties (20-22).

Nevertheless, there are plenty of critical issues that need to be overcome to push the development of these materials for practical applications. Regarding the origin of polar nature, the structure and the mechanism of the phase formation are unclear and under debate, accompanying the uncertainty of the structural evolution pathway and the relationship with the conventional nematic phase. Moreover, with only the two aforementioned polar molecular systems experimentally-proved by dielectric and optical nonlinear measurements, it is not sure if this polar nematic occurs occasionally or can be a generally accessible liquid crystal phase. To dissolve the erotetic arguments, we here establish a generic molecular guideline towards the rational design of giant-$\kappa$ polar fluids with the aid of machine learning and make full analyses of polarizing light microscopy (POM), second harmonic generation (SHG) and dielectric spectroscopy. We conclude that the polar nematic can widely exist in rod-shaped molecules under the proper conditions discussed below. We reveal the evolution of topological structures and unravel the family of the materials provides a combination of outstanding performance of dielectricity (giant-$\varepsilon$) and second-order nonlinear optical (NLO) properties (high nonlinear coefficient). Furthermore, adopting this strategy to high-molecular-weight systems additionally extends the novel polar material category from monomer to polymer. These results would promote the quicker development of a soft matter platform for high-performance fluidic dielectrics and nonlinear optics. Hereafter, for convenience and due to the common nature for all the observed polar nematics, we unify the terms, splay nematic and polar nematic, to be ‘polar nematic (PN) phase’.

## Results and Discussion

### Establishment of molecular library of the polar nematics guided by machine learning.

We synthesized more than 40 new rod-shaped molecules with dipole moments ($\mu$) ranging from 4–13 D (Fig. 1). To quantitatively extract the general dictating rules during the synthesis, we have utilized a home-made machine learning algorism (next section and Discussion S5) to efficiently optimize the molecular structure for stabilizing the PN phase. Electron-withdrawing groups such as ester, fluoride, fluorocarbon ether, nitro, nitrile, or trifluorocarbon are arranged sequentially in the same direction to create large dipole moments. After careful examination by POM, SHG and dielectric studies, these compounds can be divided into three categories according to the polar feature (labeled by red-, blue- and green-colors in Fig. 1). For the green-color labeled compounds, the PN phase could be stable at some temperatures upon the cooling process, with persistent existence of the typical PN texture, large dielectric permittivity, and high SH signal. Two phase transitions (Iso→N→PN) could also be identified by DSC measurement (Table S1). On the contrary, the compounds labeled with the red color show either crystal or the traditional nonpolar N phase with extremely weak SH signal and low dielectric constant ($\varepsilon$ < 15). The compounds belong to the third category (labeled with blue-color in Fig. 1) sits in between and exhibit the metastable PN phase. The fast crystallization within this category usually overwhelms the occurrence of the PN phase, resulting in the neglect of the PN phase under the typical observation condition of slow cooling. No PN phase or related phase transitions could be observed in DSC with typical scanning rates of 1–30 K/min. However, we could reopen the observation window of the PN phase by suppressing crystal nucleation or decreasing the crystallization rate with appropriate treatment. For instance, quickly quenching to room temperature from the isotropic state is very useful to help capture the PN phase in most of these polar compounds, enabling the observation of characteristic POM texture and strong SH signal in a short time window (several seconds to tens of minutes). Overall, more than half of the synthesized materials (23/43) were detected to exhibit strong spontaneous polar ordering that results in huge dielectric permittivity, $\varepsilon$ over 10,000 at all the frequencies below about 1-10 kHz. This strongly suggests that the PN phase could be a general liquid crystalline state with a common structure and properties as discussed later. Worth noting, in contrast to the common ferroelectrics and relaxors that exhibit giant-$\varepsilon$ in a narrow-frequency and -temperature ranges, our synthesized materials present the giant dielectric permittivity in a wide range of frequency (< 1-10 kHz) and temperature.

**Molecular features for the polar nematics.**

Taking some previously-reported molecules sharing similar chemical structures or properties into account (Fig. S11 and Table S2), the total number of molecules for machine learning is expanded to ca. 70. The essential molecular features are summarized in Fig. 2A. The machine learning calculation has been made based on both the linear Pearson's correlation method (23) and the nonlinear analyses combining the random forest regression with a modified SHapley Additive exPlanations (SHAP) analysis (24). Figures 2B and 2C demonstrate the Pearson's correlation coefficients and SHAP values for the addressed parameters. The parameters carrying high Pearson's correlation coefficients to the stability of the PN phase show wider distributions and higher absolute strength of SHAP value, confirming the importance of the parameters as below: molecular dipole > geometrical aspect ratio > angle of dipole moment, and the rest of parameters has negligible contributions. Additionally, SHAP analysis provides a quantitative evaluation of the degree of positive or negative contribution of each parameter to the stability of the PN phase. That is, the positive and negative values correspond to the stabilization and the destabilization effects on the PN phase, respectively. Figures 2D-2G plot the SHAP values as a function of the molecular dipole, the geometrical aspect ratio and the angle of dipole moment. The results offer a clear statistical guideline for the molecular design of the PN materials: (1) the dipole moment, $\mu$ > 9 D; (2) the geometrical aspect ratio (AR) defined as the length per the width, AR < 2.5; (3) the angle of the dipole should be in a moderate range, offering an oblique dipole (~20°). Worth noting, all the conditions should be satisfied to guarantee the materials with both the high polarity and the liquid crystalline nature. For example, if the dipole moment is lower than the threshold of 9 D, the traditional nonpolar nematic phase appears. This situation is supported by the prediction by Born in 1916, that a ferroelectric anisotropic fluid would exist if the molecular dipole is large enough to overcome the thermal disturbance (25). Meanwhile, the geometrical aspect ratio plays a crucial role in either avoiding the crystallization of the materials before transiting into the PN phase or expanding the temperature range of the PN phase towards lower temperatures. Indeed, the introduction of the side chain is synonymous with increasing the aspect ratio of overall mesogens, which proves efficiently stabilize the PN phase (Fig. 1, e.g. compare end-chain substitution in **1f** and **1g** with side-chain substitution in **1d** and **1e**). The longer the side chain length, the lower temperature PN phase is stabilized in material series **1**, **2** and **6**. The too long side chain may impede the parallel polar packing among rod-like mesogens and interrupt PN formation as in **4d** and **6e**. The inclination of the dipole would work as the driving force for preventing the dimerization of the molecules. Otherwise, the molecules processing the dipole along the long axis universally show the traditional nematic phase, where the net dipole is zero through the dimerization. Noticeably, all the nitrile-containing molecules (e.g. **2d-f**, **11a-b**) fail to form the PN phase, although with large dipole values (> 10 D) and side groups. We speculate that the linear shape and small cross-section area of the nitrile group facilitate the dimerization between large dipoles, in strong contrast to the planar nitro group.

## Structure, dielectricity and polarity of the polar nematics.

Revealing the topological structure of this new phase is crucial for understanding structure-property relationships. To date, the structure of the PN phase remains contradictory between previous models (18,19), and how the structure is developed from the high-temperature N phase with the uniform director field is unclear. Worth noting is that the previous works (17,18) have already triggered intensive following theoretical studies based on one of the models (20-22), the true structure should be further confirmed. To establish a systematic and a general correlation between the structures and the emergence of the polarity, we made simultaneous measurements of POM, SHG and dielectric measurements for the synthesized materials upon varying temperature. While the dielectric permittivity provides information about the dynamic polarity where the polarization switches the direction following the electric field, the SH signal clarifies the static polarity where the polarization "freezes" during the incoming light field oscillates.

Figures 3 and S3 demonstrate the systematic observations of POM, SHG and dielectricity on three PN materials including **5b**, **5c** and **1a**. Notably, though all the synthesized PN molecules have distinct chemical nature, the structural features based on the POM and differential scanning calorimetry (DSC, Table S1) are common. They exhibit a similar structural evolution on cooling via 4 steps (Figs. 3B-3F, 3I-3N): Schlieren texture (N phase)→stripe texture (N phase)→defect-free band texture (N phase)→band texture with line disclination (PN phase). Generally, from the isotropic state down to the middle of the N phase, the materials show relatively low dielectricity (e.g. $\varepsilon < 100$) and vanishingly small SH signal as like the traditional N materials (Figs. 3A,3B,3H-3J). However, as further decreasing temperature towards the N-PN transition, both the dielectric permittivity and the SH signal grow continuously and drastically, confirming the emergence of the polar nature in the N phase as a pretransitional phenomenon. Curiously, the temperature window for the increase of the polarity is more than ca. 15 °C in most of the PN materials. In this range, we observe the structural transformation from the Schlieren to the stripe texture, which runs parallel to the average molecular orientation (i.e. nematic director **n**). The stripe texture, attributed to the flexoelectric-effect-induced splay undulation of the director field, enhances its contrast upon approaching the N–PN transition (Figs. 3C and 3J, Figs. 4D-4F, Fig. S6, Discussion S2). Right above the N-PN transition, the stripe texture is replaced by the band texture without defect (Figs. 3D,3E,3K-3M), where filamentous regimes with some periodicity accompanied by different interference colors than the surroundings appear. This state is characterized by the sharping of the Schlieren dark brushes and the continuous increase of both the dielectric permittivity and SH signal, confirming the enhancement of both the structural ordering and polarity. Upon the transition to the PN phase, a swarm of line defects similarly observed pop up, seeded by the filamentous domains. Each domain separated by the defect is ferroelectric and the adjacent domains exhibit opposite polarity (19). Worth noting, the strong polarity in the PN phase already develops at the stage of the defect-free band texture in the N phase, manifested by the drastic increase of the dielectricity and SH signal (Fig. 3). Dissimilar to the extinction of the N texture when the rubbing direction is parallel to either the polariser or analyzer of the microscope, the domains of the PN phase does not show complete distinction (Figs. 4P-4S, Figs. S5 and S8). Considering the overall structural evolution, the band texture in the PN is attributed to a Splay-Néel type I director field accompanied by local splay deformation in each domain, as discussed below and in Discussions S2 and S3. The corresponding director field is also directly visualized by carefully observing the structural transition from the PN phases to the N phase upon heating (Fig. 3G). We have analyzed the PN phase texture and structure evolution in more detail in Discussions S1, S2 and S3.

**Free energy comparison.**

To quantitatively discuss the stability of the phase structures, we compare the free energies for several plausible director models that do not seemingly violate the experimental observation: Bloch-type, two Néel-type defects, and defect wall models. Figures 4A-4O summarizes the 3D model structures, along with the corresponding textural features, dielectricity and SHG characteristics. Considering the elastic deformation and the polarity effect, the total free energy is written as the volume integration of the sum of the elastic energy density ($f_{\mathrm{elastic}}$), flexoelectric energy density ($f_{\mathrm{flexo}}$), local polarity energy density ($f_{\mathrm{polar}}$) and the energy of the defect line or wall ($f_{\mathrm{defect}}$):

$$F = \int (f_{\mathrm{elastic}} + f_{\mathrm{flexo}} + f_{\mathrm{polar}} + f_{\mathrm{defect}})\mathrm{d}V + c\langle \mathbf{P}\rangle_V.$$

$$f_{\mathrm{elastic}} = \frac{1}{2}K_{11}(\mathrm{div}\,\mathbf{n})^2 + \frac{1}{2}K_{22}\big(\mathbf{n}\cdot(\mathrm{curl}\,\mathbf{n})\big)^2 + \frac{1}{2}K_{33}\big(\mathbf{n}\times(\mathrm{curl}\,\mathbf{n})\big)^2.$$

$$f_{\mathrm{flexo}} = -\gamma\mathbf{n}(\mathrm{div}\,\mathbf{n})\cdot\mathbf{P}.$$

$$f_{\mathrm{polar}} = \frac{1}{2}t(\mathbf{P}\cdot\mathbf{P})^2 + \frac{1}{2}b(\mathrm{grad}\,\mathbf{P})^2.$$

Here, the term of $f_{\mathrm{polar}}$ includes only the lowest square term since the further considerations of the higher-order terms barely change the calculation results. Also, to consider the effect that antiferroelectric order is favored at macroscopic though the local polar order exists in each domain, we add the last term of the free energy, $c\langle \mathbf{P}\rangle_V$, where $c$ is positive. This describes the overall excess polarity energy in the volume. Figure 4T demonstrates the temperature dependencies of the free energies for the Splay model with local polarity and Splay-Néel Type I PN model as well as the homogenous N state with local emerging polarity in molecule **1a**. All the parameters used in the calculation except the values of $b$ and the defect energy are obtained experimentally (Discussion S4). Consistent with the experimentally observed structural sequence, the lowest free energy state changes from the homogenous N state to the splay state at about 136 °C, then transit to the Splay-Néel type I PN phase at about 128 °C (Fig. 4T). The defect wall-type configurations, replacing the Néel region by wall defects, has considerably higher free energy than others due to the excess free energy from the large area of the defect walls. Importantly, the band texture in the PN phase cannot be explained by the symmetric splay structure, i.e. Splay-Néel Type II structure, because the distinction area in the middle of each polarization domain cannot be observed in the experiment (Fig. S8). The simulated POM images of the Splay-Néel Type I structure well reproduce the POM textures under various conditions (Figs. 4P-4S, Fig. S8). With the above-discussed experimental evidence, we conclude the most plausible model of the PN phase should be in the form of the Splay-Néel Type I director field. Some possible detailed orientational states under several-hundred nanometer-scale might be missing here and should be explored in the future in the community. In general, the achieved clearer illustration of the packing structure in the PN phase help better explore a broad range of physical effects, including the response to external fields and topological defect change to diverse surface anchoring.

**From polar nematic monomer to polymer.**

So far, most of the small-molecule PN phases are metastable at room temperature and tend to crystallize, losing the high intrinsic polarity. Therefore, we consider the possibility of further extending the materials category to high-molecular-weight side-chain polymers, which makes crystallization unlikely to occur and also provides other mechanical properties. Based on the aforementioned knowledge, in particular, the aspect ratio crucially affecting the probability of the induction of the PN phase, we synthesize a side-chain polymer, PLC (Fig. 1). As expected, PLC not only displays the common PN-like texture (Fig. 5C) but also emitting a strong SH signal (Fig. 5A). Besides, we successfully fabricate free-standing polymer fiber and film exhibiting a similar PN phase (Fig. S13). In the fiber, the polar molecules spontaneously orient along the elongation direction.

Finally, we highlight that the polar nematic materials exhibit giant dielectric permittivity and high nonlinear coefficients. Figures 3O and 3P display, respectively, the dielectric permittivity *vs* $\tan\delta^{-1}$ and *vs* NLO coefficient for various types of materials including inorganic-, organic-, hybrid-solid as well as composite materials (see Tables S3 and S4). The PN materials (**1a**, **1b**, **5b**, **5c**, *et al.*) have highly superior properties of dielectric permittivity ($\varepsilon$ over $10^4$ below 10 kHz) and low $\tan\delta^{-1}$ ($0.05 < \tan\delta < 0.20$), which represent the state of the art functionality in soft matters. The giant dielectricity in the PN materials is even comparable to well-known inorganic ferroelectric or relaxor materials (*e.g.* barium titanate crystals, see Tables S3 and S4). Besides, for the PN materials, one can notice that the high NLO coefficient ranging between 1 and 10 pm $V^{-1}$ which are comparable to some ferroelectric organic- and hybrid-solids. Note that those excellent NLO materials based on inorganics usually show quite low dielectric permittivity below 10. Hence, the excellent combination of giant dielectricity and NLO properties in the PN materials, together with the fluidity or flexibility, is rather unique and offers vast technological potentials for high-performance supercapacitors, nonlinear optical elements, memory device and electrooptic functional devices.

## Materials and Methods

**Material synthesis.** Refer to the Materials and Methods in the Supplementary Materials.

**SHG measurement.** We use a fundamental beam from a Q-switched pulsed laser (MPL-III-1064-20μJ) with a central wavelength of 1064 nm, maximum power of 200 mW, pulse duration of 5 ns, and 100 Hz repetition. The p-polarised fundamental beam was directed at LC cells. The SH light is detected at p-polarization from the transmission direction by a photomultiplier tube (DH-PMT-D100V, Daheng Optics). The time evolution of the SH signal is read by an oscilloscope and the data is transferred to a computer. The SH signal is recorded every 1°C under the control of a home-made Labview program. The schematics of the optical system are drawn in Fig. S9. The nonlinear coefficient is calculated by the maker fringe method. The symmetry of the PN phase is $C_{\infty v}$, giving spontaneous polarization mainly along the long molecular axis. The second-order nonlinear optical tensor is given by

$$\boldsymbol{d} = \begin{pmatrix} 0 & 0 & 0 & 0 & d_{15} & 0 \\ 0 & 0 & 0 & d_{24} & 0 & 0 \\ d_{31} & d_{32} & d_{33} & 0 & 0 & 0 \end{pmatrix}.$$

The second order nonlinear polarization s ($P_x$, $P_y$ and $P_z$) are calculated as below.

$$P_x(2\omega) = 2\varepsilon_0 d_{15} E_x(\omega) E_z(\omega)$$

$$P_y(2\omega) = 2\varepsilon_0 d_{24} E_y(\omega) E_z(\omega)$$

$$P_z(2\omega) = \varepsilon_0 d_{31} E_x(\omega)^2 + \varepsilon_0 d_{32} E_y(\omega)^2 + \varepsilon_0 d_{33} E_z(\omega)^2$$

$\omega$, $\varepsilon_0$, $E_x$, $E_y$ and $E_z$ represent the frequency of the light, free-space permittivity and three components of the electric field in $x$, $y$ and $z$ directions. In the polarization combinations of (polarization of excitation light, polarization of detected light) = (p,s), (s,s), (s,p) and (p,p), the probed polarization s are

$P_y(2\omega) = 0$ in (p,s)-combination,

$P_y(2\omega) = 0$ in (s,s)-combination,

$P_x(2\omega) = 0$, $P_z(2\omega) = \varepsilon_0 d_{32} E_y(\omega)^2$ in (s,p)-combination,

$P_x(2\omega) = 2\varepsilon_0 d_{15} E_x(\omega) E_z(\omega)$, $P_z(2\omega) = \varepsilon_0 d_{31} E_x(\omega)^2$ in (p,p)-combination.

In accordance with the above analyses, we observe strong SH signals in (s,p)- and (p,p)-combinations and the rest only show negligible signals.

**Machine learning.** The code for the machine learning is written in Python by combining the following libraries and packages: Pandas data analysis library, NumPy extension module for numerical calculation, Matplotlib and Seaborn for statistical data visualization, SHAP package, and scikit-learn machine learning library. Refer to Discussion S5 and Fig. S10 for the detailed algorism.

**Berreman 4×4 matrix optical simulation.** We perform the Berreman 4×4 matrix optical simulation (26) to find the adequate model structure that matches the real POM observation. Numerical simulations are made based on the director structure as shown in Fig. S8. We use the refractive indices of **5b** with ordinary and extraordinary refractive indices of $n_0$ = 1.485 and $n_e$ = 1.697 without dispersion. The liquid crystal material is assumed to be composed of a stack of 100 nm thick thin layers. To reproduce the color of the illumination Halogen lamp, we use a measured spectrum of a Halogen lamp and create 60 discrete different wavelengths evenly distributed between 380-780 nm. The simulation results are obtained as RGB images by using color-matching functions for the 60 wavelengths.

## References

1. B. Wul, Dielectric Constants of Some Titanates. *Nature* **156**, 480 (1945).

2. W. Jackson, W. Reddish, High Permittivity Crystalline Aggregates. *Nature* **156**, 717 (1945).
3. P. R. Coursey, K. G. Brand, Dielectric Constants of Some Titanates. *Nature* **157**, 297–298 (1946).
4. M. G. Harwood, P. Popper, D. F. Rushman, Curie Point of Barium Titanate. *Nature* **160**, 58–59 (1947).
5. B. T. Matthias, Dielectric Constant and Piezo-Electric Resonance of Barium Titanate Crystals. *Nature* **161**, 325-326 (1948).
6. M. Osada, Y. Ebina, H. Funakubo, S. Yokoyama, T. Kiguchi, K. Takada, T. Sasaki, High-$\kappa$ Dielectric Nanofilms Fabricated from Titania Nanosheets. *Adv. Mater.* **18**, 1023–1027 (2006).
7. Y. Y. Illarionov, T. Knobloch, M. Jech, M. Lanza, D. Akinwande, M. I. Vexler, T. Mueller, M. C. Lemme, G. Fiori, F. Schwierz, T. Grasser, Insulators for 2D nanoelectronics: the gap to bridge. *Nat. Commun.* **11**, 3385 (2020).
8. T. Li, T. Tu, Y. Sun, H. Fu, J. Yu, L. Xing, Z. Wang, H. Wang, R. Jia, J. Wu, C. Tan, Y. Liang, Y. Zhang, C. Zhang, Y. Dai, C. Qiu, M. Li, R. Huang, L. Jiao, K. Lai, B. Yan, P. Gao, H. Peng, A native oxide high-$\kappa$ gate dielectric for two-dimensional electronics. *Nat. Electron.* **3**, 473–478 (2020).
9. H. Pan, F. Li, Y. Liu, Q. Zhang, M. Wang, S. Lan, Y. Zheng, J. Ma, L. Gu, Y. Shen, P. Yu, S. Zhang, L.-Q. Chen, Y.-H. Lin, C. -W. Nan, Ultrahigh–energy density lead-free dielectric films via polymorphic nanodomain design. *Science* **365**, 578-582 (2019).
10. S. K. Kim, S. W. Lee, J. H. Han, B. Lee, S. Han, C. S. Hwang, Capacitors with an equivalent oxide thickness of <0.5 nm for nanoscale electronic semiconductor memory. *Adv. Funct. Mater.* **20**, 2989–3003 (2010).
11. D. Ji, T. Li, Y. Zou, M. Chu, K. Zhou, J. Liu, G. Tian, Z. Zhang, X. Zhang, L. Li, D. Wu, H. Dong, Q. Miao, H. Fuchs, W. Hu, Copolymer dielectrics with balanced chain-packing density and surface polarity for high-performance flexible organic electronics. *Nat. Commun.* **9**, 2339 (2018).
12. J. Robertson, High dielectric constant gate oxides for metal oxide Si transistors. *Rep. Prog. Phys.* **69**, 327–396 (2006).
13. C.-H. Lee, S.-H. Hur, Y.-C. Shin, J.-H. Choi, D.-G. Park, K. Kim, Charge-trapping device structure of SiO2/SiN/high-*k* dielectric Al2O3 for high-density flash memory. *Appl. Phys. Lett.* **86**, 152908 (2005).
14. H. Nishikawa, K. Shiroshita, H. Higuchi, Y. Okumura, Y. Haseba, S. Yamamoto, K. Sago, H. Kikuchi, A Fluid Liquid-Crystal Material with Highly Polar Order. *Adv. Mater.* **20**, 1702354 (2017).
15. R. J. Mandle, S. J. Cowling, J. W. Goodby, A nematic to nematic transformation exhibited by a rod-like liquid crystal. *Phys. Chem. Chem. Phys.* **19**, 11429–11435 (2017).
16. R. J. Mandle, S. J. Cowling, J. W. Goodby, Rational design of rod-like liquid crystals exhibiting two nematic phases. *Chemistry* **23**, 14554–14562 (2017).
17. A. Mertelj, L. Cmok, N. Sebastián, R. J. Mandle, R. R. Parker, A. C. Whitwood, J. W. Goodby, M. Čopič, Splay nematic phase. *Phys. Rev. X* **8**, 41025 (2018).

18. N. Sebastián, L. Cmok, R. J. Mandle, M. R. de la Fuente, I. D. Olenik, M. Čopič, A. Mertelj, Ferroelectric-Ferroelastic Phase Transition in a Nematic Liquid Crystal, *Phys. Rev. Lett.* **124**, 037801 (2020).

19. X. Chen, E. Korblova, D. Dong, X. Wei, R. Shao, L. Radzihovsky, M. A. Glaser, J. E. Maclennan, D. Bedrov, D. M. Walba, N. A. Clark, First-principles experimental demonstration of ferroelectricity in a thermotropic nematic liquid crystal: Polar domains and striking electro-optics. *Proc. Natl. Acad. Sci. U.S.A.* **117**, 14021-14031 (2020).

20. N. Chaturvedi, R. D. Kamien, Mechanisms to splay-bend nematic phases. *Phys. Rev. E* **100**, 022704 (2019).

21. M. Copic, A. Mertelj, Q-tensor model of twist-bend and splay nematic phases. *Phys. Rev. E* **101**, 022704 (2020).

22. M. P. Rosseto, J. V. Selinger, Theory of the splay nematic phase: Single vs. double splay. *Phys. Rev. E* **101**, 052707 (2020).

23. S. M. Stigler, Francis Galton's account of the invention of correlation, *Stat. Sci.* **4**, 73 (1989).

24. S. M. Lundber, S. -I. Lee, A unified approach to interpreting model predictions, *Proceedings of the 31st International Conference on Neural*, **30**, 4768 (2017).

25. M. Born, Über anisotrope Flüssigkeiten. Versuch einer Theorie der flüssigen Kristalle und des elektrischen Kerr-Effekts in Flüssigkeiten. Sitzungsber. *Preuss. Akad Wiss.* **30**, 614 (1916).

26. D. W. Berreman, Optics in Stratified and Anisotropic Media: 4×4-Matrix Formulation, *J. Opt. Soc. Am.* **62**, 502-510 (1972).

27. M., Gil. Preparation of bis(benzenecarboxamides) as calcium channel blockers. PCT Int. Appl., 2007068754 (2007).

28. S. Kawashita. K. Aoyagi, H. Yamanaka, R. Hantani, S. Naruoka, A. Tanimoto, Y. Hori, Y. Toyonaga, K. Fukushima, S. Miyazaki, Y. Hantani. Symmetry-based ligand design and evaluation of small molecule inhibitors of programmed cell death-1/programmed death-ligand 1 interaction. *Bioorg. Med. Chem. Lett* **29**, 2464-2467 (2019).

29. A. K. Al-Lami. Preparation and Mesomorphic Characterization of Supramolecular Hydrogen-Bonded Dimer Liquid Crystals. *Polycycl. Aromat. Compd.* **36**, 197-212 (2016).

30. D. Ndaya, R. Bosirea, R. M. Kasi. Cholesteric–azobenzene liquid crystalline copolymers: design, structure and thermally responsive optical properties. *Polym. Chem.* **10**, 3868-3878 (2019).

31. J. Pecyna, P. Kaszynski, B. Ringstrand, M. Bremer. Investigation of high Δε derivatives of the [closo-1-CB9H10]− anion for liquid crystal display applications. *J. Mater. Chem. C* **2**, 2956-2964 (2014).

32. M.-J. Gim, D. A. Beller, D. K. Yoon. Morphogenesis of liquid crystal topological defects during the nematic-smectic A phase transition. *Nat. Commun.* **8**, 15453 (2017).

33. A. Liaw, M. Wiener. Classification and Regression by random Forest. *R News*, ISSN 1609-3631.

34. C. Wu, X. Huang, X. Wu, J. Yu, L. Xie, P. Jiang. TiO2-nanorod decorated carbon nanotubes for high-permittivity and low-dielectric-loss polystyrene composites. *Compos. Sci. Technol*. **72**, 521–527 (2012).

35. Y. Chen, Q. Zhuang, X. Liu, J. Liu, S. Lin, Z. Han. Preparation of thermostable PBO/graphene nanocomposites with high dielectric constant. *Nanotechnology* **24**, 245702 (2013).

36. K. Hayashida. Dielectric properties of polymethacrylate-grafted carbon nanotube composites. *RSC Adv.* **3**, 221–227 (2013).

37. L. J. Romasanta, M. Hernández, M. A. López-Manchado, R. Verdejo. Functionalised graphene sheets as effective high dielectric constant fillers. *Nanoscale Res. Lett.* **6**, 1–6 (2011).

38. M. Panda, V. Srinivas, A. K. Thakur. On the question of percolation threshold in polyvinylidene fluoride/nanocrystalline nickel composites. *Appl. Phys. Lett.* **92**, 12–15 (2008).

39. J. Yuan, S. Yao, W. Li, A. Sylvestre, J. Bai. Vertically aligned carbon nanotube arrays on SiC microplatelets: A high figure-of-merit strategy for achieving large dielectric constant and low loss in polymer composites. *J. Phys. Chem. C* **118**, 22975–22983 (2014).

40. A. B. Dichiara, J. Yuan, S. Yao, A. Sylvestre, L. Zimmer, J. Bai. Effective synergistic effect of Al2O3 and SiC microparticles on the growth of carbon nanotubes and their application in high dielectric permittivity polymer composites. *J. Mater. Chem. A* **2**, 7980–7987 (2014).

41. C. Han, A. Gu, G. Liang, L. Yuan. Carbon nanotubes/cyanate ester composites with low percolation threshold, high dielectric constant and outstanding thermal property. *Compos. Part A Appl. Sci. Manuf.* **41**, 1321–1328 (2010).

42. J. Y. Kim, T. Y. Kim, J. W. Suk, H. Chou, J.-H. Jang, J. H. Lee, I. N. Kholmanov, D. Akinwande, R. S. Ruoff. Enhanced dielectric performance in polymer composite films with carbon nanotube-reduced graphene oxide hybrid filler. *Small* **10**, 3405–3411 (2014).

43. G. Tian, J. Song, J. Liu, S. Qi, D. Wu. Enhanced dielectric permittivity and thermal stability of graphene-polyimide nanohybrid films. *Soft Mater* **12**, 290–296 (2014).

44. C. R. Yu, D.-M. Wu, Y. Liu, H. Qiao, Z.-Z. Yu, A. Dasari, X. Du, Y.-W. Mai. Electrical and dielectric properties of polypropylene nanocomposites based on carbon nanotubes and barium titanate nanoparticles. *Compos. Sci. Technol.* **71**, 1706–1712 (2011).

45. A. Ameli, S. Wang, Y. Kazemi, C. B. Park, P. Pötschke. A facile method to increase the charge storage capability of polymer nanocomposites. *Nano Energy* **15**, 54–65 (2015).

46. P. Kim, N. M. Doss, J. P. Tillotson, P. J. Hotchkiss, M.-J. Pan, S. R. Marder, J. Li, J. P. Calame, J. W. Perry. High energy density nanocomposites based on surface-modified BaTiO3 and a ferroelectric polymer. *ACS Nano* **3**, 2581–2592 (2009).

47. Yu, K., Wang, H., Zhou, Y., Bai, Y. & Niu, Y. Enhanced dielectric properties of BaTiO3/poly(vinylidene fluoride) nanocomposites for energy storage applications. *J. Appl. Phys*. **113**, 034105 (2013).

48. K. S. Lam, Y. W. Wong, L. S. Tai, Y. M. Poon, F. G. Shin. Dielectric and pyroelectric properties of lead zirconate titanate/polyurethane composites. *J. Appl. Phys.* **96**, 3896–3899 (2004).

49. L. Zhang, X. Shan, P. Wu, Z. Y. Cheng. Dielectric characteristics of CaCu3Ti4O12/P(VDF-TrFE) nanocomposites. *Appl. Phys. A Mater. Sci. Process.* **107**, 597–602 (2012).

50. H. Tang, Y. Lin, H. A. Sodano. Synthesis of high aspect ratio batio3 nanowires for high energy density nanocomposite capacitors. *Adv. Energy Mater.* **3**, 451–456 (2013).

51. M. Arbatti, X. Shan, Z. Cheng. Ceramic-polymer composites with high dielectric constant. *Adv. Mater.* **19**, 1369–1372 (2007).

52. Y. Jin, N. Xia, R. A. Gerhardt. Enhanced dielectric properties of polymer matrix composites with BaTiO3 and MWCNT hybrid fillers using simple phase separation. *Nano Energy* **30**, 407–416 (2016).

53. S. Luo, S. Yu, R. Sun, C. P. Wong. Nano Ag-deposited BaTiO3 hybrid particles as fillers for polymeric dielectric composites: Toward high dielectric constant and suppressed Loss. *ACS Appl. Mater. Interfaces* **6**, 176–182 (2014).

54. M. A. Subramanian, A. W. Sleight. ACu3Ti4O12 and ACu3Ru4O12 perovskites: High dielectric constants and valence degeneracy. *Solid State Sci.* **4**, 347–351 (2002).

55. R. Kumari, R. Seera, A. De, R. Ranjan, T. N. Guru Row. Organic Multifunctional Materials: Second Harmonic, Ferroelectric, and dielectric properties in N-benzylideneaniline analogues. *Cryst. Growth Des.* **19**, 5934–5944 (2019).

56. S. Horiuchi, F. Ishii, R. Kumai, Y. Okimoto, H. Tachibana, N. Nagaosa, Y. Tokura. Ferroelectricity near room temperature in co-crystals of nonpolar organic molecules. *Nat. Mater.* **4**, 163–166 (2005).

57. M. J. Weber. Handbook of Optical Materials 1.9.3 Second Harmonic Generation Coefficients (CRC Press, Boca, Raton, London, New York, Washington, D.C., 2003).

58. J. Link, J. Fontanella, C. G. Andeen. Temperature variation of the dielectric properties of bismuth germanate and bismuth germanium oxide. *J. Appl. Phys.* **51**, 4352–4355 (1980).

59. M. Delfino, G. M. Loiacono, W. A. Smith. Thermal and dielectric properties of LiKSO4 and LiCsSO4. *J. Solid State Chem.* **31** 131–134 (1980).

60. A. Chevy, A. Segura, V. Muñoz. Effects of pressure and temperature on the dielectric constant of gas, gase, and inse: role of the electronic contribution. *Phys. Rev. B - Condens. Matter Mater. Phys.* **60,** 15866–15874 (1999).

61. D. Berlincourt, H. Jaffe, L. R. Shiozawa. Electroelastic properties of the sulfides, selenides, and tellurides of zinc and cadmium. *Phys. Rev.* **129**, 1009–1017 (1963).

62. R. R. Neurgaonkar, W. K. Cory. Progress in photorefractive tungsten bronze crystals. *J. Opt. Soc. Am. B* **3**, 274 (1986).

63. A. U. Sheleg, V. G. Hurtavy. The Influence of Electron Irradiation on the Dielectric Characteristics of Single Crystals of AgGaSe2. *Phys. Solid State* **61**, 1695–1698 (2019).

64. R. J. Pollina, C. W. Garland. Dielectric and ultrasonic measurements in CsH2AsOs. *Phys. Rev. B* **12**, 362–367 (1975).

65. A. De, K. V. Rao. Dielectric properties of synthetic quartz crystals. *J. Mater. Sci.* **23**, 661–664 (1988).

66. C. R. Raja, R. Gobinathan, F. D. Gnanam. Dielectric Properties of Beta Barium Borate and Potassium Pentaborate Single Crystals. *Cryst. Res. Technol.* **28**, 737–743 (1993).

67. S. W. Ko, D. A. Mourey, T. Clark, S. Trolier-Mckinstry. Synthesis, characterisation, and dielectric properties of β-Gd2(MoO4)3 thin films prepared by chemical solution deposition. *J. Sol-Gel Sci. Technol.* **54**, 269–275 (2010).

68. M. Zgonik, R. Schlesser, I. Biaggio, E. Voit, J. Tscherry, P. Günter. Materials constants of KNbO3 relevant for electro- and acousto-optics. *J. Appl. Phys.* **74**, 1287–1297 (1993).

69. K. Gesi. Electrical properties of NaNO2 single crystal in the vicinity of the ferroelectric curie temperature. *J. Phys. Soc. Japan* **22**, 979–986 (1967).

70. A. El Ghandouri, S. Sayouri, T. Lamcharfi, A. Elbasset. Structural, microstructural and dielectric properties of Ba1-xLaxTi(1-x/4)O3 prepared by sol gel method. *J. Adv. Dielectr.* **9**, 1–19 (2019).

71. K. F. Young, H. P. R. Frederikse. Compilation of the Static Dielectric Constant of Inorganic Solids. *J. Phys. Chem. Ref. Data* **2**, 313–410 (1973).

72. M. Anis, S. P. Ramteke, M. D. Shirsat, G. G. Muley, M. I. Baig. Novel report on γ-glycine crystal yielding high second harmonic generation efficiency. *Opt. Mater.* **72**, 590–595 (2017).

73. A. N. Holden, W. J. Merz, J. P. Remeika, B. T. Matthias. Properties of guanidine aluminum sulfate hexahydrate and some of its isomorphs. *Phys. Rev.* **101**, 962–966 (1956).

74. J. W. Williams, C. H. Schwrzngel. The dielectric constants of binary mixtures. VI The electric moments of certain nitro derivatives of benzene and toluene. *J. Am. Chem. Soc.* **50**, 362–368 (1928).

75. G. Q. Mei, H. Y. Zhang, W. Q. Liao. A symmetry breaking phase transition-triggered high-temperature solid-state quadratic nonlinear optical switch coupled with a switchable dielectric constant in an organic-inorganic hybrid compound. *Chem. Commun.* **52**, 11135–11138 (2016).

76. J. Zhang, S. Han, X. Liu, Z. Wu, C. Ji, Z. Sun, J. Luo. A lead-free perovskite-like hybrid with above-room-temperature switching of quadratic nonlinear optical properties. *Chem. Commun.* **54**, 5614–5617 (2018).

77. W. Q. Liao, J. X. Gao, X. N. Hua, X. G. Chen, Y. Lu. Unusual two-step sequential reversible phase transitions with coexisting switchable nonlinear optical and dielectric behaviors in [(CH3)3NCH2Cl]2[ZnCl4]. *J. Mater. Chem. C* **5**, 11873–11878 (2017).

78. S. P. Ramteke, M. I. Baig, M. Shkir, S. Kalainathan, M. D. Shirsat, G. G. Muley, M. Anisa. Novel report on SHG efficiency, Z-scan, laser damage threshold, photoluminescence, dielectric and surface microscopic studies of hybrid inorganic ammonium zinc sulphate hydrate single crystal. *Opt. Laser Technol.* **104**, 83–89 (2018).

79. I. Khan, M. Anis, U. Bhati. Influence of L-lysine on optical and dielectric traits of cadmium thiourea acetate complex crystal. *Optik.* **170**, 43–47 (2018).

## Acknowledgments

This work is supported by Guangdong Provincial Key Laboratory of Functional and Intelligent Hybrid Materials and Devices (No. 2019B121203003), The Recruitment Program of Guangdong (No. 2016ZT06C322), the National Science Foundation of China for Young Scientists of China (NSFC No. 51890871), International (Regional) Cooperation And Exchange Project (NSFC No. 12050410231), the Fundamental Research Funds for the Central Universities (No. 2019JQ05).

**Author contributions:** S.A. and M.H. designed and directed the research. J.L. and M.H. synthesized all the materials. J.Z., H.N. and S.A. made structural and viscoelastic analyses. H.N., J.K. and S.A. made measurements and analyses of dielectric properties. J.K. and S.A. built measuring system for second harmonic generation and analyzed the data. X.Z. and S.A. made density-functional theory calculation. Y.H. and S.A. wrote the codes for machine learning and made analyses of chemical structures. J.K. and S.A. wrote the codes for generating director field, and calculated the free energy and made optical simulation. S.A. and M.H. wrote the manuscript. All the authors discussed and amended about the manuscript.

**Competing interests:** There are no competing interests.

**Data and materials availability:** All data that support the findings in this study are available in the article and in Supplementary Materials. Additional information is available from the corresponding author upon reasonable request.

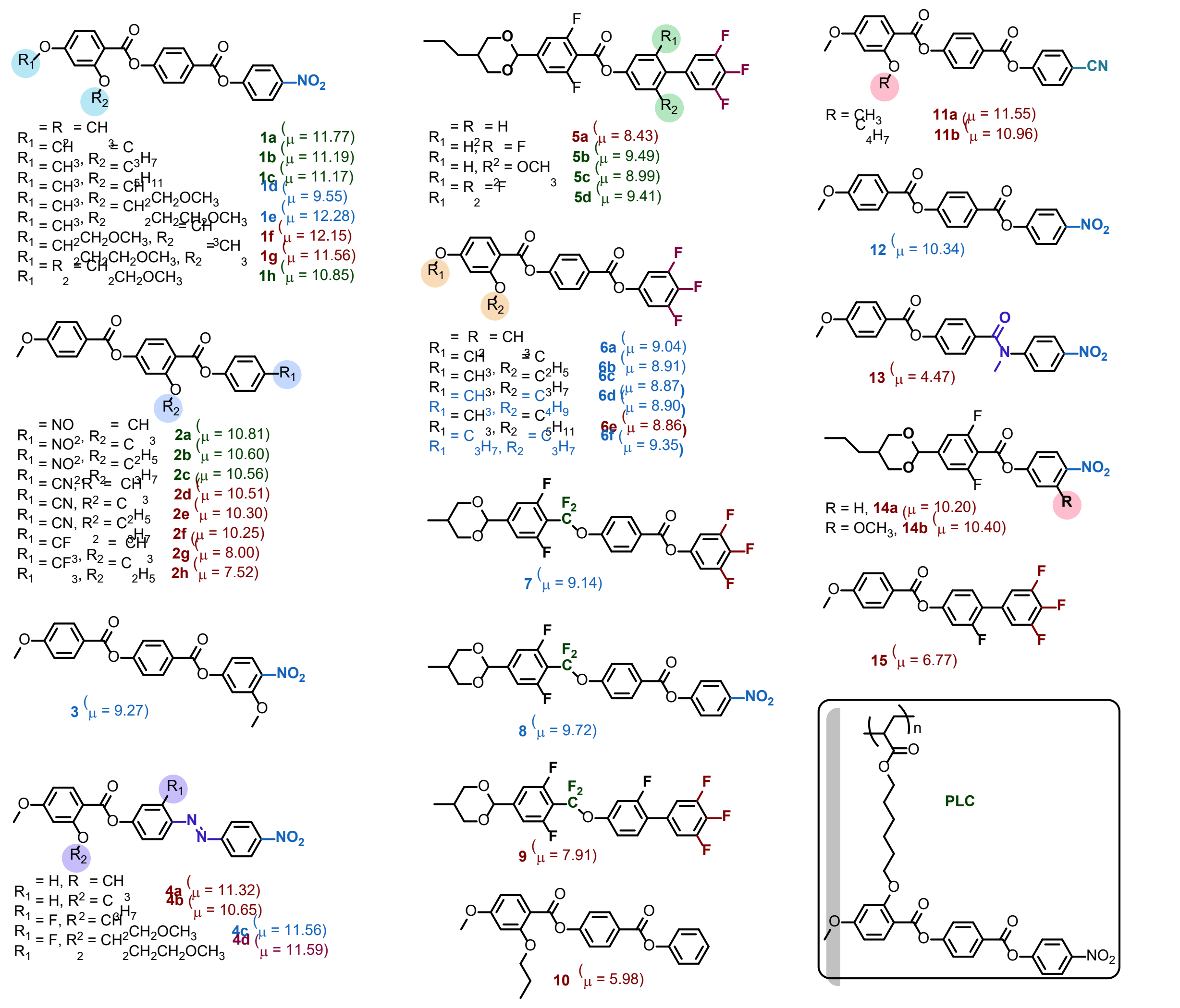

**Fig. 1 Synthetic molecular library.** All the synthesized materials are shown. The dipole moment values calculated by DFT are denoted. The compounds labeled by green color show stable PN phase upon cooling process; the blue color label indicates observed metastable PN phase; the red color label indicated no PN phase observed at all temperatures upon cooling.

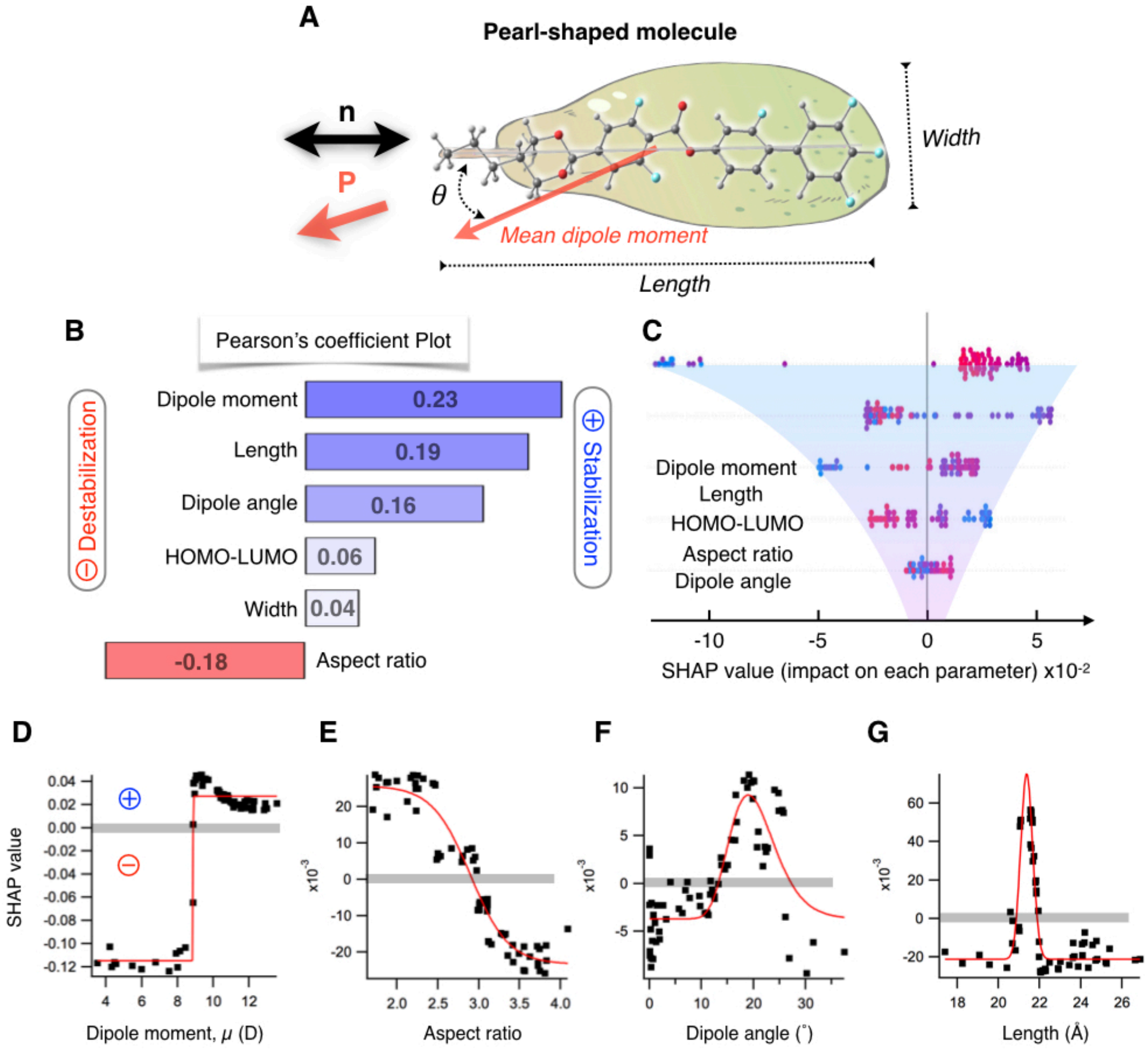

**Fig. 2 Machine-learning-driven understanding of molecular parameters stabilizing the PN phase.** (A) Molecular geometrical properties. As the model molecule, **5b** is shown. (B) Comparison of Pearson's correlation coefficient to the incidence of the PN phase. Higher positive values mean higher probability to stabilize the PN phase, and vice versa. (C) The scattered plots of SHAP values for each molecular parameter. (D-G) The SHAP values as a function of the magnitude of each molecular parameter. The red lines are fitting curves by Hill equation. The crossover values from the negative to the positive regimes indicate the threshold values for stabilizing the PN phase.

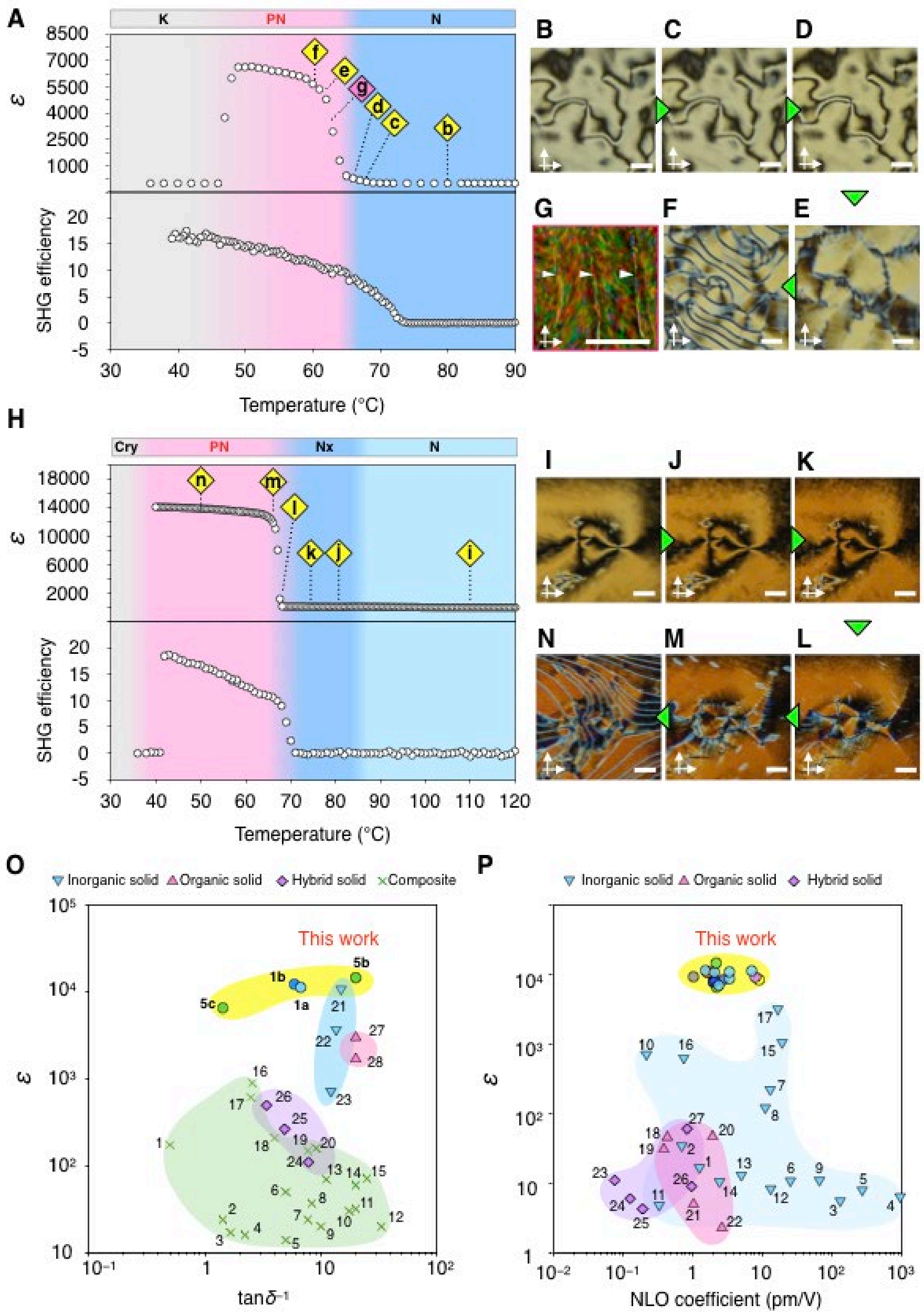

**Fig. 3 Systematic studies of structure and polarity of the PN phase POM, SHG and dielectric spectroscopy.** The corresponding POM images for **5c** (A-G) and **5b** (H-N) taken during cooling at various temperatures, indicated in (A) and (H), respectively. The striped pattern in the band structure in (G), observed on heating, is considered as the embodiment of the Splay-Néel type-I PN structure. SH efficiency is defined as the SH intensity ratio of the PN materials to that of the quartz substrate. The cell thickness is 2.7 µm. Scale bars, 20 µm. Roadmaps of both the dielectric constant *vs* the loss tangent, $\tan\delta^{-1}$ (O) and dielectric constant *vs* nonlinear optical properties (P) are shown, respectively.

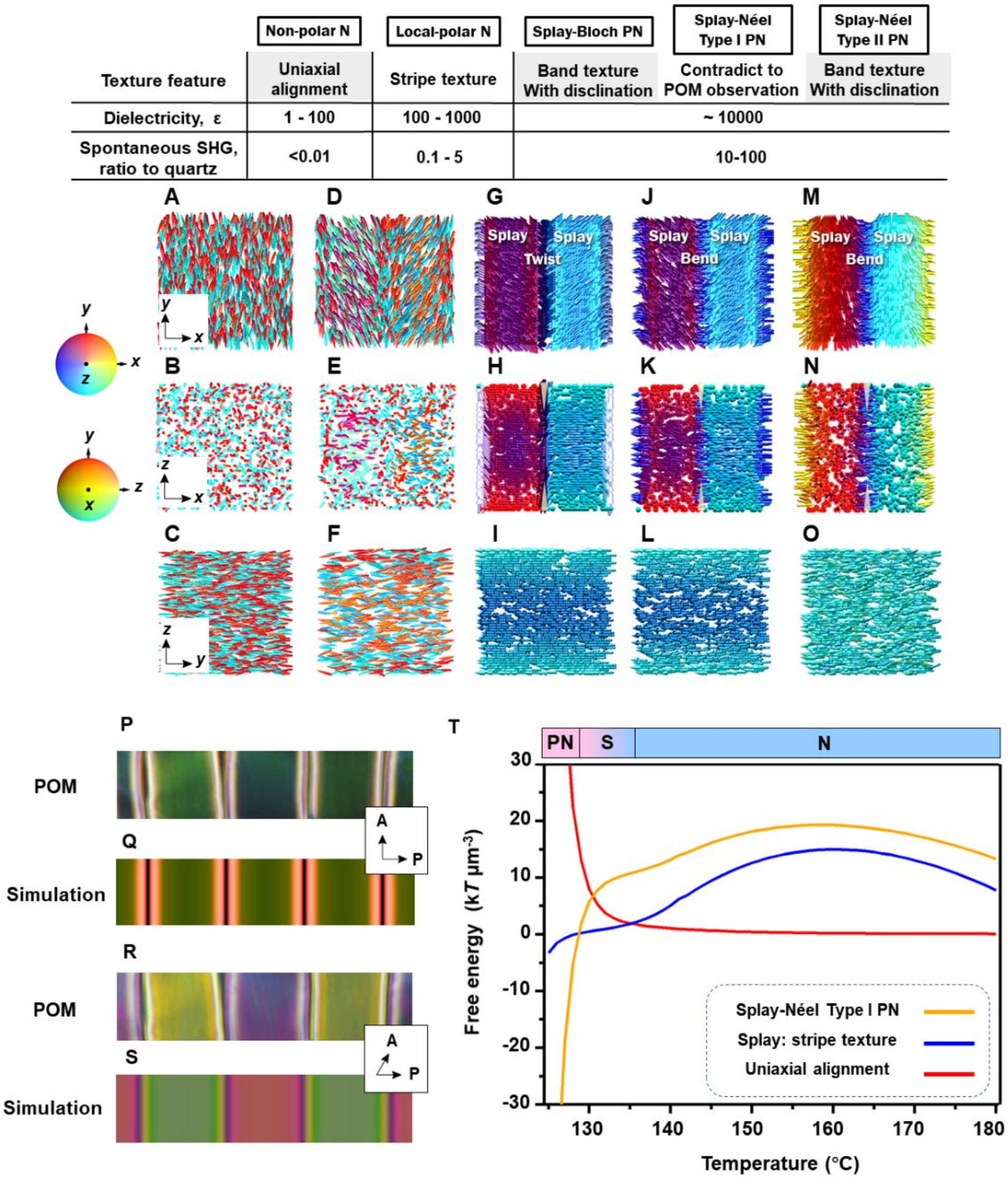

**Fig. 4 Model and the free energy of the PN phase.** (A-O) The cross sections of the director fields of the homogeneous traditional N phase without polarity, the locally-polarized N phase with the splay texture, and three possible structural models for the PN phase in *xy*, *yz* and *xz* planes. *y* axis is parallel to the rubbing direction. The elongated triangles in (H-N) represent the defect walls near the surfaces. The textural feature, dielectricity and SHG signal as the ratio to the SH signal of the quartz crystal are shown. (P-S) The comparison of the experimental and calculated POM images of the Splay-Néel type I PN structure under crossed and decrossed polarizers. The POM observation is made in a 5-μm cell. Image width, 145 μm. (T) The free energies of different packing models for **1a** as a function of temperature. The symbols, N, S and PN mean the nematic, splay and the polar nematic states.

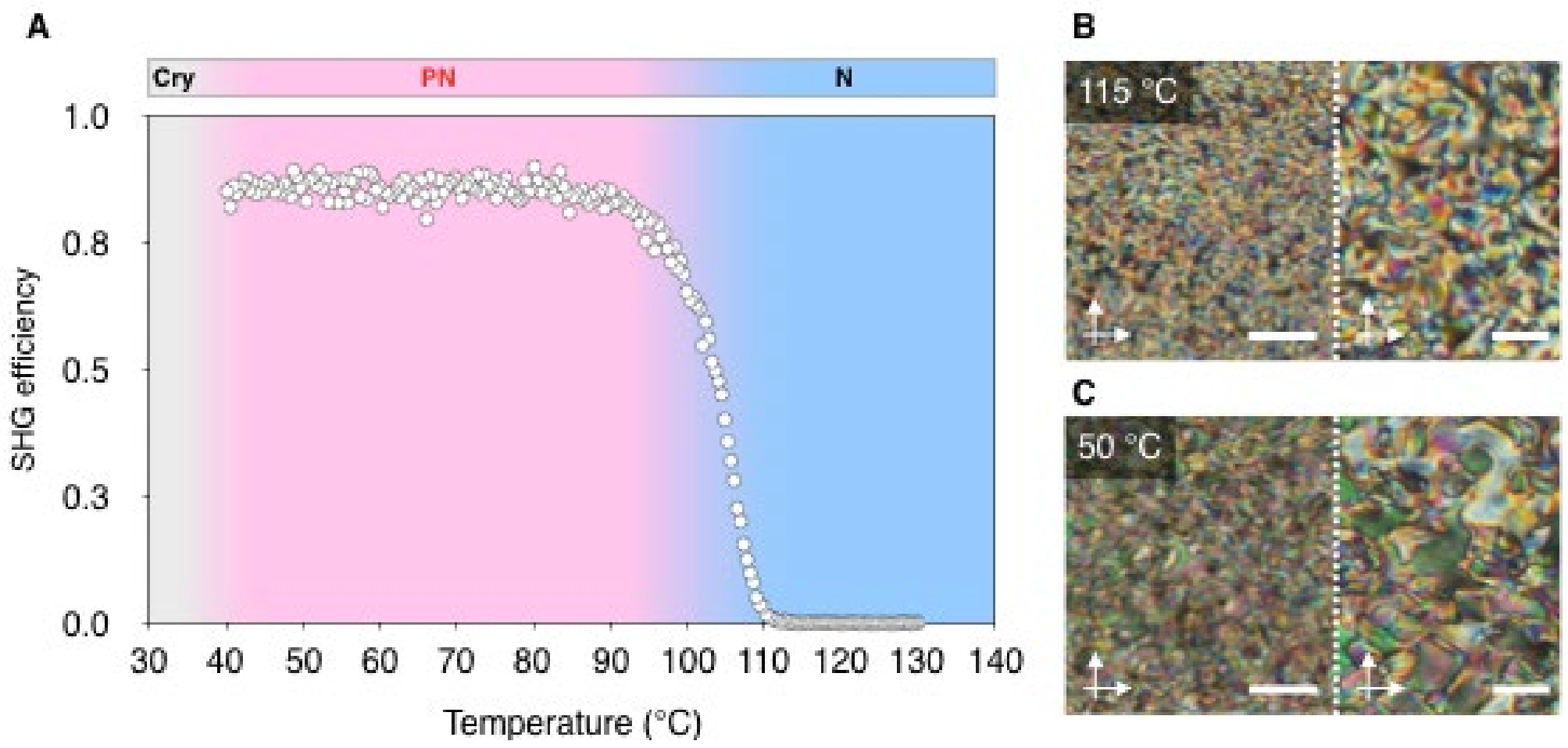

**Fig. 5 Polar nematic polymer material.** (A) SH efficiency as a function of temperature. POM texture of polar polymer at 115 °C in the N phase (B) and at 50 °C PN phases (C) under crossed polarizers. Scale bar, 50 µm (left), 20 µm (right).

# Supplementary Materials for

## Development of polar nematic fluids with giant-κ dielectric properties

Jinxing Li[1]†, Hiroya Nishikawa[2]†, Junichi Kougo[1]†, Junchen Zhou[1], Shuqi Dai[1], Wentao Tang[1], Xiuhu Zhao[1], Yuki Hisai[3], Mingjun Huang[1]*, Satoshi Aya[1]*.

Correspondence to: huangmj25@scut.edu.cn (M.H.) and satoshiaya@scut.edu.cn (S.A.)

**This PDF file includes:**

Materials and Methods
Discussions S1 to S5
Figs. S1 to S13
Tables S1 to S4
References

Materials and Methods

Materials

All commercial chemicals and solvents were used as received, unless stated otherwise. diethyl 2-propylmalonate, benzaldehyde, p-toluenesulfonic acid (p-TsOH) were obtained from Acros. 3,4-dihydro-2H-pyran, N-(3-dimethylaminopropyl)-N′-ethylcarbodiimide hydrochloride (EDC·HCl), and 4-dimethylaminopyridine (DMAP) were obtained from Sigma-Aldrich. Tetrahydrofuran (THF, Energy Chemical). Dichloromethane (DCM, Energy Chemical), Petroleum ether (PE, Energy Chemical) Ethyl acetate (EA, Energy Chemical), Methanol (MeOH, Energy Chemical, reagent grade), N,N-Dimethylformamide (DMF, Energy Chemical, reagent grade), diethyl 2-propylmalonate (98%, TCI), benzaldehyde (Sigma-Aldrich), 2-propylpropane-1,3-diol (98%, Energy Chemical), 1-bromo-2-fluoro-4-methoxybenzene (98%, Energy Chemical), (3,4,5-trifluorophenyl)boronic acid (98%, Innochem), Boron tribromide (17% in Methylene chloride, Adamas), 4-bromo-3-methoxyphenol (99%, TCI), Tetrabutylammonium bromide (TBAB), palladium on carbon (10wt%, Adamas), 4-hydroxybenzoic acid (99%, Adamas), p-toluenesulfonic acid monohydrate (p-TsOH, 98.5%, Aldrich), pyridiniumtoluenesulfonate (PPTS, 98%, Innochem), methyl 2-hydroxy-4-methoxybenzoate (99%, Adamas), methyl 2,4-dimethoxybenzoate (98.5%, Adamas), 1-bromopropane, 1-bromobutane, 1-bromopentane were obtained from Adamas. 4-hydroxybenzoic acid (99%, Adamas), 2,4-dihydroxybenzoic acid (99%, Adamas), 4-nitroaniline (99.5%, Aladdin), phenol (99%, Adamas), methyl 2-hydroxy-4-methoxybenzoate (99.5%, TCI), Triethylamine ($Et_3N$, Energy Chemical, reagent grade), Dimethyl sulfoxide (DMSO, Energy Chemical, reagent grade), 2,4-dihydroxybenzaldehyde (99%, Adamas), 2-hydroxy-4-methoxybenzaldehyde (99%, Adamas), Iodomethane (99.5%, Aladdin), Diisopropylcarbodiimide solution (DIPC, 99%, Adamas), 6-bromohexan-1-ol (98%, Energy Chemical), Acryloyl chloride (99%, Adamas), 2,4-dihydroxybenzaldehyde (99%, Adamas).

Methods

**General Williamson etherification method**: The substituted phenol, halogenated alkyl ether (> 1 equiv), potassium carbonate (2 equiv) and sodium iodide (1 equiv) were suspended in DMF and heated at 100 °C with vigorous stirring until complete consumption of the phenol as judged by TLC (18 - 24 h). The reaction was then cooled. Water was added to the mixture, extracted with EA, and washed by brine. The organic phase was dried by anhydrous $MgSO_4$. After the solvent was removed, the residue was collected as product without further purification.

**General ester hydrolysis method**: The ester (1 equiv) and potassium hydroxide (> 2 equiv) were dissolved into ethanol/water (highest water concentration without precipitation) and heated under reflux until the complete consumption of the ester (TLC). The reaction was cooled, acidified with 2M HCl. The precipitate was collected and dried. If required the precipitate was recrystallized from an appropriate solvent system. General yield ~96%.

**General aldehyde oxidation method**: Sodium dihydrogen phosphate (4 equiv) and sodium chlorite (3.5 equiv) were added to a stirred solution of aldehyde (1 equiv) in a mixture of Vol (DMSO)/Vol ($H_2O$) = 4/3 at 0°C. The mixture was allowed to warm to room temperature and stir for 6 hours. The mixture was diluted with water and solid $NaHCO_3$ was added to adjust the pH of the solution to 8. The solution was washed with ethyl acetate, then the pH was adjusted to 4 by the addition of 1M HCl solution and extracted with ethyl acetate. The combined organic extracts ware washed with brine, dried ($NaSO_4$) and evaporated under reduced pressure to give the compound. General yield 95%.

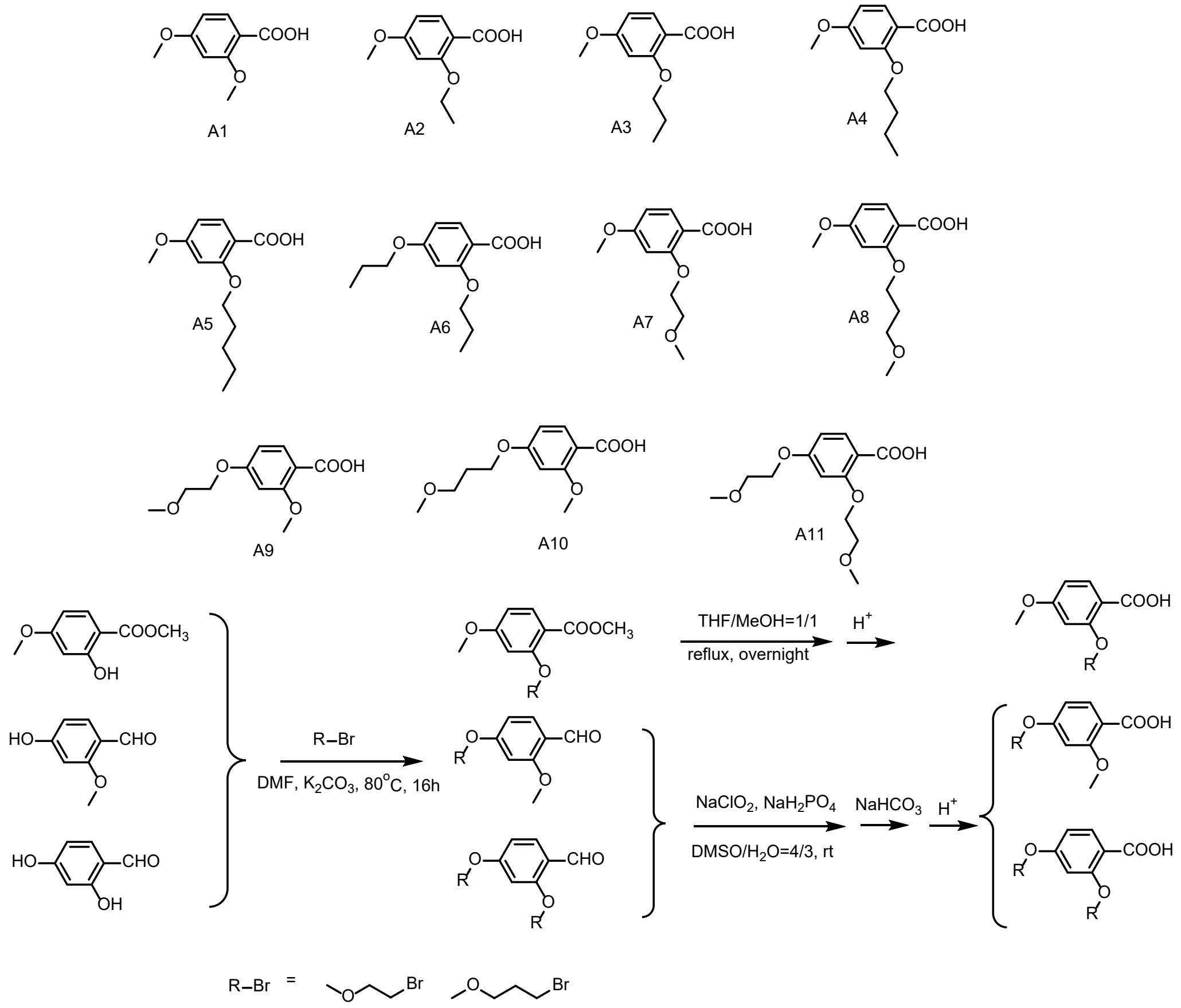

Among them, compound A1, A2, A3, A4 were purchased from Beijing J&K Scientific. Compound A5 was synthesized with reference to a patent (*27*). Compound A6 was purchased from Jiangsu Aikon Biopharmaceutical. The benzoate precursor form of A7 and A8 were synthesized by general Williamson etherification method. After hydrolysis using general ester hydrolysis method, A7 and A8 were acquired. A9, A10, and A11 were synthesized by first preparing their benzaldehyde form using general Williamson etherification method, followed by the general aldehyde oxidation method to make the carboxylic acid form.

**Methyl 4-methoxy-2-(2-methoxyethoxy)benzoate**: $^1$H NMR (500 MHz, Chloroform-*d*) δ 7.88 – 7.77 (m, 1H), 6.53 – 6.40 (m, 2H), 4.14 (dd, *J* = 7.0, 2.5 Hz, 2H), 3.81 (dt, *J* = 8.5, 2.0 Hz, 6H), 3.52 – 3.42 (m, 3H). $^{13}$C NMR (126 MHz, Chloroform-*d*) δ 166.12, 164.12, 160.62, 133.83, 112.88, 105.36, 70.90, 68.87, 59.41, 55.47, 51.61.
**4-methoxy-2-(2-methoxyethoxy)benzoic acid (A7)**: $^1$H NMR (500 MHz, Chloroform-*d*) δ 8.11 (d, *J* = 8.8 Hz, 1H), 6.64 (dd, *J* = 8.8, 2.3 Hz, 1H), 6.51 (d, *J* = 2.3 Hz, 1H), 4.36 – 4.27 (m, 2H), 3.86 (s, 3H), 3.82 – 3.74 (m, 2H), 3.45 (s, 3H). $^{13}$C NMR (126 MHz, Chloroform-*d*) δ 165.35, 164.89, 158.80, 135.43, 111.19, 107.19, 100.09, 69.86, 69.15, 59.16, 55.74.

**Methyl 4-methoxy-2-(3-methoxypropoxy)benzoate**: $^1$H NMR (500 MHz, Chloroform-*d*) δ 7.90 – 7.76 (m, 1H), 6.48 (dd, *J* = 6.7, 2.2 Hz, 2H), 4.15 – 4.02 (m, 2H), 3.88 – 3.77 (m, 6H), 3.61 (td, *J* = 6.0, 1.7 Hz, 2H), 3.38 – 3.30 (m, 3H), 2.12 – 2.05 (m, 2H).$^{13}$ C NMR (126 MHz, Chloroform-*d*) δ 166.25, 164.18, 160.76, 133.76, 112.49, 112.47, 104.92, 99.80, 69.06, 65.65, 58.70, 55.43, 51.55, 29.48.
**4-Methoxy-2-(3-methoxypropoxy)benzoic acid (A8)**: $^1$H NMR (500 MHz, Chloroform-*d*) δ 8.14 (d, *J* = 8.8 Hz, 1H), 6.63 (dd, *J* = 8.8, 2.3 Hz, 1H), 6.51 (d, *J* = 2.3 Hz, 1H), 4.30 (t, *J* = 5.9 Hz, 2H), 3.87 (s, 3H), 3.67 – 3.53 (m, 2H), 3.37 (s, 3H), 2.16 (p, *J* = 5.8 Hz, 2H). $^{13}$C NMR (126 MHz, Chloroform-*d*) δ 165.46, 164.96, 158.94, 135.62, 110.73, 106.42, 99.00, 70.02, 68.13, 58.90, 55.71, 29.16.

**2-Methoxy-4-(2-methoxyethoxy)benzaldehyde**: $^1$H NMR (400 MHz, Chloroform-*d*) δ 10.28 (s, 1H), 7.79 (d, *J* = 8.6 Hz, 1H), 6.60 – 6.45 (m, 2H), 4.23 – 4.16 (m, 2H), 3.89 (s, 3H), 3.80 – 3.74 (m, 2H), 3.46 (s, 3H). $^{13}$C NMR (126 MHz, Chloroform-*d*) δ 188.38, 165.39, 163.56, 130.69, 130.67, 119.16, 119.14, 106.03, 98.78, 98.76, 70.71, 67.62, 59.25, 55.63.
**2-Methoxy-4-(2-methoxyethoxy)benzoic acid (A9)**: $^1$H NMR (400 MHz, Chloroform-*d*) δ 10.49 (s, 1H), 8.13 (d, *J* = 8.7 Hz, 1H), 6.70 – 6.57 (m, 2H), 4.24 – 4.16 (m, 2H), 4.03 (s, 3H), 3.81 – 3.74 (m, 2H), 3.46 (s, 3H). $^{13}$C NMR (126 MHz, Chloroform-*d*) δ 165.19, 164.30, 159.41, 135.60, 110.60, 106.79, 99.54, 70.69, 67.75, 59.29, 56.65, 42.69.

**2-Methoxy-4-(3-methoxypropoxy)benzaldehyde**: $^1$H NMR (400 MHz, Chloroform-*d*) δ 10.25 (s, 1H), 7.76 (d, *J* = 8.7 Hz, 1H), 6.52 (d, *J* = 8.7 Hz, 1H), 6.44 (s, 1H), 4.11 (t, *J* = 6.4 Hz, 2H), 3.87 (s, 3H), 3.53 (t, *J* = 6.0 Hz, 2H), 3.33 (s, 3H), 2.04 (p, *J* = 6.2 Hz, 2H). $^{13}$C NMR (126 MHz, Chloroform-*d*) δ 188.34, 165.70, 163.65, 130.68, 118.92, 106.43, 98.23, 68.88, 65.32, 58.75, 55.60, 29.44.
**2-Methoxy-4-(3-methoxypropoxy)benzoic acid (A10)**: $^1$H NMR (400 MHz, Chloroform-*d*) δ 10.47 (s, 1H), 8.12 (d, *J* = 8.7 Hz, 1H), 6.69 – 6.57 (m, 2H), 4.23 – 4.16 (m, 2H), 4.03 (s, 3H), 3.81 – 3.74 (m, 2H), 3.46 (s, 3H). $^{13}$C NMR (126 MHz, Chloroform-*d*) δ 165.27, 164.61, 159.50, 135.54, 110.23, 107.29, 98.89, 68.83, 65.47, 58.78, 56.62, 42.69, 29.41.

**2,4-Bis(2-methoxyethoxy)benzaldehyde**: $^1$H NMR (500 MHz, Chloroform-*d*) δ 10.30 (s, 1H), 7.76 (d, *J* = 8.7 Hz, 1H), 6.52 (d, *J* = 10.8 Hz, 1H), 6.48 (d, *J* = 2.2 Hz, 1H), 4.20 – 4.10 (m, 4H), 3.79 – 3.69 (m, 4H), 3.41 (s, 6H). $^{13}$C NMR (126 MHz, Chloroform-*d*) δ 188.30, 165.23, 162.90, 130.27, 119.35, 106.50, 99.68, 70.72, 70.70, 68.15, 67.63, 59.32, 59.25.
**2,4-Bis(2-methoxyethoxy)benzoic acid (A11)**: $^1$H NMR (500 MHz, Chloroform-*d*) δ 8.08 (d, *J* = 8.8 Hz, 1H), 6.64 (d, *J* = 11.1 Hz, 1H), 6.58 (d, *J* = 2.3 Hz, 1H), 4.32 – 4.26 (m, 2H), 4.19 – 4.13 (m, 2H), 3.81 – 3.72 (m, 4H), 3.43 (d, *J* = 1.2 Hz, 6H). $^{13}$C NMR (126 MHz, Chloroform-*d*) δ 165.33, 164.05, 158.73, 135.39, 111.38, 107.45, 100.88, 70.68, 69.85, 69.11, 67.74, 59.27, 59.15, 42.68.

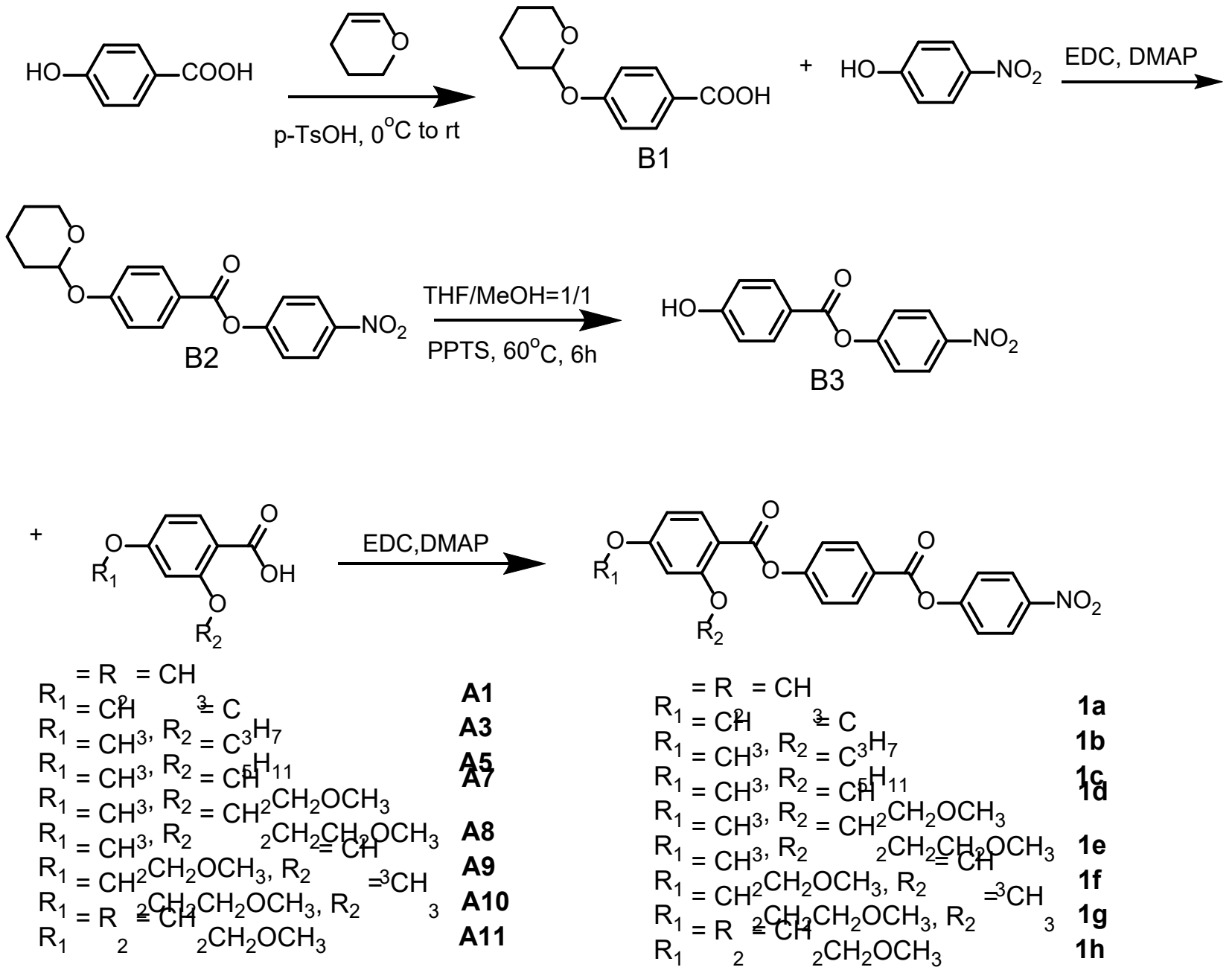

**4-((tetrahydro-2H-pyran-2-yl)oxy)benzoic acid (B1):** A round bottom flask was charged with 4-hydroxybenzoic acid (2.76 g, 0.02mol), para-toluenesulfonic acid (1.96 g, 0.0103 mol), and 20 mL diethyl ether forming a dispersion solution. The solution was degassed by bubbling nitrogen through for 5 min at ice bath. Then 3,4- Dihydro-2H-pyran (2.8 mL, 0.0307 mol) was added dropwise via injector. The solution was slowly warmed to room temperature, and stirred under nitrogen for 5-6 h. The precipitate was then collected by filtration, washed with 20 mL diethyl ether (as little as possible), and dried in a vacuum oven. Yield: 2.89 g (69.3%); appearance: white powder. $^1$H NMR (400 MHz, Chloroform-*d*) δ 8.06 (d, *J* = 8.7 Hz, 2H, ArH), 7.10 (d, *J* = 8.6 Hz, 2H, ArH), 5.53 (q, *J* = 2.8 Hz, 1H, CH), 3.86 (d, *J* = 21.0 Hz, 1H, $CH_2$), 3.63 (d, *J* = 11.2 Hz, 1H, $CH_2$), 2.07 – 1.50 (m, 6H, $CH_2$). $^{13}$C NMR (126 MHz, Chloroform-*d*) δ 132.17, 115.96, 94.70, 62.97, 30.71, 25.46, 25.34, 19.82, 19.77, 19.65.

**4-nitrophenyl 4-((tetrahydro-2H-pyran-2-yl)oxy)benzoate (B2):** A round bottom flask was charged with **B1** (10 g, 45 mmol), EDC (10.35 g, 54 mmol), N, N-dimethylaminopyridine (7.51 g, 54 mmol), then 100 mL dichloromethane solvent was added and cooled to 0 °C (reagents mostly insoluble) and flashed with nitrogen. Then 4-nitrophenol (7.51 g, 54 mmol) was added along with 100 mL dichloromethane. The solution was stirred at 0 °C for 1 h then at room temperature for 17 hr. The solution was stripped of solvent by rotary evaporation, and dry-loaded onto a silica gel column for purification using dichloromethane/ethyl acetate as eluent. Yield: 12g (76.8%); appearance: white powder. $^1$H NMR (500 MHz, Chloroform-*d*) δ 8.31 (d, *J* = 9.1 Hz, 2H, ArH), 8.12 (dd, *J* = 17.7, 8.9 Hz, 2H, ArH), 7.40 (d, *J* = 9.2 Hz, 2H, ArH), 7.05 (dd, *J* = 114.9, 8.9 Hz, 2H, ArH), 5.57 (s, 1H, CH), 4.06 – 3.82 (m, 1H, $CH_2$), 3.61 (d, *J* = 55.9 Hz, 1H, $CH_2$), 2.03-1.64 (s, 6H, $CH_2$). $^{13}$C NMR (101 MHz, Chloroform-*d*) δ 163.94, 161.97, 155.95, 145.25, 132.39, 125.20, 122.67, 121.31, 116.30, 96.16, 62.10, 30.05, 18.45.

**4-nitrophenyl 4-hydroxybenzoate (B3):** A round bottom flask was charged with **B2** (1 g, 2.9 mmol), catalyst amount PTSA, 20 mL THF, 20 mL ethanol. The solution was heated at 60 °C and monitored by TLC (6-48 h). The solvents were removed under reduced pressure. Water (80 mL) was added to the residue and extracted with DCM (3 × 100 mL). The organic phase was dried by anhydrous $MgSO_4$. The residue after removing solvent was purified by silicone chromatography, and dried in the vacuum oven. Yield: 0.72 g (95.1%); appearance: faint yellow powder. $^{1}$H NMR (400 MHz, DMSO-$d_6$) δ 10.64 (s, 1H, OH), 8.34 (d, *J* = 9.1 Hz, 2H, ArH), 8.02 (d, *J* = 8.8 Hz, 2H, ArH), 7.58 (d, *J* = 9.1 Hz, 2H, ArH), 6.95 (d, *J* = 8.8 Hz, 2H, ArH). $^{13}$C NMR (101 MHz, Chloroform-*d*) δ 168.81, 168.00, 160.87, 149.79, 137.40, 129.90, 127.51, 123.75, 120.56, 82.59, 82.27, 81.95, 44.97, 44.76, 44.55, 44.34.

**General Esterification Method**: The acid (A1, A3, A5, A7, A8, A9, A10, A11), compound **B3** (0.98 equiv), EDC (1.5 equiv) and DMAP (~ 0.05 equiv) were combined into a dried flask with DCM added. The suspension was vigorously stirred until the formation of the ester was complete (14 – 18h). The solution was stripped of solvent by rotary evaporation, and dry-loaded onto a silica gel column for purification using EA/hexane as eluent. Appearance: white powder. General yield (~80%).
**1a** (~88%)**:** $^{1}$H NMR (500 MHz, Chloroform-d) δ 8.33 (d, J = 9.1 Hz, 2H), 8.25 (d, J = 8.7 Hz, 2H), 8.10 (d, J = 8.7 Hz, 1H), 7.41 (dd, J = 19.6, 8.9 Hz, 4H), 6.62 – 6.52 (m, 2H), 3.92 (d, J = 18.6 Hz, 6H). $^{13}$C NMR (126 MHz, Chloroform-*d*) δ 165.40, 163.63, 162.80, 162.54, 155.99, 155.71, 145.44, 134.65, 131.89, 125.57, 125.30, 122.66, 122.52, 110.37, 105.00, 99.03, 56.06, 55.64.

**1b** (~85 %)**:** $^{1}$H NMR (500 MHz, Chloroform-*d*) δ 8.38 – 8.31 (m, 2H), 8.26 (d, *J* = 8.7 Hz, 2H), 8.06 (d, *J* = 8.8 Hz, 1H), 7.46 – 7.41 (m, 2H), 7.39 (d, *J* = 8.7 Hz, 2H), 6.57 (dd, *J* = 8.8, 2.3 Hz, 1H), 6.53 (d, *J* = 2.2 Hz, 1H), 4.03 (t, *J* = 6.4 Hz, 2H), 3.89 (s, 3H), 1.88 (h, *J* = 7.2 Hz, 2H), 1.07 (t, *J* = 7.4 Hz, 3H). $^{13}$C NMR (126 MHz, Chloroform-*d*) δ 165.26, 163.64, 163.21, 161.95, 156.11, 155.71, 145.44, 134.64, 131.95, 125.53, 125.30, 122.49, 110.64, 105.00, 99.71, 70.43, 55.61, 22.52, 10.62.

**1c** (~82%)**:** $^{1}$H NMR (400 MHz, Chloroform-*d*) δ 8.33 (d, *J* = 9.1 Hz, 2H), 8.26 (d, *J* = 8.7 Hz, 2H), 8.06 (d, *J* = 8.7 Hz, 1H), 7.41 (dd, *J* = 17.2, 8.9 Hz, 4H), 6.59 – 6.54 (m, 1H), 6.52 (d, *J* = 2.2 Hz, 1H), 4.06 (t, *J* = 6.5 Hz, 2H), 3.89 (s, 3H), 1.86 (dt, *J* = 14.5, 6.6 Hz, 2H), 1.49 (dt, *J* = 14.7, 7.1 Hz, 2H), 1.37 (dt, *J* = 14.9, 7.2 Hz, 2H), 0.89 (t, *J* = 7.3 Hz, 3H). $^{13}$C NMR (101 MHz, Chloroform-*d*) δ 165.26, 163.65, 163.27, 161.92, 156.13, 155.72, 145.45, 134.66, 131.94, 125.54, 125.30, 122.67, 122.49, 110.69, 105.00, 99.73, 68.98, 55.61, 28.81, 28.15, 22.39, 14.00.

**1d** (~89%)**:** $^{1}$H NMR (400 MHz, Chloroform-*d*) δ 8.31 – 8.23 (m, 2H), 8.23 – 8.16 (m, 2H), 8.00 (d, *J* = 8.8 Hz, 1H), 7.42 – 7.29 (m, 4H), 6.53 (dd, *J* = 8.8, 2.3 Hz, 1H), 6.49 (d, *J* = 2.3 Hz, 1H), 4.21 – 4.11 (m, 2H), 3.82 (s, 3H), 3.79 – 3.70 (m, 2H), 3.37 (s, 3H). $^{13}$C NMR (101 MHz, Chloroform-*d*) δ 165.22, 163.64, 161.64, 156.10, 155.72, 145.47, 134.62, 131.94, 125.58, 125.31, 122.66, 122.50, 111.07, 105.64, 100.33, 70.86, 68.76, 59.41, 55.64.

**1e** (~78%)**:** $^{1}$H NMR (400 MHz, Chloroform-*d*) δ 8.32 – 8.23 (m, 2H), 8.23 – 8.16 (m, 2H), 8.00 (d, *J* = 8.6 Hz, 1H), 7.41 – 7.26 (m, 4H), 6.54 – 6.45 (m, 2H), 4.10 (t, *J* = 6.2 Hz, 2H), 3.82 (s, 3H), 3.53 (t, *J* = 6.1 Hz, 2H), 3.25 (s, 3H), 2.04 (p, *J* = 6.1 Hz, 2H). $^{13}$C NMR (101 MHz, Chloroform-*d*) δ 165.32, 163.62, 163.06, 161.89, 156.10, 155.71, 145.47, 134.61, 131.98, 125.59, 125.31, 122.66, 122.49, 110.56, 105.33, 99.70, 68.95, 65.72, 58.72, 55.62, 29.46.

**1f** (~82%)**:** $^{1}$H NMR (400 MHz, Chloroform-*d*) δ 8.40 – 8.33 (m, 2H), 8.32 – 8.25 (m, 2H), 8.11 (d, *J* = 8.7 Hz, 1H), 7.51 – 7.36 (m, 4H), 6.69 – 6.55 (m, 2H), 4.29 – 4.18 (m, 2H), 3.96 (s, 3H), 3.87 – 3.77 (m, 2H), 3.50 (s, 3H). $^{13}$C NMR (101 MHz, Chloroform-*d*) δ 164.57, 163.64, 162.80, 162.47, 155.99, 155.72, 145.46, 134.63, 131.90, 125.60, 125.31, 122.67, 122.52, 110.62, 105.25, 99.88, 70.77, 67.64, 59.33, 56.09.

**1g** (~75%)**:** $^{1}$H NMR (400 MHz, Chloroform-*d*) δ 8.31 – 8.23 (m, 2H), 8.22 – 8.14 (m, 2H), 8.01 (d, *J* = 8.7 Hz, 1H), 7.43 – 7.28 (m, 4H), 6.57 – 6.43 (m, 2H), 4.09 (t, *J* = 6.3 Hz, 2H), 3.87 (s, 3H), 3.51 (t, *J* = 6.1 Hz, 2H), 3.31 (s, 3H), 2.02 (p, *J* = 6.2 Hz, 2H). $^{13}$C NMR (101 MHz, Chloroform-*d*) δ 164.89, 163.65, 162.81, 162.58, 156.03, 155.72, 145.46, 134.61, 131.89, 125.56, 125.31, 122.67, 122.53, 110.22, 105.64, 99.37, 68.92, 65.34, 58.80, 56.07, 29.49.

**1h** (~80%)**:** $^{1}$H NMR (400 MHz, Chloroform-*d*) δ 8.29 – 8.23 (m, 2H), 8.21 – 8.16 (m, 2H), 7.98 (dd, *J* = 8.6, 1.9 Hz, 1H), 7.37 – 7.29 (m, 4H), 6.53 (d, *J* = 8.7 Hz, 2H), 4.22 – 4.04 (m, 4H), 3.73 (dt, *J* = 9.7, 4.6 Hz, 4H), 3.38 (d, *J* = 14.2 Hz, 6H). $^{13}$C NMR (126 MHz, Chloroform-*d*) δ 164.37, 163.63, 162.96, 161.54, 156.06, 155.70, 145.45, 134.60, 131.94, 125.31, 122.66, 122.49, 111.20, 105.87, 101.00, 70.82, 70.75, 68.68, 67.63, 59.40, 59.33.

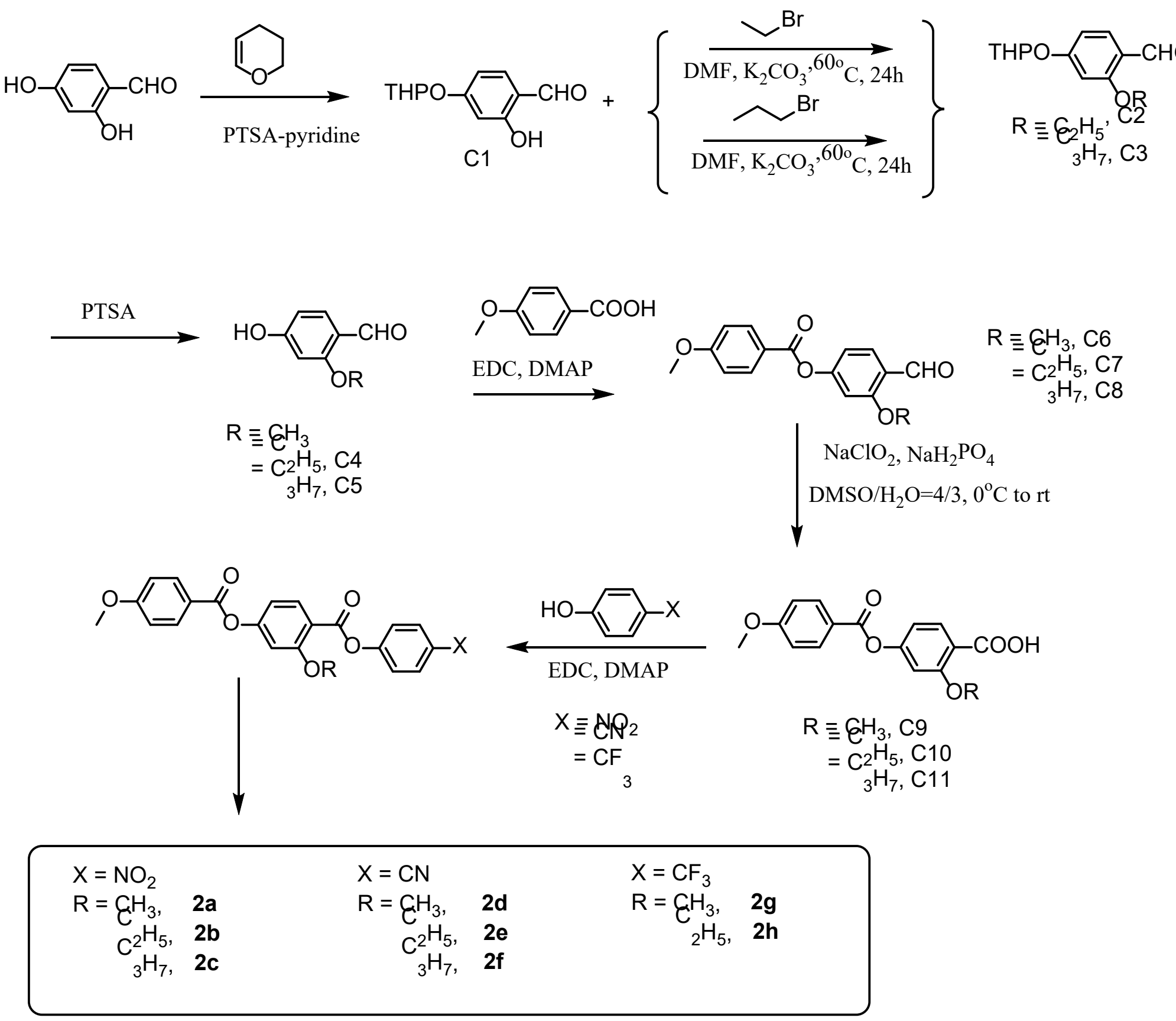

**2-hydroxy-4-((tetrahydro-2H-pyran-2-yl)oxy)benzaldehyde (C1):** The compound was synthesized according to the literature (*28*).

**2-ethoxy-4-hydroxybenzaldehyde (C4):** the compound **C1**, 1-bromoethane (> 1 equiv), potassium carbonate (2 equiv) and sodium iodide (1 equiv) were suspended in DMF and heated at 90 °C with vigorous stirring (18 - 24 h). The reaction was then cooled, followed by adding water into the mixture, extracted with EA, and washed with water and brine. The organic phase was dried by anhydrous $MgSO_4$. After the solvent was removed, the residue **C2** was collected and directly used for next deprotection. A round bottom flask was charged with crude product **C2** (1 equiv), PPTS (1 equiv), THF/ethanol=1/1. The solution was heated at 60 °C (6 - 48 h). The solvents were removed under reduced pressure. Water was added to the residue and extracted with DCM. The organic phase was dried by anhydrous $MgSO_4$, and the solvent-removed residue was purified by chromatography and dried in the vacuum oven. General yield: 95%. $^1$H NMR (400 MHz, Chloroform-*d*) δ 10.27 (d, *J* = 5.4 Hz, 1H), 9.83 (d, *J* = 14.8 Hz, 1H), 7.98 – 7.61 (m, 1H), 6.45 (dd, *J* = 24.8, 7.1 Hz, 2H), 4.36 – 4.01 (m, 2H), 1.68 – 1.40 (m, 3H). $^{13}$C NMR (101 MHz, Chloroform-*d*) δ 188.30, 165.03, 163.59, 130.15, 117.78, 108.55, 63.90, 40.40, 40.20, 39.76, 14.53.

**4-hydroxy-2-propoxybenzaldehyde (C5):** synthesized in similar procedure as in **C4**. $^1$H NMR (400 MHz, Chloroform-*d*) δ 10.29 (d, *J* = 9.7 Hz, 1H), 9.89 (d, *J* = 21.9 Hz, 1H), 7.70 (t, *J* = 8.8 Hz, 1H), 6.46 (dd, *J* = 20.3, 9.9 Hz, 2H), 4.31 – 3.85 (m, 2H), 2.13 – 1.67 (m, 2H), 1.30 – 0.92 (m, 3H). $^{13}$C NMR (101 MHz, Chloroform-*d*) δ 188.13, 165.12, 163.76, 130.02, 117.78, 99.34, 69.69, 40.19, 39.94, 39.71, 22.30, 10.49.

**4-formyl-3-methoxyphenyl 4-methoxybenzoate (C6), 3-ethoxy-4-formylphenyl 4-methoxybenzoate (C7), 4-formyl-3-propoxyphenyl 4-methoxybenzoate (C8)** were synthesized by the General Esterification Method as in the synthetic procedures in **1a-1f**. They are directly used for next oxidation step. C9, C10, and C11 were acquired by the General Aldehyde Oxidation Method used for **A7-A11** synthesis, with Sodium dihydrogen phosphate (4 equiv) and sodium chlorite (3.5 equiv) as the oxidant condition. Two step yield is about 70%.
**C9:** $^1$H NMR (400 MHz, Chloroform-*d*) δ 7.98 (d, *J* = 8.8 Hz, 2H), 7.88 (d, *J* = 8.4 Hz, 1H), 6.85 (d, *J* = 8.8 Hz, 2H), 6.82 – 6.72 (m, 2H), 3.79 (d, *J* = 27.1 Hz, 6H). $^{13}$C NMR (101 MHz, Chloroform-*d*) δ 166.30, 164.12, 159.90, 133.65, 132.29, 121.04, 114.07, 106.04, 56.43, 55.51, 40.19, 39.98, 39.56.

**C10:** $^1$H NMR (500 MHz, Chloroform-*d*) δ 8.28 (d, *J* = 8.6 Hz, 1H), 8.16 (d, *J* = 9.0 Hz, 3H), 7.08 – 6.93 (m, 5H), 4.36 (q, *J* = 7.0 Hz, 2H), 3.93 (s, 4H), 1.60 (t, *J* = 7.0 Hz, 4H). $^{13}$C NMR (101 MHz, Chloroform-*d*) δ 164.12, 158.82, 155.42, 133.71, 132.28, 120.98, 114.43, 113.92, 107.02, 65.59, 55.51, 39.99, 14.49.

**C11:** $^{1}$H NMR (500 MHz, Chloroform-*d*) δ 8.29 (d, *J* = 8.9 Hz, 1H), 8.17 (d, *J* = 8.9 Hz, 2H), 7.03 (d, *J* = 9.5 Hz, 4H), 4.25 (t, *J* = 6.6 Hz, 2H), 3.94 (s, 3H), 1.99 (h, *J* = 7.3 Hz, 3H), 1.13 (t, *J* = 7.4 Hz, 3H). $^{13}$C NMR (101 MHz, Chloroform-*d*) δ 164.72, 164.35, 164.10, 158.32, 156.17, 134.93, 132.48, 120.96, 115.63, 115.11, 114.06, 106.85, 72.13, 55.60, 22.24, 10.38.

**2a-2h** were synthesized by the General Esterification Method used for synthesis of **1a-1f**.
**2a** (~85%)**:** $^{1}$H NMR (500 MHz, Chloroform-*d*) δ 8.40 – 8.31 (m, 2H), 8.24 – 8.10 (m, 3H), 7.49 – 7.38 (m, 2H), 7.11 – 6.93 (m, 4H), 3.97 (d, *J* = 23.7 Hz, 6H).$^{13}$C NMR (101 MHz, Chloroform-*d*) δ 164.17, 162.29, 161.47, 156.46, 155.72, 145.09, 133.53, 132.31, 125.02, 122.72, 114.80, 113.94, 106.41, 56.25, 55.50, 40.19, 39.98.

**2b** (~82%)**:** $^{1}$H NMR (500 MHz, Chloroform-*d*) δ 8.35 – 8.29 (m, 2H), 8.19 – 8.14 (m, 2H), 8.12 – 8.08 (m, 1H), 7.45 – 7.39 (m, 2H), 7.01 (dd, *J* = 9.4, 2.4 Hz, 2H), 6.93 (d, *J* = 7.4 Hz, 2H), 4.17 (q, *J* = 7.0 Hz, 2H), 3.92 (s, 3H), 1.49 (t, *J* = 7.0 Hz, 3H). $^{13}$C NMR (101 MHz, Chloroform-*d*) δ 164.28, 164.19, 162.71, 160.99, 156.45, 155.92, 145.26, 133.65, 132.46, 125.22, 122.72, 121.16, 115.25, 114.03, 113.73, 107.33, 64.99, 55.60, 14.61.

**2c** (~75%)**:** $^{1}$H NMR (500 MHz, Chloroform-*d*) δ 8.36 – 8.29 (m, 2H), 8.19 – 8.14 (m, 2H), 8.10 (d, *J* = 8.3 Hz, 1H), 7.45 – 7.37 (m, 2H), 7.04 – 6.97 (m, 2H), 6.97 – 6.89 (m, 2H), 4.06 (t, *J* = 6.4 Hz, 2H), 3.92 (s, 3H), 1.88 (h, *J* = 7.2 Hz, 2H), 1.06 (t, *J* = 7.4 Hz, 3H). $^{13}$C NMR (101 MHz, Chloroform-*d*) δ 164.28, 164.21, 161.12, 156.46, 155.94, 145.26, 133.74, 132.47, 125.25, 122.71, 121.17, 115.20, 114.03, 113.66, 107.23, 70.75, 55.60, 22.46, 10.57.

**2d** (~82%)**:** $^{1}$H NMR (500 MHz, Chloroform-*d*) δ 8.33 – 8.00 (m, 3H), 7.90 – 7.64 (m, 2H), 7.45 – 7.36 (m, 2H), 7.10 – 7.00 (m, 2H), 6.97 (d, *J* = 7.7 Hz, 2H), 3.96 (d, *J* = 19.7 Hz, 6H). $^{13}$C NMR (101 MHz, Chloroform-*d*) δ 169.01, 166.21, 138.37, 137.19, 127.91, 118.82, 111.30, 82.46, 61.11, 60.38, 45.27, 45.06, 44.85, 44.43.

**2e** (~78%)**:** $^{1}$H NMR (500 MHz, Chloroform-*d*) δ 8.22 – 8.12 (m, 2H), 8.09 (d, *J* = 9.2 Hz, 1H), 7.73 (d, *J* = 8.6 Hz, 2H), 7.37 (d, *J* = 8.6 Hz, 2H), 7.01 (d, *J* = 8.8 Hz, 2H), 6.92 (d, *J* = 7.4 Hz, 2H), 4.16 (q, *J* = 7.0 Hz, 2H), 3.91 (s, 3H), 1.48 (t, *J* = 7.0 Hz, 3H). $^{13}$C NMR (101 MHz, Chloroform-*d*) δ 164.27, 164.20, 162.83, 160.93, 156.37, 154.41, 133.68, 133.60, 132.46, 123.05, 121.17, 118.42, 115.39, 114.03, 113.71, 109.55, 107.32, 64.98, 55.60, 14.61.

**2f** (~85%)**:** $^{1}$H NMR (500 MHz, Chloroform-*d*) δ 8.20 – 8.13 (m, 2H), 8.09 (d, *J* = 8.3 Hz, 1H), 7.73 (d, *J* = 8.7 Hz, 2H), 7.36 (d, *J* = 8.7 Hz, 2H), 7.01 (d, *J* = 8.9 Hz, 2H), 6.95 – 6.87 (m, 2H), 4.05 (t, *J* = 6.4 Hz, 2H), 3.91 (s, 3H), 1.87 (h, *J* = 7.2 Hz, 2H), 1.06 (t, *J* = 7.4 Hz, 3H). $^{13}$C NMR (101 MHz, Chloroform-*d*) δ 164.27, 164.22, 162.94, 161.06, 156.38, 154.44, 133.70, 132.46, 123.04, 121.18, 118.42, 115.34, 114.02, 113.63, 109.55, 107.21, 70.74, 55.60, 22.46, 10.57.

**2g** (~88%)**:** $^{1}$H NMR (500 MHz, Chloroform-*d*) δ 8.20 – 8.15 (m, 2H), 8.15 – 8.11 (m, 1H), 7.69 (d, *J* = 8.5 Hz, 2H), 7.36 (d, *J* = 8.4 Hz, 2H), 7.04 – 6.98 (m, 2H), 6.95 (d, *J* = 7.4 Hz, 2H), 3.94 (d, *J* = 20.1 Hz, 6H). $^{13}$C NMR (101 MHz, Chloroform-*d*) δ 169.01, 166.21, 138.37, 137.19, 127.91, 118.82, 111.30, 82.46, 61.11, 60.38, 45.27, 45.06, 44.85, 44.43.

**2h** (~85%)**:** $^{1}$H NMR (500 MHz, Chloroform-*d*) δ 8.24 – 8.16 (m, 2H), 8.16 – 8.10 (m, 1H), 7.72 (d, *J* = 8.6 Hz, 2H), 7.39 (d, *J* = 8.5 Hz, 2H), 7.07 – 7.01 (m, 2H), 6.95 (d, *J* = 7.2 Hz, 2H), 4.19 (q, *J* = 7.0 Hz, 2H), 3.94 (s, 3H), 1.51 (t, *J* = 7.0 Hz, 3H). $^{13}$C NMR (101 MHz, Chloroform-*d*) δ 164.24, 164.22, 163.27, 160.82, 156.17, 153.59, 133.53, 132.45, 126.80, 126.76, 122.39, 121.24, 115.82, 114.01, 113.65, 107.32, 64.98, 55.59, 14.62.

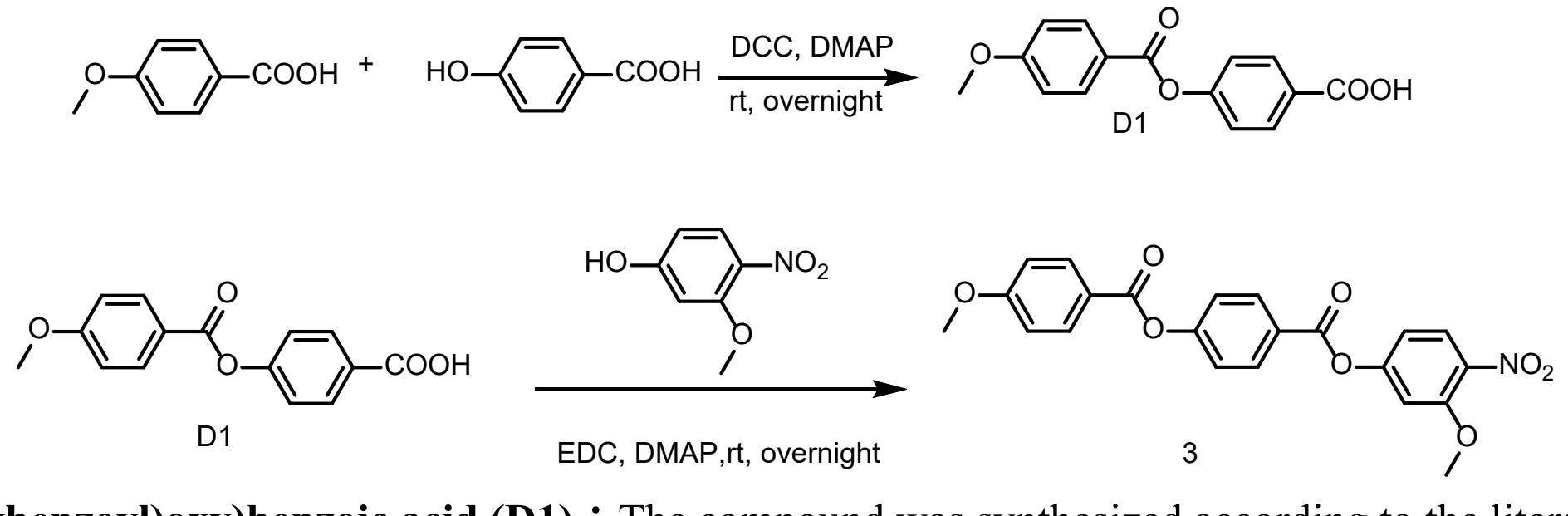

**4-((4-methoxybenzoyl)oxy)benzoic acid (D1)：**The compound was synthesized according to the literature (*29*).

**3：**synthesized by the general esterification method as in the synthetic procedures of **1a-1f**. Appearance: white powder, 2.56g, yield: 85.6%. $^{1}$H NMR (500 MHz, Chloroform-*d*) δ 8.32 – 8.23 (m, 2H), 8.22 – 8.12 (m, 2H), 8.01 (d, *J* = 8.9 Hz, 1H), 7.44 – 7.36 (m, 2H), 7.08 – 6.97 (m, 3H), 6.94 (dd, *J* = 8.9, 2.3 Hz, 1H), 3.95 (d, *J* = 36.8 Hz, 6H). $^{13}$C NMR

(126 MHz, Chloroform-*d*) δ 164.29, 164.23, 163.58, 155.85, 155.28, 154.49, 137.00, 132.48, 131.98, 127.24, 125.91, 122.36, 121.10, 114.03, 113.65, 107.51, 56.81, 55.60.

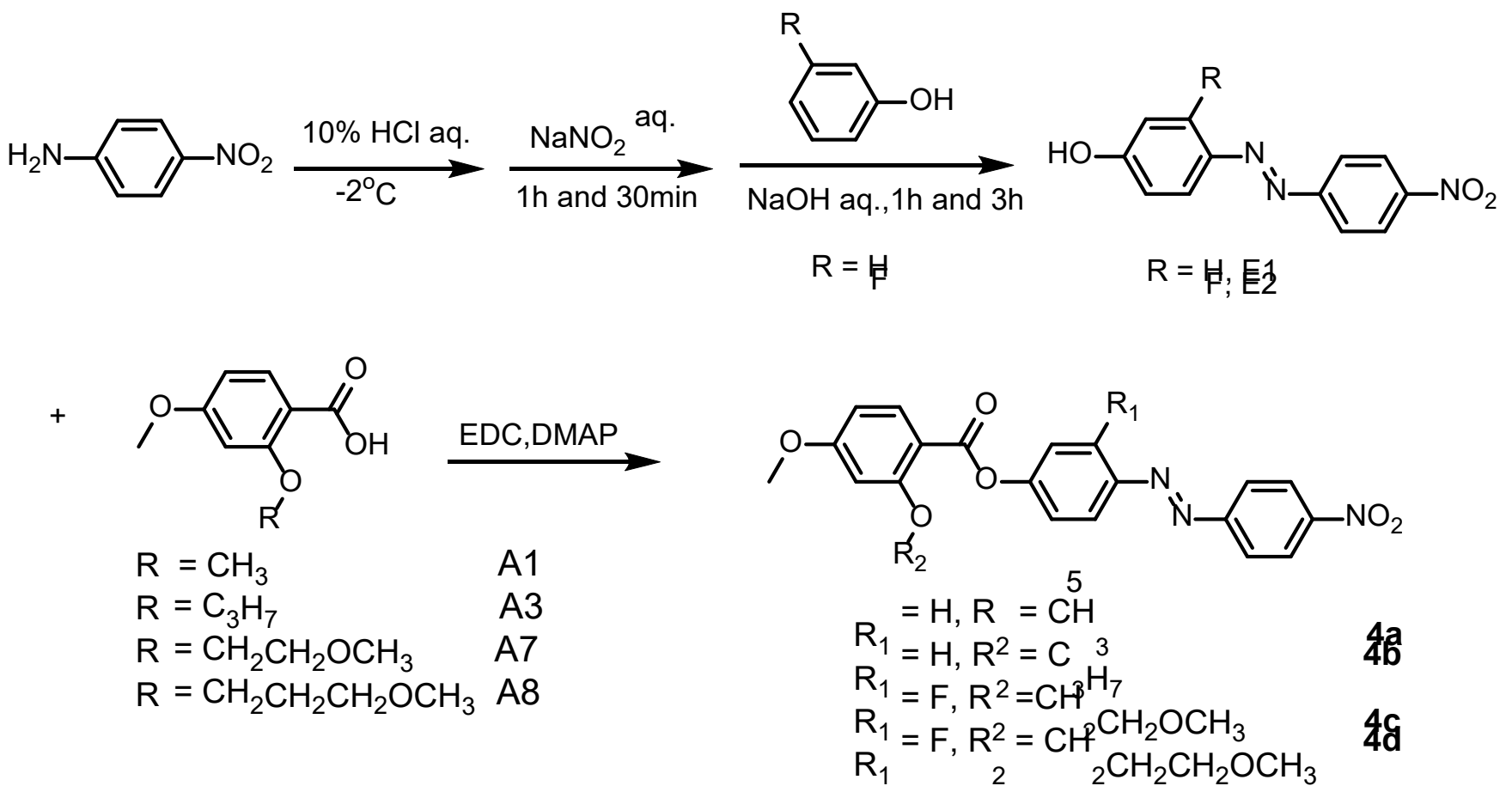

**(E)-4-((4-nitrophenyl)diazenyl)phenol (E1, E2):** These two compounds were synthesized according to the literature (*30*).
(E)-3-fluoro-4-((4-nitrophenyl)diazenyl)phenol (**E2**): $^1$H NMR (500 MHz, DMSO-$d_6$) δ 11.14 (s, 1H), 8.41 (d, *J* = 9.0 Hz, 2H), 8.01 (d, *J* = 9.0 Hz, 2H), 7.77 (d, *J* = 17.8 Hz, 1H), 6.85 (d, *J* = 15.0 Hz, 1H), 6.79 (d, *J* = 11.4 Hz, 1H). $^{13}$C NMR (126 MHz, Chloroform-*d*) δ 161.22, 156.03, 148.52, 124.76, 123.41, 119.01, 112.19, 104.30, 104.11.

**4a,4b, 4c, 4d:** synthesized by the general esterification method as in the synthetic procedures of **1a-1f**. Appearance: yellow to red powder.
**4a** (85%)**:** $^1$H NMR (400 MHz, Chloroform-*d*) δ 8.43 – 8.36 (m, 2H), 8.11 (d, *J* = 8.7 Hz, 1H), 8.08 – 7.99 (m, 4H), 7.44 – 7.38 (m, 2H), 6.63 – 6.52 (m, 2H), 3.93 (d, *J* = 15.2 Hz, 6H). $^{13}$C NMR (101 MHz, Chloroform-*d*) δ 165.30, 163.06, 162.47, 155.69, 154.44, 149.80, 148.66, 134.60, 124.73, 123.42, 122.90, 110.61, 104.97, 99.03, 56.05, 55.63.

**4b** (83%)**:** $^1$H NMR (400 MHz, Chloroform-*d*) δ 8.41 – 8.34 (m, 2H), 8.12 – 7.97 (m, 5H), 7.45 – 7.35 (m, 2H), 6.68 – 6.46 (m, 2H), 4.03 (t, *J* = 6.4 Hz, 2H), 3.89 (s, 3H), 1.89 (h, *J* = 7.3 Hz, 2H), 1.08 (t, *J* = 7.4 Hz, 3H). $^{13}$C NMR (101 MHz, Chloroform-*d*) δ 165.15, 163.45, 161.88, 155.71, 154.55, 149.80, 148.66, 134.60, 124.73, 123.43, 122.86, 110.89, 104.97, 99.73, 70.44, 55.59, 22.53, 10.63.

**4c** (78%)**:** $^1$H NMR (400 MHz, Chloroform-*d*) δ 8.48 – 8.36 (m, 2H), 8.14 – 8.01 (m, 3H), 7.90 (dd, *J* = 8.9, 8.2 Hz, 1H), 7.29 (dd, *J* = 11.0, 2.3 Hz, 1H), 7.16 (ddd, *J* = 8.9, 2.3, 1.2 Hz, 1H), 6.67 – 6.48 (m, 2H), 4.06 (t, *J* = 6.4 Hz, 2H), 3.92 (s, 3H), 1.92 (dtd, *J* = 13.7, 7.4, 6.4 Hz, 2H), 1.11 (t, *J* = 7.4 Hz, 3H). $^{13}$C NMR (101 MHz, Chloroform-*d*) δ 165.39, 162.89, 162.04, 155.78, 148.88, 134.68, 124.77, 123.70, 118.42, 118.38, 118.09, 111.58, 111.35, 110.36, 105.06, 99.70, 70.45, 55.62, 22.53, 10.63.

**4d** (85%)**:** $^1$H NMR (400 MHz, Chloroform-*d*) δ 8.46 – 8.37 (m, 2H), 8.14 – 8.03 (m, 3H), 7.91 (dd, *J* = 8.9, 8.2 Hz, 1H), 7.30 (d, *J* = 2.4 Hz, 1H), 7.16 (ddd, *J* = 9.0, 2.4, 1.2 Hz, 1H), 6.59 (dd, *J* = 8.8, 2.3 Hz, 1H), 6.55 (d, *J* = 2.3 Hz, 1H), 4.09 (t, *J* = 6.5 Hz, 2H), 1.96 – 1.84 (m, 2H), 1.56 – 1.47 (m, 2H), 1.45 – 1.34 (m, 2H), 0.93 (t, *J* = 7.3 Hz, 3H). $^{13}$C NMR (101 MHz, Chloroform-*d*) δ 165.37, 162.95, 162.00, 155.79, 148.89, 134.69, 124.77, 123.71, 118.40, 118.37, 118.08, 111.57, 111.35, 110.41, 105.04, 99.71, 68.98, 55.62, 28.81, 28.16, 22.39, 14.00.

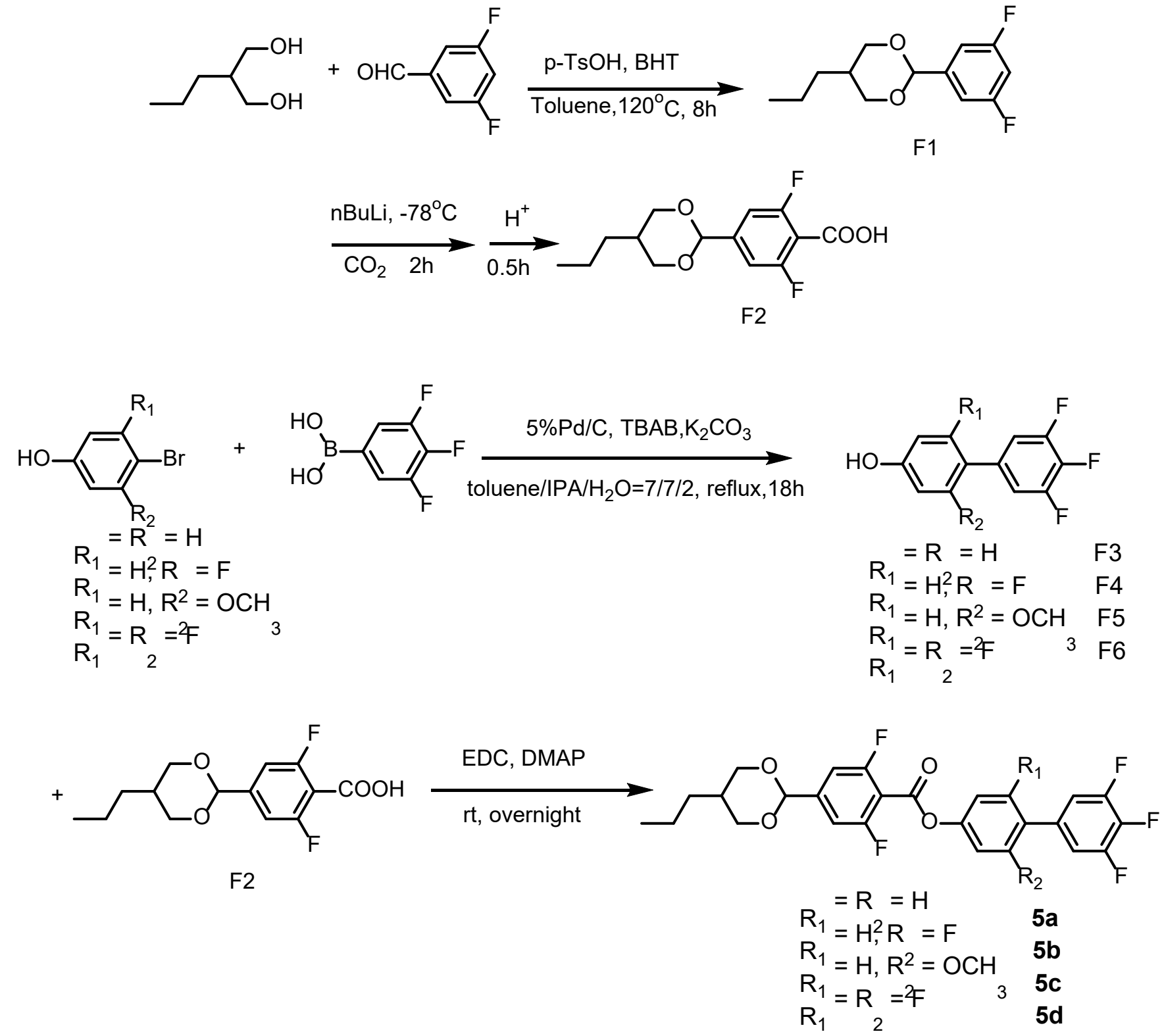

**2-phenyl-5-propyl-1,3-dioxane (F1):** 2-propylpropane-1,3-diol (5.0 g, 42.3 mmol), 3,5-difluorobenzoic acid (5.57 g, 35.3 mmol), BHT (0.116 g, 0.53 mmol), p-TsOH (0.30 g, 1.76 mmol) were added to 150 mL toluene in a 250 mL flask and reflux at 125°C over 8h. Then the mixture was cooled to room temperature, extracted by DCM, washed with brine, dried over anhydrous $Na_2SO_4$. The organic phase was concentrated under vacuum, purified by flash chromatograph and recrystallized from EtOH as the pale-yellow compound **F1** (6.41 g, 75%). $^1$H NMR (400 MHz, Chloroform-*d*) δ 7.02 (d, $J$ = 6.1 Hz, 2H, ArH), 6.81 – 6.71 (m, 1H, ArH), 5.35 (s, 1H, CH), 4.22 (dd, $J$ = 11.8, 4.6 Hz, 2H, $CH_2O$), 3.51 (t, $J$ = 11.5 Hz, 2H, $CH_2O$), 2.19 – 2.05 (m, 1H, CH), 1.37 – 1.28 (m, 2H, $CH_2$), 1.13 – 1.03 (m, 2H, $CH_2$), 0.95 (d, $J$ = 4.2 Hz, 3H, $CH_3$). $^{13}$C NMR (101 MHz, Chloroform-*d*) δ 164.20, 161.61, 142.28, 109.41, 103.97, 99.65, 72.56, 70.64, 34.03, 31.62, 30.29, 20.57, 19.55, 14.19.

**4-(5-propyl-1,3-dioxan-2-yl)benzoic acid (F2):** A 50 mL three-neck flask was loaded with **F1** (1.0 g, 4.13 mmol) and dry THF 25 mL, then the mixture was flashed with $N_2$ for 5 minutes. The solution was cooled to -78 °C, n-BuLi (2.5 M hexane solution, 1.98 mL, 4.95 mmol) was injected to the mixture over 10 min. After stirring 1 h, excess amount of dried ice was input into the solution under $N_2$ atmosphere and stirred for 1h. After that the reaction mixture was quenched by 1M HCl aq. THF solvent was removed. The residue was dissolved in EA, washed with water, dried by $MgSO_4$. Then the solvent was removed and the crude product was recrystallized from EtOH as white solid (1.05 g, 86%). $^1$H NMR (500 MHz, DMSO-$d_6$) δ 13.97 (s, 1H, COOH), 7.18 (d, $J$ = 8.7 Hz, 2H, ArH), 5.49 (s, 1H, ArH), 4.15 (d, $J$ = 16.3 Hz, 2H, ArH), 3.52 (t, $J$ = 11.4 Hz, 2H, $CH_2O$), 1.97 (t, $J$ = 11.3 Hz, 1H, CH), 1.31 – 1.23 (m, 2H, $CH_2$), 1.09 – 0.99 (m, 2H, $CH_2$), 0.91 – 0.83 (m, 3H, $CH_3$). $^{13}$C NMR (101 MHz, Chloroform-*d*) δ 166.51, 162.49, 159.92, 110.38, 110.11, 98.88, 98.86, 72.55, 70.62, 33.86, 31.50, 30.19, 19.50, 14.16.

**3',4',5'-trifluoro-[1,1'-biphenyl]-4-ol (F3) or 2,3',4',5'-tetrafluoro-[1,1'-biphenyl]-4-ol (F4) or 3',4',5'-trifluoro-2-methoxy-[1,1'-biphenyl]-4-ol (F5) or 2,3',4',5',6-pentafluoro-[1,1'-biphenyl]-4-ol (F6):** In a 50 ml flask, (3,4,5-trifluorophenyl)boronic acid (1.0 g, 5.69 mmol), 4-bromophenol (0.89 g, 5.17 mmol) or 4-bromo-3-fluorophenol (0.98 g, 5.17 mmol) or 4-bromo-3-methoxyphenol (1.05 g, 5.17 mmol), TBAB (0.426 g, 1.29 mmol), $K_2CO_3$ (1.43 g, 10.34 mmol), and palladium on carbon (10 wt%, 0.20 g) were added, then the mixed solvent of volume ratio 7/7/2 (tolunen/isopropanol/$H_2O$) was added to the bottle. After refluxed for 6 h, the system was cooled to room temperature and poured into water, followed by extracting with toluene, washing with saturated brine (50 mL× 2), drying over anhydrous $MgSO_4$. Then the solvent was vacuum removed. The crude was purified by silica gel chromatography and the colorless solids were obtained.

**F3** (85.6%)**:** The compound was synthesized according to the literature (*31*).

**F4** (81.6%)**:** $^1$H NMR (400 MHz, Chloroform-*d*) δ 8.68 (s, 1H, OH), 7.76 (s, 1H, ArH), 3.76 (d, $J$ = 7.1 Hz, 2H, ArH), 1.26 (s, 2H, ArH). $^{13}$C NMR (101 MHz, Chloroform-*d*) δ 161.40, 158.92, 157.54, 131.77, 130.86, 118.50, 112.89, 112.08, 112.04, 104.07, 103.81, 68.12, 25.61.

**F5** (86.8%)**:** $^{1}$H NMR (500 MHz, Chloroform-*d*) δ 7.17 – 7.06 (m, 3H, ArH), 6.55 – 6.44 (m, 2H, ArH), 3.80 (s, 3H, $OCH_3$). $^{13}$C NMR (101 MHz, Chloroform-*d*) δ 157.49, 157.02, 131.15, 113.24, 107.52, 99.47, 55.58.

**F6** (82.3%)**:** $^{1}$H NMR (400 MHz, Chloroform-*d*) δ 7.12 – 7.01 (m, 2H), 6.56 – 6.45 (m, 2H). $^{13}$C NMR (126 MHz, Chloroform-*d*) δ 155.90, 152.45, 152.42, 152.37, 152.34, 150.43, 150.39, 150.36, 139.84, 137.84, 136.94, 136.91, 131.01, 130.99, 128.23, 115.97, 110.58, 110.54, 110.45, 110.41.

**5a, 5b, 5c, 5d** were synthesized by the general esterification method as in the synthetic procedures of **1a-1f**. Appearance: white solids.
**5a** (85%): $^{1}$H NMR (500 MHz, Chloroform-*d*) δ 7.58 (d, *J* = 8.7 Hz, 2H), 7.37 (d, *J* = 8.7 Hz, 2H), 7.25 – 7.14 (m, 4H), 5.43 (s, 1H), 4.28 (dd, *J* = 11.8, 4.6 Hz, 2H), 3.57 (t, *J* = 11.5 Hz, 2H), 2.23 – 2.10 (m, 1H), 1.43 – 1.30 (m, 3H), 1.18 – 1.07 (m, 2H), 0.97 (t, *J* = 7.3 Hz, 3H). $^{13}$C NMR (101 MHz, Chloroform-*d*) δ 162.18, 159.72, 150.56, 145.33, 128.08, 122.30, 111.25, 110.34, 98.88, 98.86, 72.59, 33.90, 30.23, 19.53, 14.19.

**5b** (88%): $^{1}$H NMR (500 MHz, Chloroform-*d*) δ 7.43 (t, *J* = 8.6 Hz, 1H, ArH), 7.22 – 7.14 (m, 6H, ArH), 5.40 (s, 1H, CH), 4.25 (d, *J* = 16.3 Hz, 2H, $CH_2$), 3.54 (t, *J* = 11.4 Hz, 2H, $CH_2$), 2.14 (s, 1H, CH), 1.35 (d, *J* = 7.8 Hz, 2H, $CH_2$), 1.17 – 1.04 (m, 2H, $CH_2$), 0.92 (s, 3H, $CH_3$). $^{13}$C NMR (101 MHz, Chloroform-*d*) δ 162.18, 159.72, 150.56, 145.33, 128.08, 122.30, 111.25, 110.34, 98.88, 98.86, 72.59, 33.90, 30.23, 19.53, 14.19.

**5c** (81%)**:** $^{1}$H NMR (400 MHz, Chloroform-*d*) δ 7.30 (d, *J* = 8.3 Hz, 1H, ArH), 7.22 – 7.10 (m, 4H, ArH), 6.94 (dd, *J* = 8.3, 2.2 Hz, 1H, ArH), 6.88 (d, *J* = 2.1 Hz, 1H, ArH), 5.40 (s, 1H, CH), 4.26 (dd, *J* = 11.8, 4.6 Hz, 2H, $CH_2$), 3.85 (s, 3H), 3.54 (t, *J* = 11.5 Hz, 2H, $CH_2$), 2.23 – 2.02 (m, 1H, CH), 1.53 (s, 1H, CH), 1.39 – 1.29 (m, 3H, $CH_3$), 1.14 – 1.09 (m, 2H, $CH_2$), 0.94 (t, *J* = 7.4 Hz, 3H, $CH_3$). $^{13}$C NMR (101 MHz, Chloroform-*d*) δ 159.71, 157.01, 151.32, 130.91, 113.92, 110.39, 110.12, 105.42, 98.85, 72.59, 55.86, 33.91, 19.54, 14.19.

**5d** (81%)**:** $^{1}$H NMR (400 MHz, Chloroform-*d*) δ 7.17 – 7.10 (m, 2H), 7.05 (ddt, *J* = 8.5, 7.4, 1.2 Hz, 2H), 6.99 – 6.89 (m, 2H), 5.33 (s, 1H), 4.28 – 4.13 (m, 2H), 3.57 – 3.39 (m, 2H), 2.07 (tddd, *J* = 11.4, 9.2, 6.9, 4.6 Hz, 1H), 1.35 – 1.22 (m, 2H), 1.10 – 0.98 (m, 2H), 0.87 (t, *J* = 7.3 Hz, 3H). $^{13}$C NMR (101 MHz, Chloroform-*d*) δ 162.30, 161.02, 160.94, 158.88, 158.53, 158.45, 145.95, 114.94, 110.51, 110.48, 110.28, 110.24, 106.69, 106.41, 106.39, 98.77, 98.75, 98.73, 72.60, 33.90, 30.22, 19.53, 14.19.

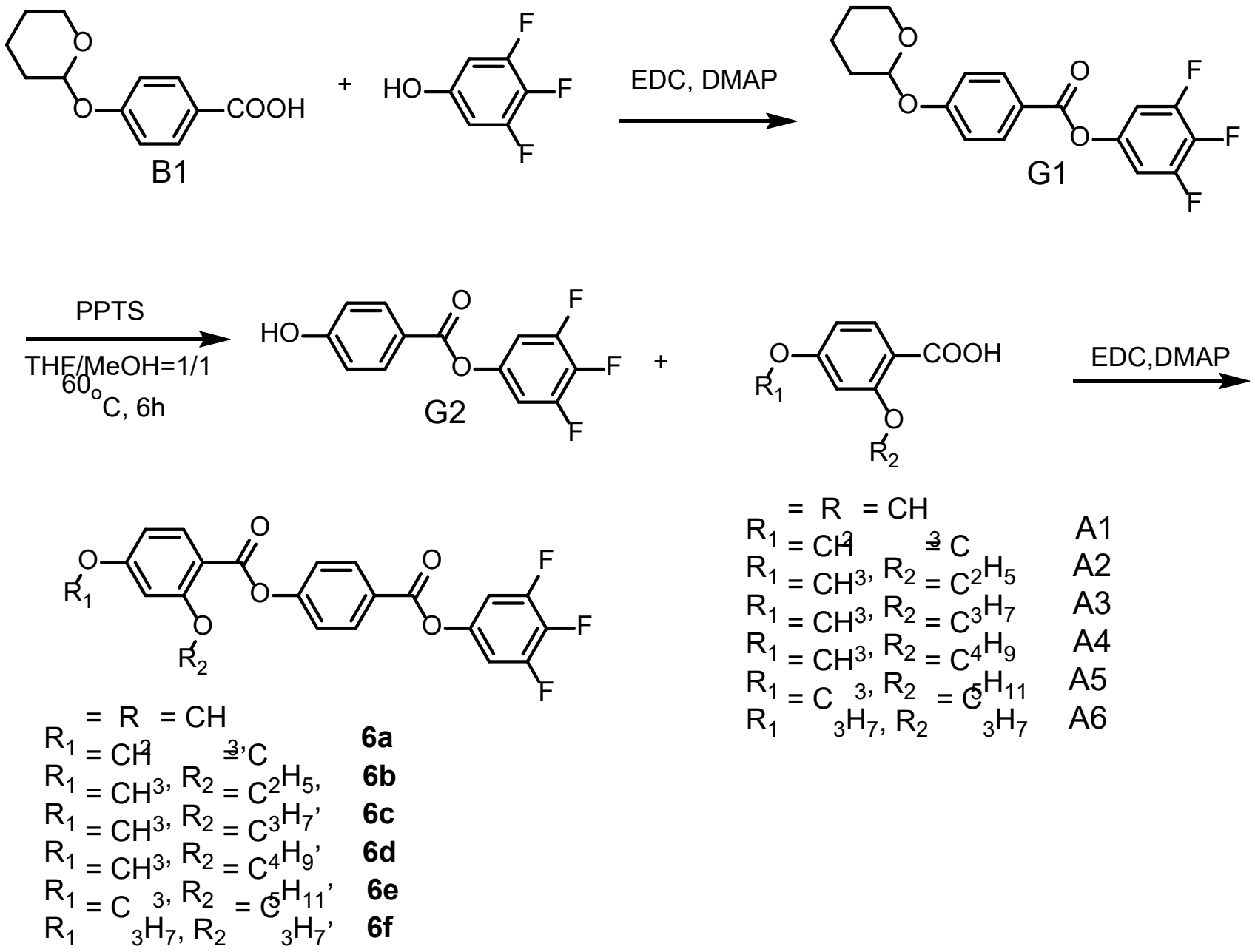

**3,4,5-trifluorophenyl 4-((tetrahydro-2H-pyran-2-yl)oxy)benzoate (G1):** A round bottom flask was charged with **B1** (2.0 g, 9.0 mmol), EDC (2.6 g, 13.5 mmol), DMAP (0.020 g, 0.45 mmol), and 3,4,5-trifluorophenol (1.31 g, 8.82 mmol). The solution was stirred at room temperature for 12 hr. The solution was stripped of solvent by rotary evaporation, and dry-loaded onto a silica gel column for purification. The chromatography was run using dichloromethane/ethyl acetate eluent. Yield: 2.39 g (75.4%); appearance: white powder. $^{1}$H NMR (500 MHz, Chloroform-*d*) δ 8.07 (d, *J* = 8.8 Hz, 2H), 6.93 – 6.89 (m, 4H), 4.94 (d, *J* = 27.7 Hz, 1H), 3.94 – 3.82 (m, 1H), 3.54 (t, *J* = 15.6 Hz, 1H), 1.79 – 1.68 (m, 2H),

1.57 (d, $J$ = 17.3 Hz, 4H). $^{13}$C NMR (101 MHz, Chloroform-*d*) δ 161.90, 132.30, 121.28, 116.26, 107.35, 96.15, 62.08, 30.07, 25.04, 18.45.

**3,4,5-trifluorophenyl 4-hydroxybenzoate (G2):** A round bottom flask was charged with **G1** (2.0 g, 5.68 mmol), PPTS (0.57 g, 5.68 mmol), 20 mL THF, and 20 mL ethanol. The solution was heated at 60 °C and monitored by TLC (6-48 h). The solvent was removed under reduced pressure. Water (40 mL) was added to the residue, and the product was extracted with DCM (3 × 50 mL). The organic phase was dried by anhydrous $MgSO_4$. The solvent was removed and the residue was purified by chromatography. The product was then concentrated and dried in the vacuum oven. Yield: 1.41 g (93.8%); appearance: white powder. $^{1}$H NMR (400 MHz, Chloroform-*d*) δ 8.07 (d, $J$ = 8.9 Hz, 2H, ArH3F), 7.01 – 6.84 (m, 4H, ArHOH). $^{13}$C NMR (101 MHz, Chloroform-*d*) δ 168.90, 123.61, 120.49, 112.16, 82.35, 82.03, 44.94, 44.73.

**6a, 6b, 6c, 6d, 6e, 6f** were synthesized by the general esterification method as in the synthetic procedures of **1a-1f**. Appearance: white solids.
**6a** (81%): $^{1}$H NMR (400 MHz, Chloroform-*d*) δ 8.21 (d, $J$ = 8.8 Hz, 2H), 8.10 (d, $J$ = 8.7 Hz, 1H), 7.42 – 7.34 (m, 2H), 7.05 – 6.88 (m, 2H), 6.65 – 6.51 (m, 2H), 3.93 (d, $J$ = 11.9 Hz, 6H). $^{13}$C NMR (101 MHz, Chloroform-*d*) δ 165.41, 163.75, 162.53, 155.95, 134.63, 131.78, 125.47, 122.48, 110.33, 107.35, 107.11, 105.02, 56.01, 55.60.

**6b** (80%)**:** $^{1}$H NMR (400 MHz, Chloroform-*d*) δ 8.26 – 8.17 (m, 2H), 8.07 (d, $J$ = 8.8 Hz, 1H), 7.42 – 7.33 (m, 2H), 7.00 – 6.89 (m, 2H), 6.63 – 6.47 (m, 2H), 4.14 (q, $J$ = 7.0 Hz, 2H), 3.89 (s, 3H), 1.49 (t, $J$ = 7.0 Hz, 3H). $^{13}$C NMR (101 MHz, Chloroform-*d*) δ 165.25, 163.78, 163.06, 161.86, 156.03, 134.56, 131.84, 125.46, 122.46, 110.69, 107.35, 107.29, 107.18, 107.11, 105.05, 99.89, 64.67, 55.59, 14.66.

**6c** (85%): $^{1}$H NMR (400 MHz, Chloroform-*d*) δ 8.25 – 8.18 (m, 2H), 8.06 (d, $J$ = 8.7 Hz, 1H), 7.41 – 7.33 (m, 2H), 6.95 (ddd, $J$ = 9.1, 4.4, 3.0 Hz, 2H), 6.62 – 6.48 (m, 2H), 4.03 (t, $J$ = 6.4 Hz, 2H), 3.89 (s, 3H), 1.88 (dtd, $J$ = 13.8, 7.4, 6.3 Hz, 2H), 1.07 (t, $J$ = 7.4 Hz, 3H). $^{13}$C NMR (101 MHz, Chloroform-*d*) δ 165.26, 161.95, 156.07, 134.64, 131.87, 125.46, 122.46, 121.98, 110.65, 107.35, 105.01, 99.71, 70.43, 55.59, 10.62.

**6d** (81%): $^{1}$H NMR (400 MHz, Chloroform-*d*) δ 8.21 (d, $J$ = 8.8 Hz, 2H), 8.06 (d, $J$ = 8.7 Hz, 1H), 7.37 (d, $J$ = 8.8 Hz, 2H), 7.04 – 6.85 (m, 2H), 6.62 – 6.47 (m, 2H), 4.07 (t, $J$ = 6.4 Hz, 2H), 3.89 (s, 3H), 1.92 – 1.76 (m, 2H), 1.52 (dd, $J$ = 14.8, 7.4 Hz, 2H), 0.95 (t, $J$ = 7.4 Hz, 3H). $^{13}$C NMR (101 MHz, Chloroform-*d*) δ 165.25, 161.95, 156.06, 134.64, 131.87, 131.12, 125.46, 122.45, 121.97, 107.35, 104.97, 68.62, 55.60, 31.15, 19.22, 13.80.

**6e** (85%): $^{1}$H NMR (400 MHz, Chloroform-*d*) δ 8.28 – 8.17 (m, 2H), 8.05 (d, $J$ = 8.7 Hz, 1H), 7.37 (d, $J$ = 8.7 Hz, 2H), 7.03 – 6.87 (m, 2H), 6.61 – 6.46 (m, 2H), 4.06 (t, $J$ = 6.5 Hz, 2H), 3.89 (s, 3H), 1.86 (dt, $J$ = 14.5, 6.5 Hz, 2H), 1.48 (dt, $J$ = 14.6, 7.1 Hz, 2H), 1.36 (dq, $J$ = 14.3, 7.1 Hz, 2H), 0.89 (t, $J$ = 7.3 Hz, 3H). $^{13}$C NMR (101 MHz, Chloroform-*d*) δ 165.28, 161.94, 156.04, 134.66, 131.87, 131.12, 125.49, 122.45, 121.98, 110.63, 107.35, 105.00, 68.97, 55.60, 28.80, 28.14, 22.39, 13.99.

**6f** (79%): $^{1}$H NMR (400 MHz, Chloroform-*d*) δ 8.21 (d, $J$ = 8.8 Hz, 2H), 8.04 (d, $J$ = 8.7 Hz, 1H), 7.37 (d, $J$ = 8.8 Hz, 2H), 7.06 – 6.83 (m, 2H), 6.66 – 6.44 (m, 2H), 4.12 – 3.94 (m, 4H), 1.95 – 1.78 (m, 4H), 1.07 (t, $J$ = 7.4 Hz, 6H). $^{13}$C NMR (101 MHz, Chloroform-*d*) δ 164.89, 163.79, 163.25, 161.98, 156.09, 134.61, 131.87, 131.11, 125.44, 122.46, 121.99, 110.38, 107.34, 107.10, 105.50, 100.15, 70.43, 69.88, 52.15, 22.48, 10.48.

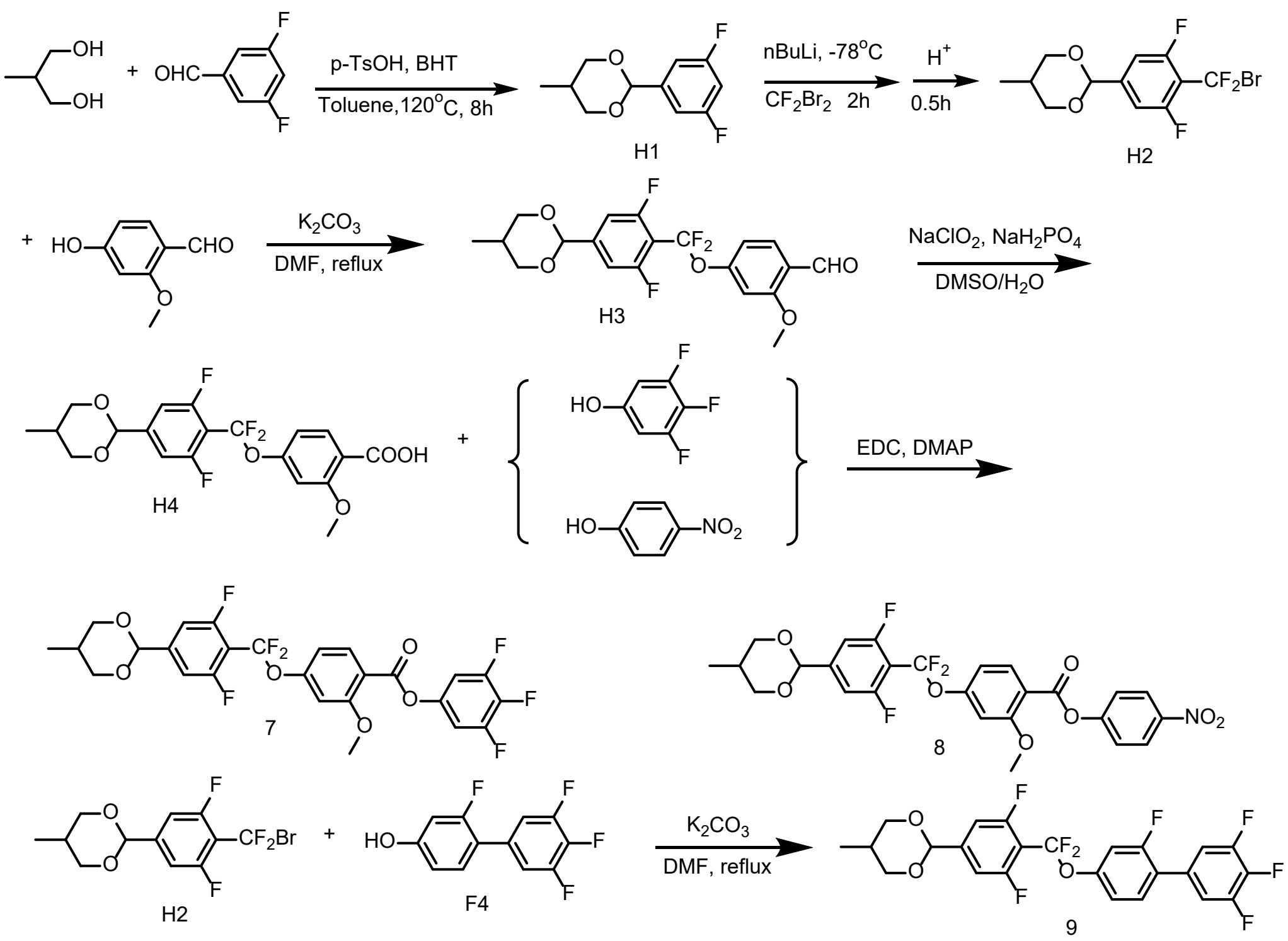

**2-(3,5-difluorophenyl)-5-methyl-1,3-dioxane (H1):** synthesized in similar procedure as in F1. Pale yellow liquid was obtained (7.56 g, 76.3%). $^{1}$H NMR (500 MHz, Chloroform-*d*) δ 7.03 (dd, *J* = 8.1, 1.8 Hz, 2H), 6.79 – 6.74 (m, 1H), 5.36 (s, 1H), 4.18 (dd, *J* = 11.8, 4.7 Hz, 2H), 3.49 (t, *J* = 11.5 Hz, 2H), 2.27 – 2.14 (m, 1H), 0.77 (d, *J* = 6.8 Hz, 3H). $^{13}$C NMR (126 MHz, Chloroform-*d*) δ 163.93, 163.83, 161.96, 161.86, 142.19, 109.39, 109.34, 109.23, 109.18, 104.18, 103.98, 103.78, 99.44, 99.41, 99.39, 77.27, 73.58, 72.20, 29.29, 15.82, 12.25.

**2-(4-(bromodifluoromethyl)-3,5-difluorophenyl)-5-methyl-1,3-dioxane (H2):** A 50 mL three neck flask was loaded with compound **H1** (5.0 g, 23.34 mmol) and dry THF 25 mL, then the mixture was flashed with $N_2$ for 5 minutes. The solution was cooled to -78 °C, n-BuLi (2.5 M hexane solution, 56 mL, 140 mmol) was injected to the mixture over 30 min. After stirring 1h, $CF_2Br_2$ (2M THF solution, 14 mL,28 mmol) was dropped into the solution over 30 min under $N_2$ atmosphere and further stirred for 1h. The reaction mixture was quenched by 1M HCl aq. THF solvent was removed. The residue was dissolved in EA, washed with water, dried by $MgSO_4$. Then the solvent was removed and the crude product was recrystallized from EtOH as white solid (1.05 g, 86%). $^{1}$H NMR (500 MHz, Chloroform-*d*) δ 7.16 – 7.11 (m, 2H), 5.35 (d, *J* = 10.4 Hz, 1H), 4.22 – 4.13 (m, 2H), 3.58 – 3.38 (m, 2H), 2.27 – 2.12 (m, 1H), 0.83 – 0.65 (m, 3H). $^{13}$C NMR (126 MHz, Chloroform-*d*) δ 159.16, 157.14, 144.86, 114.82, 112.39, 110.85, 110.82, 110.79, 110.77, 110.66, 110.63, 110.60, 110.01, 109.96, 109.83, 98.89, 98.83, 98.36, 72.24, 29.26, 15.76, 12.19.

**4-((2,6-difluoro-4-(5-methyl-1,3-dioxan-2 yl)phenyl)difluoromethoxy)benzaldeh-yde (H3):** The 4-hydroxy-2-methoxybenzaldehyde (2.13 g, 14 mmol), compound **H2** (4.0 g, 11.66 mmol), potassium carbonate (1.61 g, 11.66 mmol) were mixed in DMF and heated at 90 °C with vigorous stirring (18-24 h). The reaction was then cooled, water added, extracted with EA, and washed with water and brine. The organic phase was dried by anhydrous $MgSO_4$. Then the solvent was removed, and dry-loaded onto a silica gel column for purification using EA/hexane eluent. Appearance: white powder, yield (4.06 g, 90.6%). $^{1}$H NMR (400 MHz, Chloroform-*d*) δ 10.37 (s, 1H), 7.81 (dd, *J* = 8.6, 2.3 Hz, 1H), 7.13 (d, *J* = 10.2 Hz, 2H), 6.91 (d, *J* = 8.6 Hz, 1H), 6.85 (d, *J* = 1.8 Hz, 1H), 5.36 (s, 1H), 4.19 (dd, *J* = 11.8, 4.7 Hz, 2H), 3.92 (s, 3H), 3.49 (t, *J* = 11.5 Hz, 2H), 2.27 – 2.14 (m, 1H), 0.78 (d, *J* = 6.8 Hz, 3H). $^{13}$C NMR (126 MHz, Chloroform-*d*) δ 188.56, 162.83, 156.29, 130.07, 122.49, 113.29, 110.67, 110.65, 110.48, 110.45, 104.78, 98.49, 73.57, 72.23, 55.90, 29.25, 12.23.

**4-((2,6-difluoro-4-(5-methyl-1,3-dioxan-2-yl)phenyl)difluoromethoxy)benzoic acid (H4):** synthesized by general aldehyde oxidation method used for the **A7 - A11** synthesis. White solid, 4.06 g, yield 97%. $^{1}$H NMR (400 MHz, Chloroform-*d*) δ 10.45 (s, 1H), 8.17 (d, *J* = 8.7 Hz, 1H), 7.14 (d, *J* = 10.2 Hz, 2H), 7.03 (d, *J* = 10.8 Hz, 1H), 6.93 (d, *J* = 1.9 Hz, 1H), 5.37 (s, 1H), 4.19 (dd, *J* = 11.8, 4.7 Hz, 2H), 4.07 (s, 3H), 3.50 (t, *J* = 11.5 Hz, 2H), 2.27 – 2.14 (m, 1H), 0.78 (d, *J* = 6.8 Hz, 3H). $^{13}$C NMR (126 MHz, Chloroform-*d*) δ 164.54, 158.90, 155.34, 135.17, 115.03, 114.65, 110.71, 110.69, 110.67, 110.52, 110.49, 105.01, 98.46, 73.57, 72.24, 56.99, 42.69, 29.25, 12.23.

**7**, **8**, and **9** were synthesized by the general esterification method as in the synthetic procedures of **1a-1f**. Appearance: white solids.

**7** (79%)**:** $^{1}$H NMR (400 MHz, Chloroform-*d*) δ 7.90 (d, *J* = 8.6 Hz, 1H), 7.07 (d, *J* = 10.3 Hz, 2H), 6.92 – 6.76 (m, 4H), 5.30 (s, 1H), 4.17 – 4.02 (m, 2H), 3.86 (s, 3H), 3.53 – 3.25 (m, 2H), 2.25 – 2.04 (m, 1H), 0.71 (d, *J* = 6.8 Hz, 3H). $^{13}$C NMR (101 MHz, Chloroform-*d*) δ 162.53, 161.54, 161.20, 155.61, 133.79, 114.96, 112.64, 110.72, 110.69, 110.48, 110.44, 107.40, 107.33, 107.22, 107.16, 105.45, 98.49, 73.58, 56.29, 29.26, 12.23.

**8** (86%)**:** $^{1}$H NMR (400 MHz, Chloroform-*d*) δ 8.37 – 8.24 (m, 2H), 8.04 (dd, *J* = 8.7, 2.5 Hz, 1H), 7.44 – 7.35 (m, 2H), 7.14 (d, *J* = 10.1 Hz, 2H), 7.02 – 6.84 (m, 2H), 5.37 (s, 1H), 4.19 (dd, *J* = 11.8, 4.7 Hz, 2H), 3.94 (s, 3H), 3.50 (t, *J* = 11.5 Hz, 2H), 2.21 (td, *J* = 11.2, 6.6 Hz, 1H), 0.79 (d, *J* = 6.8 Hz, 3H). $^{13}$C NMR (101 MHz, Chloroform-*d*) δ 161.66, 155.76, 155.69, 145.31, 133.90, 125.19, 122.73, 114.97, 112.67, 110.73, 110.70, 110.49, 110.45, 105.47, 98.49, 73.58, 56.32, 29.26, 12.23.

**9** (85%)**:** $^{1}$H NMR (400 MHz, Chloroform-*d*) δ 7.34 (t, *J* = 8.7 Hz, 1H), 7.21 – 7.08 (m, 6H), 5.37 (s, 1H), 4.26 – 4.15 (m, 2H), 3.59 – 3.43 (m, 2H), 2.30 – 2.10 (m, 1H), 0.79 (d, *J* = 6.8 Hz, 3H). $^{13}$C NMR (101 MHz, Chloroform-*d*) δ 161.25, 160.56, 158.68, 158.06, 150.99, 150.88, 144.65, 130.80, 130.62, 130.58, 123.75, 120.23, 117.99, 117.96, 113.29, 113.26, 113.07, 113.03, 110.71, 110.68, 110.58, 110.47, 110.43, 110.32, 98.56, 98.54, 98.51, 73.58, 72.24, 29.26, 12.22.

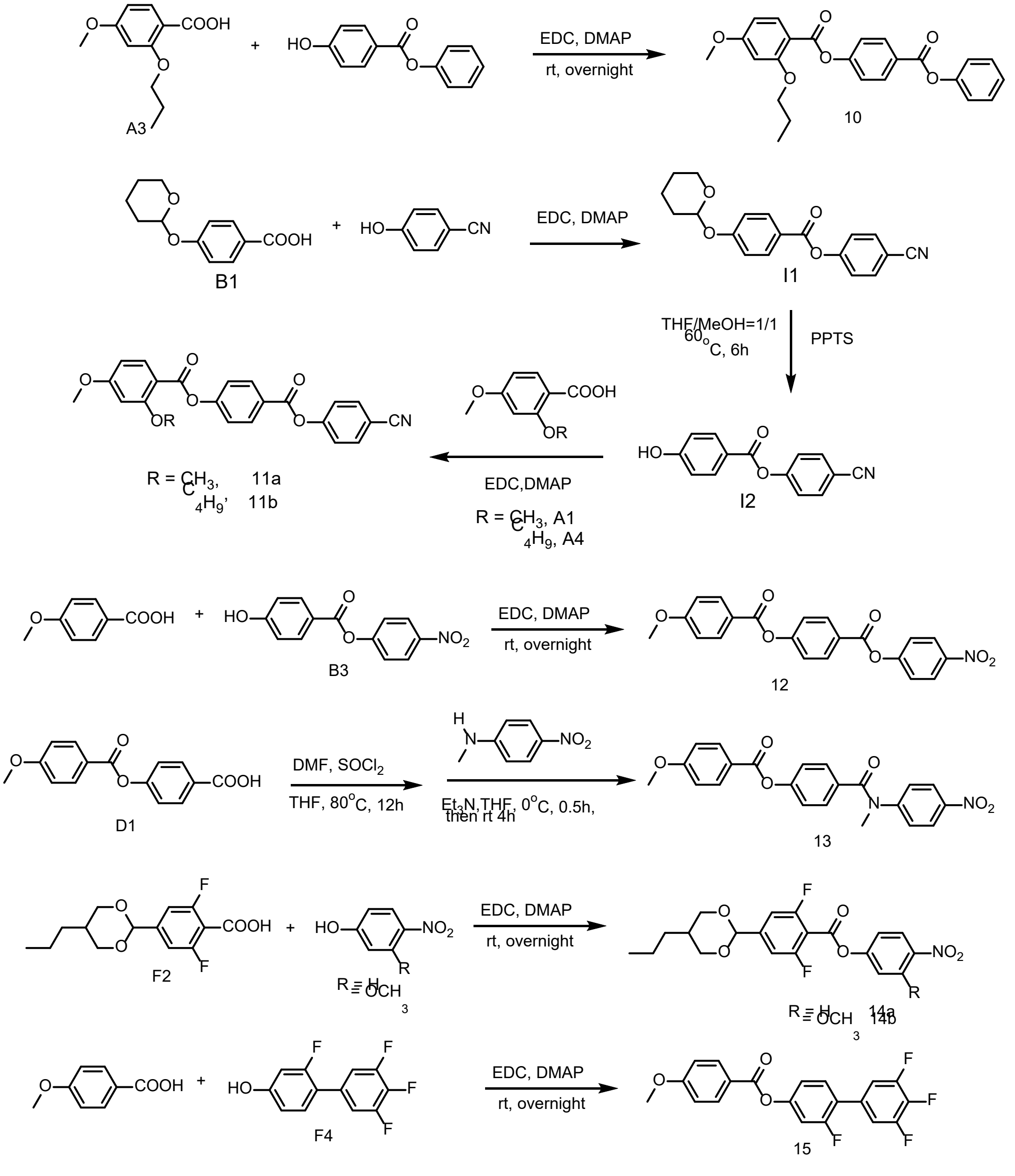

**4-cyanophenyl 4-((tetrahydro-2H-pyran-2-yl)oxy)benzoate (I1):** synthesized by similar procedure as for **G1**. Yield: 11.35 g (78%); appearance: white powder. 1H NMR (400 MHz, Chloroform-*d*) δ 8.13 (d, *J* = 8.9 Hz, 2H), 7.73 (d, *J* = 8.7 Hz, 2H), 7.35 (d, *J* = 8.7 Hz, 2H), 7.15 (d, *J* = 8.9 Hz, 2H), 5.56 (t, *J* = 3.0 Hz, 1H), 3.95 – 3.75 (m, 1H), 3.70 – 3.53 (m, 1H), 2.10 – 1.95 (m, 1H), 1.91 (t, *J* = 7.4 Hz, 2H), 1.79 – 1.58 (m, 3H). $^{13}$C NMR (101 MHz, Chloroform-*d*) δ 164.04, 161.90, 154.46, 133.69, 132.36, 123.01, 121.46, 118.38, 116.27, 109.59, 96.14, 62.09, 30.07, 25.04, 18.45.

**4-cyanophenyl 4-hydroxybenzoate (I2):** synthesized by similar procedure as for **G2**. Yield: 0.71 g (96%); appearance: faint yellow powder. $^{1}$H NMR (400 MHz, Chloroform-*d*) δ 14.75 (s, 1H), 12.69 (d, *J* = 8.7 Hz, 2H), 12.43 (d, *J* = 8.5 Hz, 2H), 12.12 – 12.01 (m, 2H), 11.61 (d, *J* = 8.7 Hz, 2H). $^{13}$C NMR (101 MHz, Chloroform-*d*) δ 159.35, 138.38, 137.35, 137.32, 137.28, 127.94, 127.92, 123.84, 123.13, 120.54, 113.93, 83.08, 82.99, 82.76, 82.67, 82.43, 82.35, 45.26, 45.05, 44.84, 44.63, 44.42.

**10**, **11a**, **11b**, **12**, **14a**, **14b** and **15** were synthesized by the general esterification method as in the synthetic procedures of **1a-1f**. Appearance: white solids.

**10** (82%)**:** $^{1}$H NMR (500 MHz, Chloroform-*d*) δ 8.29 (d, *J* = 8.7 Hz, 2H), 8.09 (d, *J* = 8.7 Hz, 1H), 7.47 (t, *J* = 7.9 Hz, 2H), 7.39 (d, *J* = 8.7 Hz, 2H), 7.31 (d, *J* = 7.4 Hz, 1H), 7.25 (d, *J* = 7.7 Hz, 2H), 6.59 (dd, *J* = 8.8, 2.3 Hz, 1H), 6.55 (d, *J* = 2.2 Hz, 1H), 4.06 (t, *J* = 6.4 Hz, 2H), 3.92 (s, 3H), 1.91 (h, *J* = 7.1 Hz, 2H), 1.10 (t, *J* = 7.4 Hz, 3H). $^{13}$C NMR (101 MHz, Chloroform-*d*) δ 165.17, 164.61, 163.38, 161.89, 155.59, 150.96, 134.63, 131.77, 129.52, 126.67, 125.92, 122.24, 121.75, 110.89, 104.98, 99.75, 70.46, 55.60, 22.54, 10.64.

**11a** (82%)**:** $^{1}$H NMR (500 MHz, Chloroform-*d*) δ 8.30 – 8.19 (m, 2H), 8.10 (d, *J* = 8.7 Hz, 1H), 7.80 – 7.69 (m, 2H), 7.45 – 7.32 (m, 4H), 6.63 – 6.51 (m, 2H), 3.92 (d, *J* = 18.1 Hz, 6H). $^{13}$C NMR (126 MHz, Chloroform-*d*) δ 165.39, 163.73, 162.82, 162.53, 155.92, 154.23, 134.65, 133.76, 131.85, 125.69, 122.96, 122.49, 118.30, 110.39, 109.86, 104.99, 99.03, 56.06, 55.64.

**11b** (80%)**:** $^{1}$H NMR (400 MHz, Chloroform-*d*) δ 8.35 – 8.18 (m, 2H), 8.13 – 7.99 (m, 1H), 7.85 – 7.65 (m, 2H), 7.38 (d, *J* = 8.2 Hz, 4H), 6.60 – 6.54 (m, 1H), 6.53 (s, 1H), 4.07 (t, *J* = 6.3 Hz, 2H), 3.89 (s, 3H), 1.83 (q, *J* = 6.9 Hz, 2H), 1.54 (dq, *J* = 14.6, 7.4 Hz, 2H), 0.95 (t, *J* = 7.4 Hz, 3H). $^{13}$C NMR (101 MHz, Chloroform-*d*) δ 165.25, 163.21, 161.95, 156.05, 154.25, 134.64, 133.77, 131.91, 125.65, 122.97, 122.46, 118.30, 110.68, 109.86, 104.99, 99.73, 68.64, 55.61, 31.15, 19.21, 13.80.

**12** (78%)**:** $^{1}$H NMR (500 MHz, Chloroform-*d*) δ 8.37 – 8.31 (m, 2H), 8.28 (d, *J* = 8.7 Hz, 2H), 8.17 (d, *J* = 8.9 Hz, 2H), 7.46 – 7.38 (m, 4H), 7.01 (d, *J* = 8.9 Hz, 2H), 3.92 (s, 3H). $^{13}$C NMR (101 MHz, Chloroform-*d*) δ 164.31, 164.24, 155.89, 155.68, 145.50, 132.50, 132.02, 125.88, 125.33, 122.66, 122.38, 121.13, 114.05, 55.60.

**14a** (75%)**:** $^{1}$H NMR (400 MHz, Chloroform-*d*) δ 8.31 (d, *J* = 9.0 Hz, 2H), 7.45 (d, *J* = 9.1 Hz, 2H), 7.19 (d, *J* = 9.4 Hz, 2H), 5.40 (s, 1H), 4.25 (dd, *J* = 11.6, 4.6 Hz, 2H), 3.54 (t, *J* = 11.4 Hz, 2H), 2.24 – 2.04 (m, 1H), 1.39 – 1.31 (m, 2H), 1.10 (q, *J* = 7.2 Hz, 2H), 0.93 (t, *J* = 7.3 Hz, 3H). $^{13}$C NMR (101 MHz, Chloroform-*d*) δ 162.31, 159.74, 158.82, 154.93, 145.70, 125.32, 122.57, 110.26, 98.73, 72.59, 70.70, 33.90, 30.21, 19.53, 14.19.

**14b** (76%)**:** $^{1}$H NMR (400 MHz, Chloroform-*d*) δ 7.91 (d, *J* = 8.9 Hz, 1H), 7.13 (dt, *J* = 10.2, 1.2 Hz, 2H), 6.96 (d, *J* = 2.3 Hz, 1H), 6.89 (dd, *J* = 8.9, 2.3 Hz, 1H), 5.33 (s, 1H), 4.25 – 4.13 (m, 2H), 3.92 (s, 3H), 3.47 (ddd, *J* = 11.6, 10.3, 1.5 Hz, 2H), 2.07 (ttt, *J* = 11.5, 7.0, 4.6 Hz, 1H), 1.35 – 1.21 (m, 2H), 1.08 – 0.97 (m, 2H), 0.87 (t, *J* = 7.3 Hz, 3H). $^{13}$C NMR (101 MHz, Chloroform-*d*) δ 162.32, 162.26, 159.74, 158.83, 154.48, 154.44, 137.27, 127.21, 113.51, 110.52, 110.49, 110.29, 110.25, 107.37, 98.77, 98.75, 72.60, 56.87, 33.90, 30.22, 19.53, 14.19.

**15** (84%)**:** $^{1}$H NMR (500 MHz, Chloroform-*d*) δ 8.21 – 8.11 (m, 2H), 7.42 (t, *J* = 8.7 Hz, 1H), 7.22 – 7.16 (m, 2H), 7.15 – 7.09 (m, 2H), 7.04 – 6.98 (m, 2H), 3.91 (s, 3H). $^{13}$C NMR (101 MHz, Chloroform-*d*) δ 164.26, 160.65, 158.16, 152.48, 151.98, 149.91, 132.46, 131.03, 130.50, 121.10, 118.39, 114.01, 113.06, 110.93, 55.58.

**13:** A mixture of compound **D1** (2.0 g, 6.62 mmol), 20 mL THF and 2~3 drop DMF were loaded in a 50 mL two-necked bottle with a condenser, then thionyl chloride (6.30 g, 3.84 mL, 52.93 mL) was slowly added for 10 minutes by an injector. Then the mixture refluxed for 16 h and concentrated under reduced pressure. The residue was collected and directly used for next step reaction. A 100 round bottom flask charged with N-methyl-4-nitroaniline (2.0 g, 13.14 mmol), triethylamine (15 mmol), and 50 mL dichloromethane solvent was cooled to 0 °C under nitrogen. Then the obtained acyl chloride dissolved in 10 mL dichloromethane was slowly added into the mixture. The solution was further stirred at room temperature for 20 h. The solution was stripped of solvent by rotary evaporation, and dry-loaded onto a silica gel column for purification. Appearance: white powder, 4.21 g, yield: 88%. $^{1}$H NMR (400 MHz, Chloroform-*d*) δ 8.18 – 8.05 (m, 4H), 7.44 – 7.37 (m, 2H), 7.25 – 7.17 (m, 2H), 7.16 – 7.08 (m, 2H), 7.01 – 6.93 (m, 2H), 3.89 (s, 3H), 3.57

(s, 3H). $^{13}$C NMR (101 MHz, Chloroform-*d*) δ 169.80, 164.27, 164.13, 152.71, 150.59, 145.20, 132.35, 132.15, 130.30, 126.59, 124.75, 121.75, 121.28, 113.94, 77.24, 55.55, 38.30.

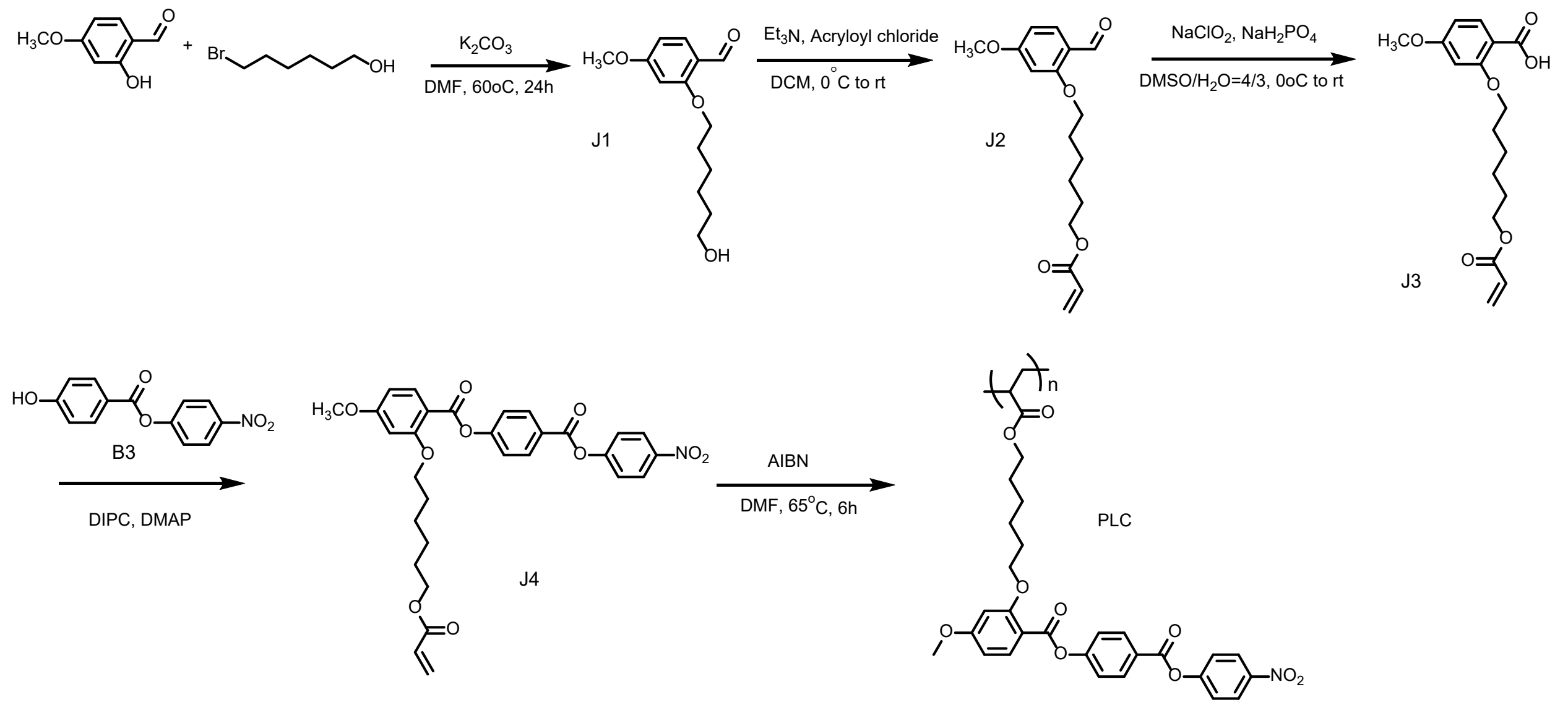

**J1:** A 50mL round bottom flask was charged with 2-hydroxy-4-methoxybenzaldehyde (2.50 g, 32.86 mmol), potassium carbonate (9.08 g, 65.73 mmol), and 30 mL DMF. The solution was degassed by bubbling nitrogen for 3 min. Then 6-bromohexan-1-ol (5.2 mL, 7.14g, 39.44 mmol) was added via injector. The solution was heated at 60°C and vigorously stirred for overnight. The mixture was cooled to room temperature, 200mL water poured, and extracted with EA. The organic phase was washed with water and brine, then dried with anhydrous $MgSO_4$. The solvent was removed by rotary evaporation, and dry-loaded onto a silica gel column for purification using EA/hexane as eluent. Yield: 7.23 g, 87.2%, appearance: colorless oil. $^1$H NMR (400 MHz, Chloroform-*d*) δ 10.33 (s, 1H), 7.81 (d, *J* = 8.7 Hz, 1H), 6.54 (d, *J* = 10.4 Hz, 1H), 6.43 (d, *J* = 2.2 Hz, 1H), 4.05 (t, *J* = 6.4 Hz, 2H), 3.87 (s, 3H), 3.67 (t, *J* = 6.5 Hz, 2H), 1.93 – 1.81 (m, 2H), 1.65 – 1.48 (m, 6H). $^{13}$C NMR (101 MHz, Chloroform-*d*) δ 188.43, 166.21, 163.33, 162.62, 130.17, 118.90, 105.92, 98.42, 68.37, 62.45, 62.43, 55.58, 36.49, 32.54, 31.40, 28.91, 25.84, 25.49.

**J2:** The compound **J1** (5.0 g, 19.8 mmol), triethylamine (3.01 g, 4.12 ml, 29.73 mmol) and 20 mL anhydrous DCM were added in a dried flask, flashed with nitrogen for 5 min. The mixture was stirred under ice bath for 5min, then acryloyl chloride (2.15 g, 1.93 mL) was added dropwisely via injector over 10 min. The solution was then slowly warmed to room temperature and stirred overnight. The organic phase was washed with water and brine, then dried with anhydrous $MgSO_4$. The solvent was removed by rotary evaporation and dry-loaded onto a silica gel column for purification using EA/hexane as eluent. Yield: 4.98 g, 82%, appearance: yellow powder. $^1$H NMR (400 MHz, Chloroform-*d*) δ 10.33 (d, *J* = 0.8 Hz, 1H, CHO), 7.80 (d, *J* = 8.7 Hz, 1H, ArH), 6.53 (ddd, *J* = 8.7, 2.2, 0.8 Hz, 1H, ArH), 6.45 – 6.35 (m, 2H, $CH_2$), 6.12 (dd, *J* = 17.3, 10.4 Hz, 1H, ArH), 5.82 (dd, *J* = 10.4, 1.5 Hz, 1H, CH), 4.17 (t, *J* = 6.6 Hz, 2H, $CH_2$), 4.04 (t, *J* = 6.3 Hz, 2H, $CH_2$), 3.86 (s, 3H, $CH_3$), 1.92 – 1.81 (m, 2H, $CH_2$), 1.78 – 1.66 (m, 2H, $CH_2$), 1.62 – 1.40 (m, 4H, $CH_2$). $^{13}$C NMR (101 MHz, Chloroform-*d*) δ 188.26, 166.27, 166.15, 163.24, 130.56, 130.23, 128.53, 119.02, 105.87, 98.49, 68.28, 64.40, 55.59, 28.88, 28.53, 25.74, 25.69.

**J3:** synthesized by general aldehyde oxidation method as in **A7** – **A11** synthesis. Yield: 4.05 g (96%); appearance: faint yellow powder. $^1$H NMR (400 MHz, Chloroform-*d*) δ 10.34 (s, 1H, COOH), 8.08 (dd, *J* = 8.8, 0.9 Hz, 1H ArH), 6.62 (dd, *J* = 8.8, 2.2 Hz, 1H ArH), 6.52 (d, *J* = 2.3 Hz, 1H ArH), 6.40 (dd, *J* = 17.3, 1.5 Hz, 1H, CH), 6.12 (dd, *J* = 17.3, 10.4 Hz, 1H, $CH_2$), 5.83 (dd, *J* = 10.4, 1.5 Hz, 1H, $CH_2$), 4.19 (dt, *J* = 16.2, 6.6 Hz, 4H, $CH_2$), 3.87 (s, 3H, $CH_3$), 1.99 – 1.86 (m, 2H, $CH_2$), 1.72 (p, *J* = 6.8 Hz, 2H, $CH_2$), 1.63 – 1.42 (m, 4H, $CH_2$). $^{13}$C NMR (101 MHz, Chloroform-*d*) δ 166.29, 165.40, 165.05, 158.97, 135.42, 130.66, 128.48, 110.45, 106.67, 99.41, 69.97, 64.27, 55.74, 28.76, 28.45, 25.62, 25.56.

**J4:** synthesized by the general esterification method as in the synthetic procedures of **1a-1f**. Appearance: white powder, 2.46g, yield 77.4%. $^1$H NMR (400 MHz, Chloroform-*d*) δ 8.42 – 8.19 (m, 4H), 8.06 (d, *J* = 8.8 Hz, 1H), 7.51 – 7.32 (m, 4H), 6.67 – 6.46 (m, 2H), 6.38 (dd, *J* = 17.3, 1.4 Hz, 1H), 6.10 (dd, *J* = 17.3, 10.4 Hz, 1H), 5.79 (dd, *J* = 10.4, 1.4 Hz, 1H), 4.09 (dt, *J* = 22.2, 6.5 Hz, 4H), 3.89 (s, 3H), 1.86 (h, *J* = 8.1, 7.3 Hz, 2H), 1.68 – 1.61 (m, 2H), 1.56 (dt, *J* = 15.0, 7.3 Hz, 2H), 1.43 (q, *J* = 8.8, 8.2 Hz, 2H). $^{13}$C NMR (101 MHz, Chloroform-*d*) δ 165.28, 163.62, 163.12, 161.87, 156.09, 155.72, 145.43, 134.67, 131.96, 130.55, 128.56, 125.57, 125.30, 122.68, 122.48, 110.62, 105.06, 99.73, 68.71, 64.46, 55.63, 29.01, 28.53, 25.69, 25.65.

**P1:** The monomer **J4** (1.0 g), AIBN (0.01 equiv) and 3 mL DMF were combined into a Schlenk flask. After freeze-pump-thaw cycles with $N_2$ for three times, the mixture was put into 65°C oil bath and stirred for 12 hr. After cooling down to room temperature, the mixture was precipitated into methanol. White solid was collected by filtration and purified by precipitation for two more times. The white powder product was obtained and dried. Yield: 50%. $^1$H NMR (400 MHz, Chloroform-*d*) δ 8.41 – 7.84 (m, 6H), 7.57 – 7.06 (m, 5H), 6.46 (p, *J* = 20.6, 18.6 Hz, 2H), 4.38 – 3.34 (m, 7H), 1.83 (d, *J* = 38.9 Hz, 2H), 1.52 (d, *J* = 37.6 Hz, 4H), 1.28 (d, *J* = 20.0 Hz, 2H).

Supplementary Text

Discussion S1: The effect of the thickness on the texture of the PN phase

Through a massive optical observation of the synthesized materials, we find the type of the texture depends on the thickness. Figs. S1 and S2 demonstrates the detailed textural evolutions at different temperatures from the N phase to the PN phase under various sample preparation conditions for **5b** and **1a**. As a general trend, as discussed in the main text, all the samples exhibit the Schlieren texture in the N phase, the stripe texture corresponding to the flexoelectric-effect-induced splay undulation appear near the N-PN phase transition, and the defect lines appear in the PN phase. While the variations of the thickness and the anchoring condition do not change much the appearance of the Schlieren (or homogeneous planar, LX-1400 alignment layer, Hitachi Chemical Co., Ltd.) and stripe textures, they sensitively affect the line-defect-mediated texture in the PN phase. When the cell thickness is small (c.a. < 5 μm), a periodic arrangement of the line defects along the rubbing or local N director direction is always observed as discussed in the main text. Unsimilar to the N phase, the domains between the two line defects do not show the extinction upon the rotation of the sample. This suggests the existence of the director rotation occur along the cell normal direction, consistent with the Splay-Néel Type I PN model (Figs. S7 and S8, Discussion S3). The line defects are stable without changing the appearance upon varying temperature. On the other hand, when the cell thickness is large (c.a. >5 μm), the line defects appear more dispersed in the cell without certain periodicity. Each domain separated by the line defects are darker than the texture in thin cells under POM as reported previously, but still not completely extinct. Fig. S3 shows the variation of texture, dielectric and SHG properties as a function of temperature in **1a**. The surface is treated planarly with an alignment agent (KPI-3000, Shenzhen Haihao Telecom Corp.). Under this surface condition, the line disclinations of the PN phase is clearly observed. The variation process of the texture is same with those shown in Figs. 3B-3G, 3I-3N.

Discussion S2: Evolution of the phase structure: experimental and model

As we mentioned in the main text, due to the novelty of the PN phase and the limited access to the material state, there is a plenty room to discuss the details of the structure. Also, the previous observations of the PN phase remains contradictory and the under debate (*18, 19*). Moreover, there are limited insights on the details of how the phase structure is developed from the high-temperature N phase to the PN phase, thereby providing no criteria of how to judge the incidence of the PN phase. As shown in in the main text, we confirm the textural evolution for all the synthesized PN materials is common, and the textural features contains many important information that previous studies have not addressed before. Crucially, our observation of POM, dielectric and SHG measurements have been made simultaneously, firmly confirming the relationship between the polarity and each textural state.

In details, by lowering temperature towards the PN phase, a sequential topological variation including three transient structures is clarified. Cooling from the Schlieren texture of the nematic phase in a nontreated planar slab, near the N-PN transition, a well-ordered stripe undulation of director field is found to be imposed on the Schlieren texture. The stripes that run parallel to the local director gradually develop the contrast with decreasing temperature within about 15 °C from the N-PN transition (Fig. S4). According to Polscope imaging technique in a thin cell of 2.1 μm, the stripe modulation is the embodiment of the periodic splay deformation as shown in Figs. 4D-4F. Worth noting, this splay undulation in the N phase is similar to the proposed structure for the splay phase of RM734 in the Ref. (*18*). According to our observation, it is difficult to find the stripe texture in unidirectionally aligned slabs like rubbed cells except near the N-PN transition temperature because the planar anchoring overwhelms the incidence. Another aspect is, Gim et al. reported a similar texture appears near the nematic-smectic phase transition in the so-called hybrid geometry, where the coexisting planar and homeotropic alignment conditions adopt director rotating by π/2 in the depth direction (*32*). They conclude the incidence of the stripes is driven by the divergence of the ratio, $K_{33}/K_{11}$, of the nematic bend to splay elastic constants based on theoretical consideration. Indeed, the splay elastic constant $K_{11}$ decreases with decreasing temperature in the N phase, confirmed in most of the current materials, leading to the dramatic increase of $K_{33}/K_{11}$. However, it should be noted that the situation still has no driving force for the stripe modulation to appear here if only $K_{33}/K_{11}$ matters. The difference compared to Ref. (*32*) arises because we start with a uniform planar-aligned N phase, where the surface anchoring can be assumed to be infinitely strong so the initial undistorted director field of the nematic phase along the depth direction should sustain and no elastic instability is expected. The plausible mechanism is the pretransitional growth of the polar ordering, evidenced by the drastic increase of dielectric permittivity and SH signal in the corresponding temperature window (Figs. 3A and 3H). The stripe disappears in the vicinity of the N-PN transition and replaced by the band texture as discussed in the main text. Sebastián et al. also reported the observable area and contrast of the stripe undulation of RM734 (**1a**) significantly decrease in the PN phase (*18*) (also see Discussion S3). Cooling from the stripe texture leads to the band texture in the N phase, 1-2°C above the N-PN transition. This state remarkably resembles the PN phase in structure and polarity. Especially, the reported high polarity in the PN phase of RM734 already develops at this stage. With further decreasing temperature, filamentous regimes with some periodicity accompanied by different interference color than the surroundings appear (e.g. Fig. 3E). Upon the transition to the PN phase, a swarm of line defects similarly observed in Ref. (*19*) pop up, seeded by the filamentous domains. Each domain separated by the defect is ferroelectric and the adjacent domains exhibit opposite polarity.

In Fig. S7, several possible structures for the PN phase is drawn. Considering the above-discussed circumstances, several plausible topological structures are possible for the line defects: Splay-Bloch-type, two types of Splay-Néel-type defects or defect wall. To judge which corresponds to the current texture of the PN phase, we start the discussion from the filamentous regimes because the evolution from the filamentous regimes to line defects provides important transient information. As stated above, the filamentous regimes exhibit different interference color than the surroundings, suggesting the existence of in-plane or out-of-plane rotation of the director. Consider the Bloch defect carries a twist deformation by $\pi$ along the direction normal to the local director. Taking the strong planar anchoring into count, while the twist deformation exists in the surface plane, the splay and bend deformations do in the surface-normal direction. No in-plane azimuthal variation of the director is accompanied so the global extinction, i.e. dark as the surroundings, is expected if the local director is parallel or perpendicular to polarizers under crossed polarizers. On the other hand, the Splay-Néel-type defect exhibits an azimuthal rotation by $\pi$ to connect the neighboring domains. Therefore, while the splay and bend deformations exist parallel to the surface plane, the twist does in the surface-normal direction. This results in the variation of light transmission intensity around the defect under crossed polarizers (Fig. S8). When we rotate the sample under POM, no extinction state for the filamentous regimes appears. The interference color of the regimes remains unchanged. This suggests that the Splay-Néel-type defect model fits the current situation, and the invariance of interference color should arise because the optical rotation of the twisted structure along the cell normal direction in the filamentous regimes. Going into the PN phase, the filamentous regimes are replaced by line-like defects. Around the line-like defects, both the longitudinal and transverse optical aberrations are serious, leading to the failure of resolving the image clearly. This indicates the significant director variation near the defects. This observation fits well to the Néel-type defect model. Comparing the two types of the Splay-Néel-type defects, type-II should show distinction in middle between two disclination lines. However, this contradicts the experimental observation that no extinction position exists between two disclination lines. Therefore, we attributed the structure of the PN phase to the Splay-Néel Type I defect with a mixed splay and bend director distributions. This is also strongly supported by the consistency between the experimental and calculated POM images (Fig. S8, Discussion S3).

Discussion S3: Comparison of the POM images between the experimental observation and optical calculation for possible models

Fig. S8 shows the three most probable director models for the PN phase. For each model, we compare the observed and calculated images under crossed-nicol and decrossed situations. Only the Splay-Néel Type I model matches the observation in both situations. In the POM images, if one carefully analyses the spatial distribution of the transmitted light, one finds some stripes with periodicity similar to that of the stripe texture remain in the PN phase near the N-PN phase transition. With decreasing the temperature, the contrast becomes lower. This is similar to what Sebastián et al. observed by SHG microscopy in Ref. (*18*). Based on our analyses discussed in the main text and the detailed discussion on the ferroelectricity of the domains by Chen et al. (*19*), our current conclusion is that each domain separated by line defects is ferroelectric and the observed stripes near the phase transition is the remnant of the stripe texture observed in the N phase.

Discussion S4: Parameters used for free energy calculation.

To calculate the temperature dependence of the free energy for each model, we use the experimental results of the elastic constants of **1a**, and other parameters as follows: $c = 600$ Jm$^{-4}$C$^{-1}$, $\gamma = 0.006$ V, $t = 1 \times 10^{8} \sim 1.1 \times 10^{11}$ JmC$^{-2}$, $b = 1.1 \times 10^{-4}$ m$^{2}$ V$^{2}$ N$^{-1}$. Worth noting that, by replacing the experimental results of the temperature dependence of the elastic constants for **1a** by that for **5b** and several other materials can reproduce the structural transition temperatures by the current calculation.

Discussion S5: Machine learning algorism.

The machine learning method consists of three major steps (Fig. S10). First, after selecting training data from the molecular library, a random selection of molecules from the training data is made based on bootstrap sampling, i.e., randomly selecting molecules (datasets) with allowing duplication for $i$ datasets. Second, the bootstrapped molecules are classified into detailed categories based on the random forest regression (RFR) algorism (*33*). Here, a decision tree is grown for each of the bootstrapped molecular datasets by branching the molecules to different categories, judging from randomly selected branching condition of involved physicochemical or structural parameters. As a result, by averaging the output of decision tree regressions, a subsequent prediction of new data feature can be generated. Third, to interpret the RFR output as a degree of the contribution of a certain parameter to the incidence of the PN phase, a modified SHapley Additive exPlanations (SHAP) analysis (*24*) is made, which grabs "hidden" information and predict which parameters and how much the parameters influence the stability through the nonlinear inter-parameter relationships. This enables the visualization of importance of each molecular physicochemical or structural parameters on the incidence of the PN phase. Note that negative SHAP values do not mean that the PN cannot exist at all but suggest the probability to stabilise the PN decreases.

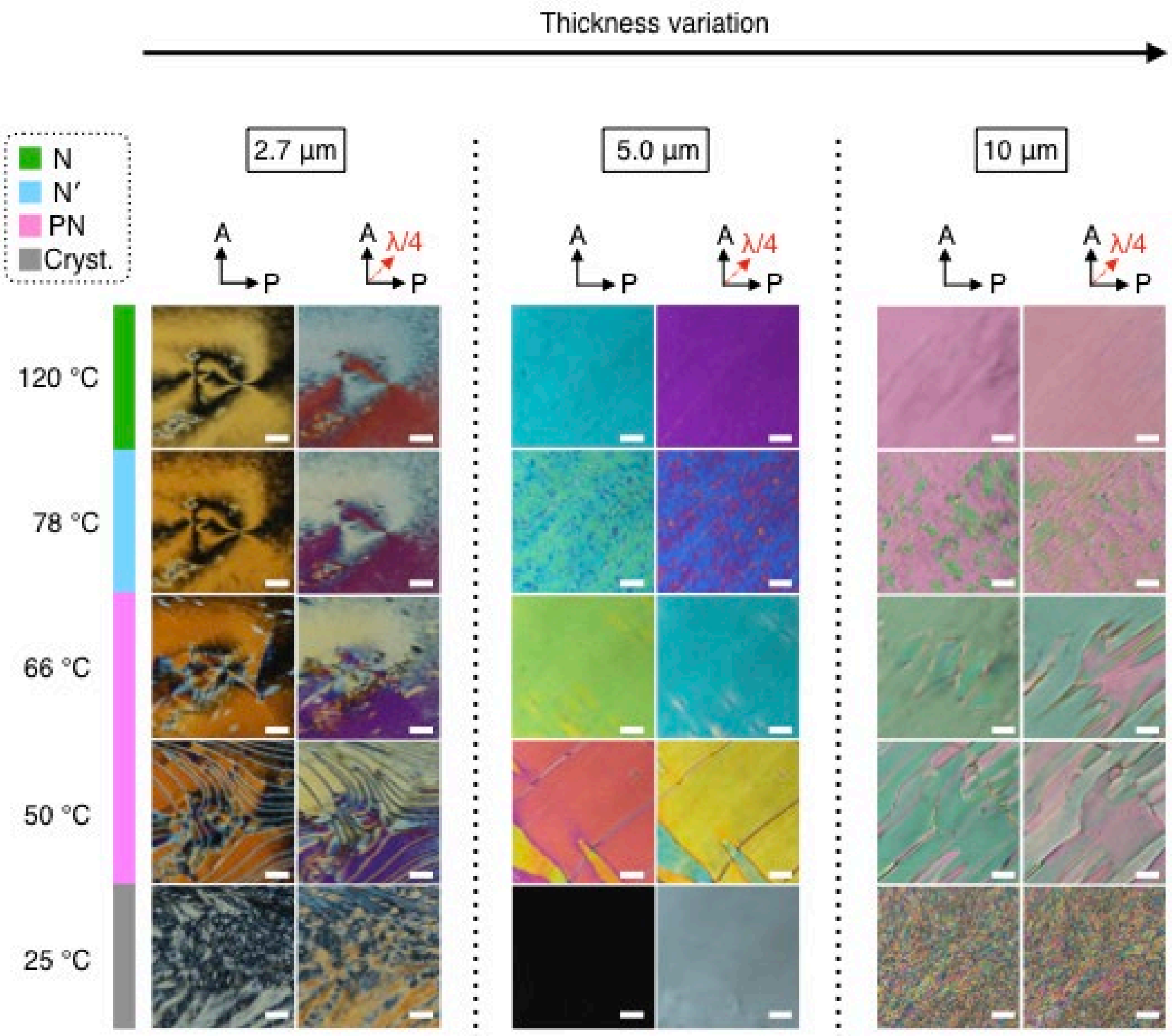

Fig. S1 Texture of 5b at various thickness conditions. Scale bar, 20 µm.

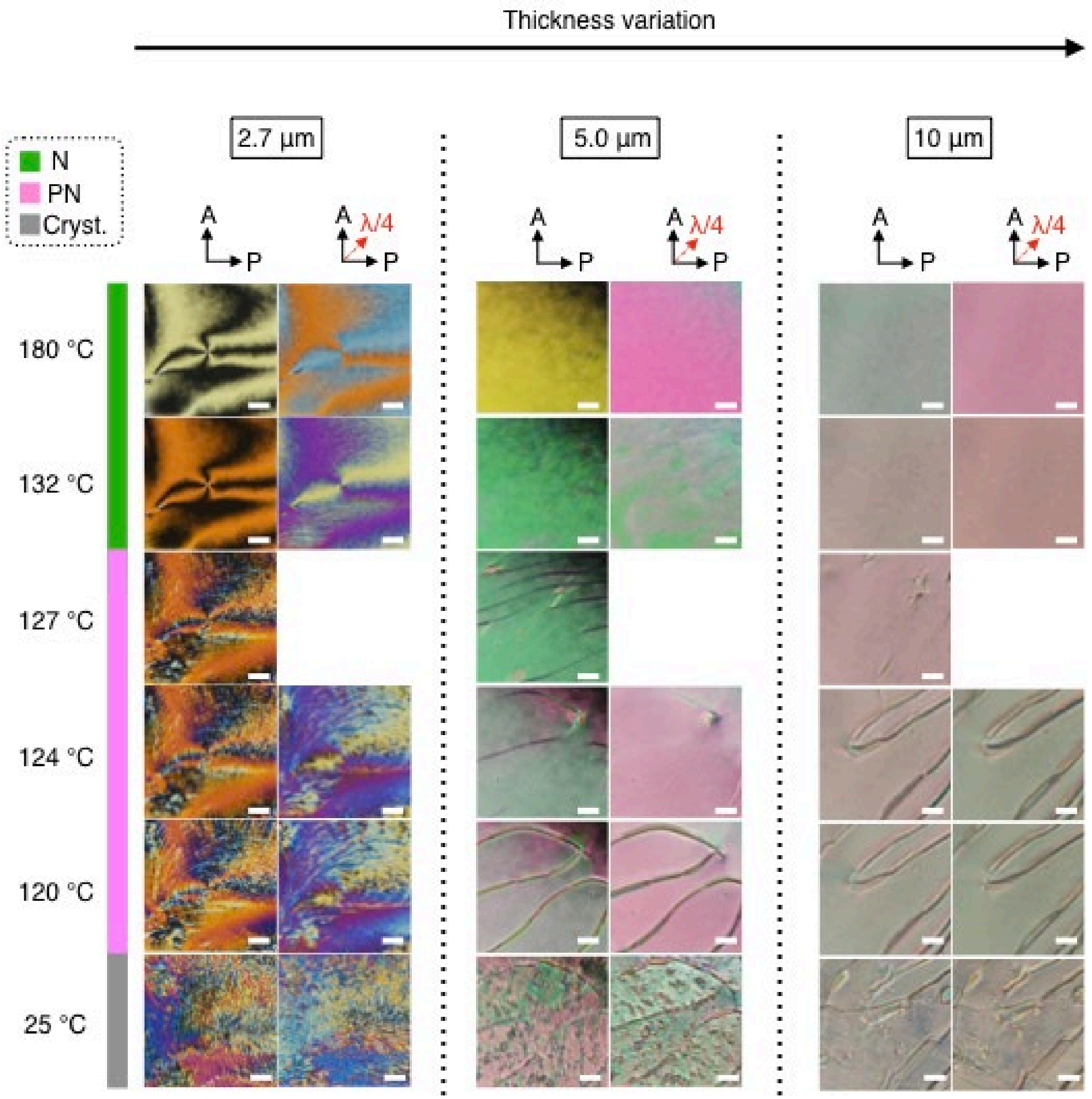

Fig. S2 Texture of 1a at various thickness conditions. Scale bar, 20 μm.

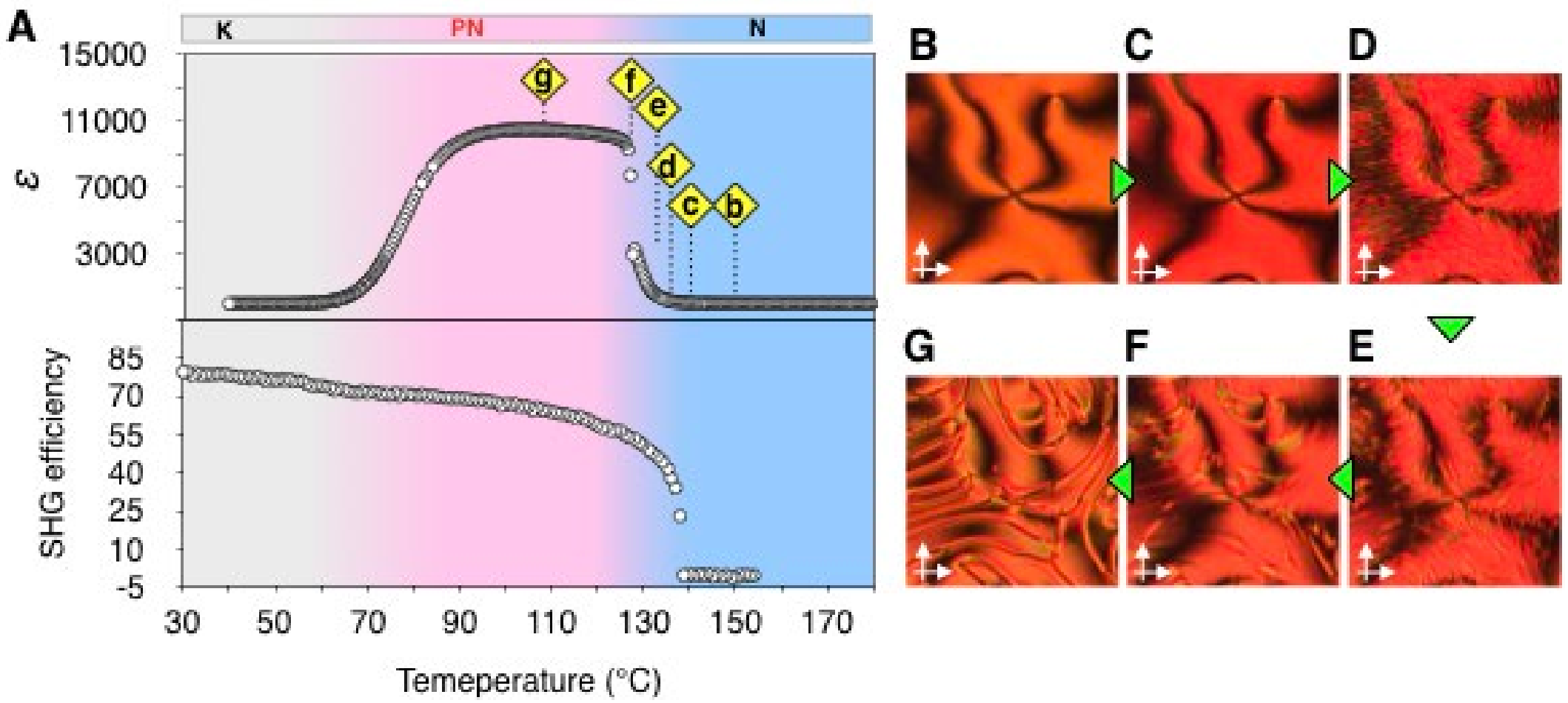

Fig. S3 Systematic studies of structure and polarity of the PN phase: SHG (A), dielectric spectroscopy (A) and POM (B-G) for 1a. Image height, 715 μm.

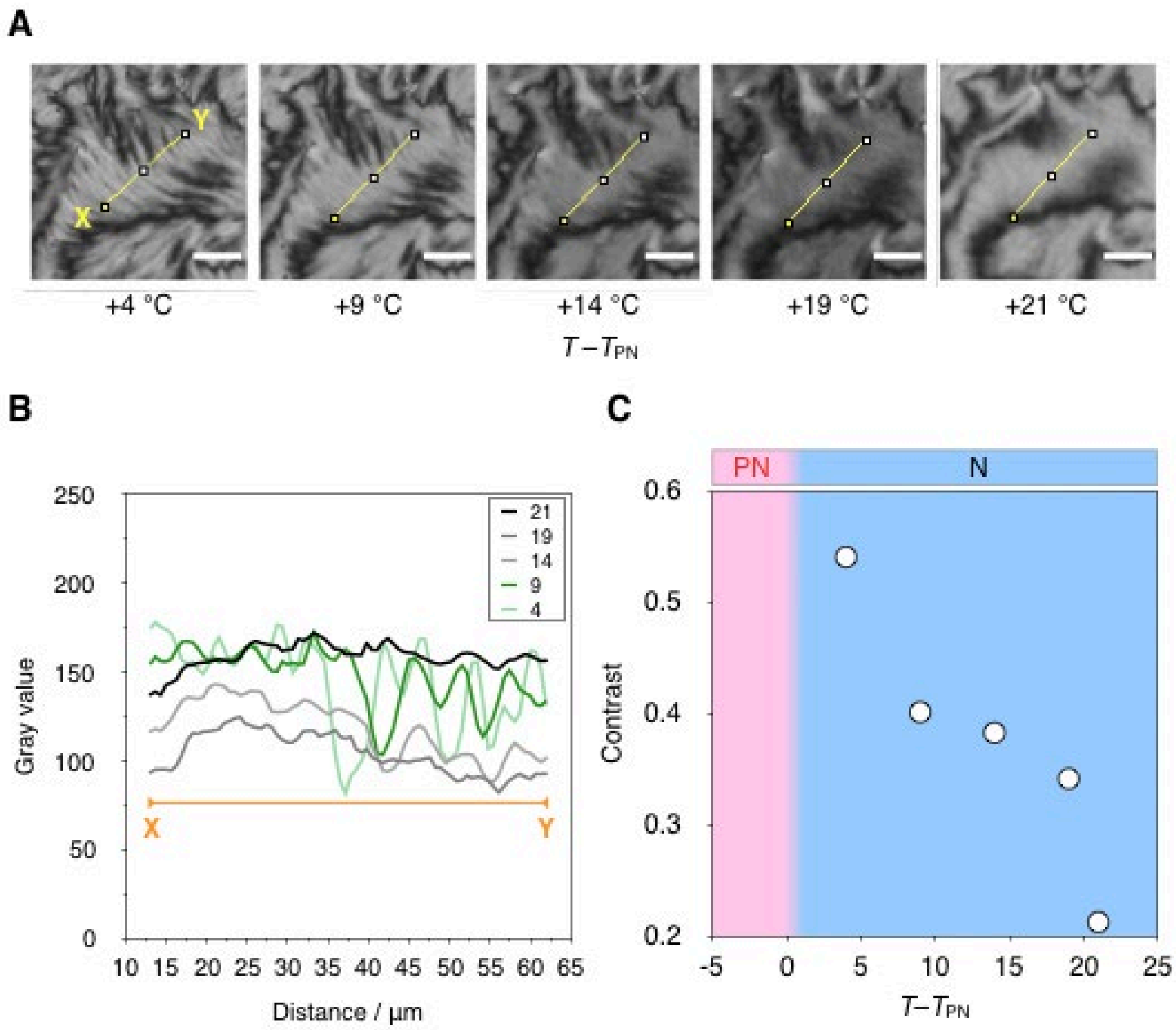

Fig. S4 (A) Stripe texture of 5c at various temperatures. Scale bar, 20 µm. (B) Contrast of the stripe texture as a function of the temperature. About 15°C above the N-PN transition, the stripe texture manifests itself. The contrast increases with decreasing temperature towards the N-PN transition.

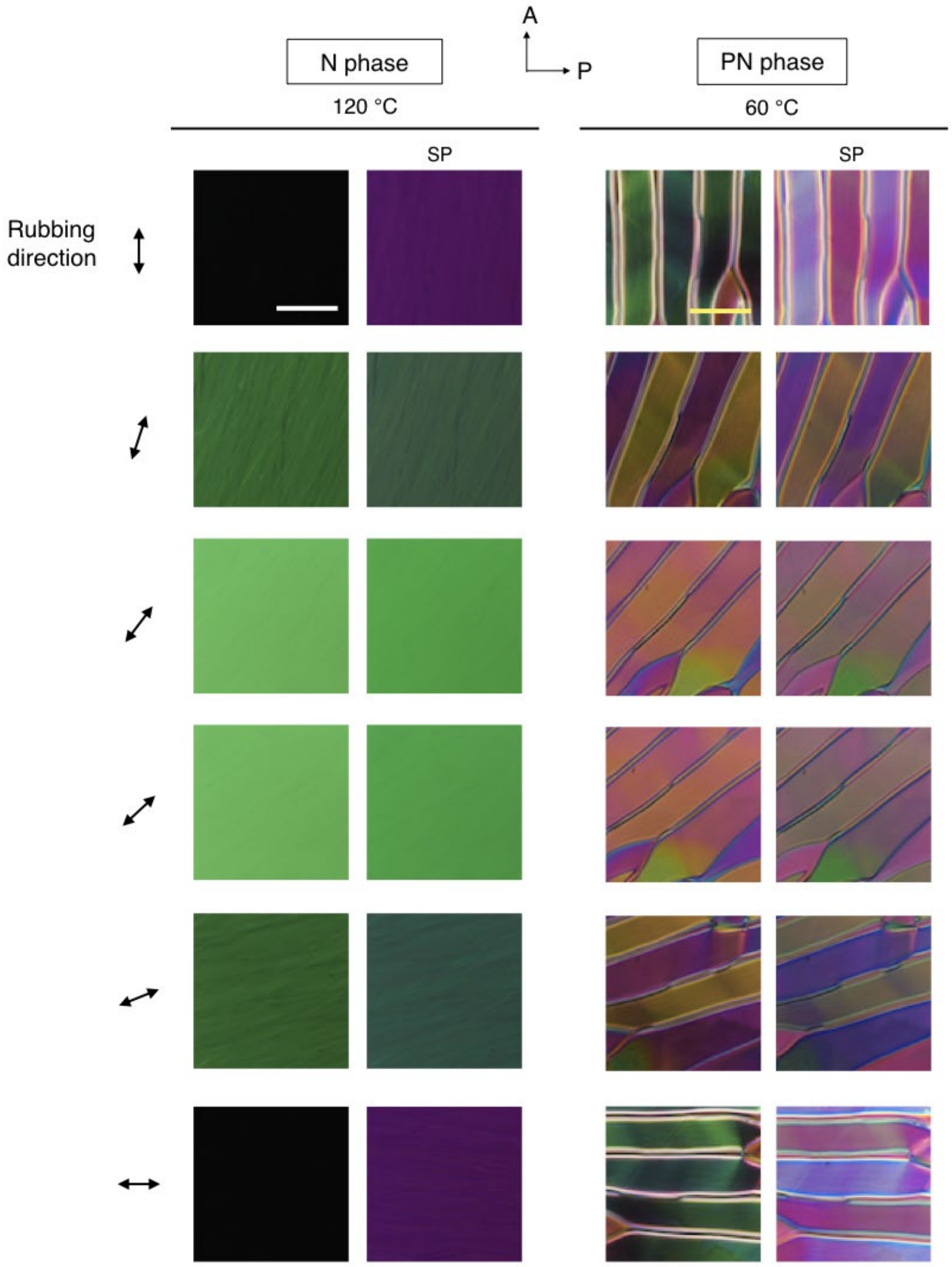

Fig. S5 Texture of 5b in a 5-μm rubbed cell. The sample is rotated under the crossed or decrossed polarisers. Importantly, no distinction position is found in the PN phase. SP indicates a tint plate is inserted. Scale bar, 50 μm.

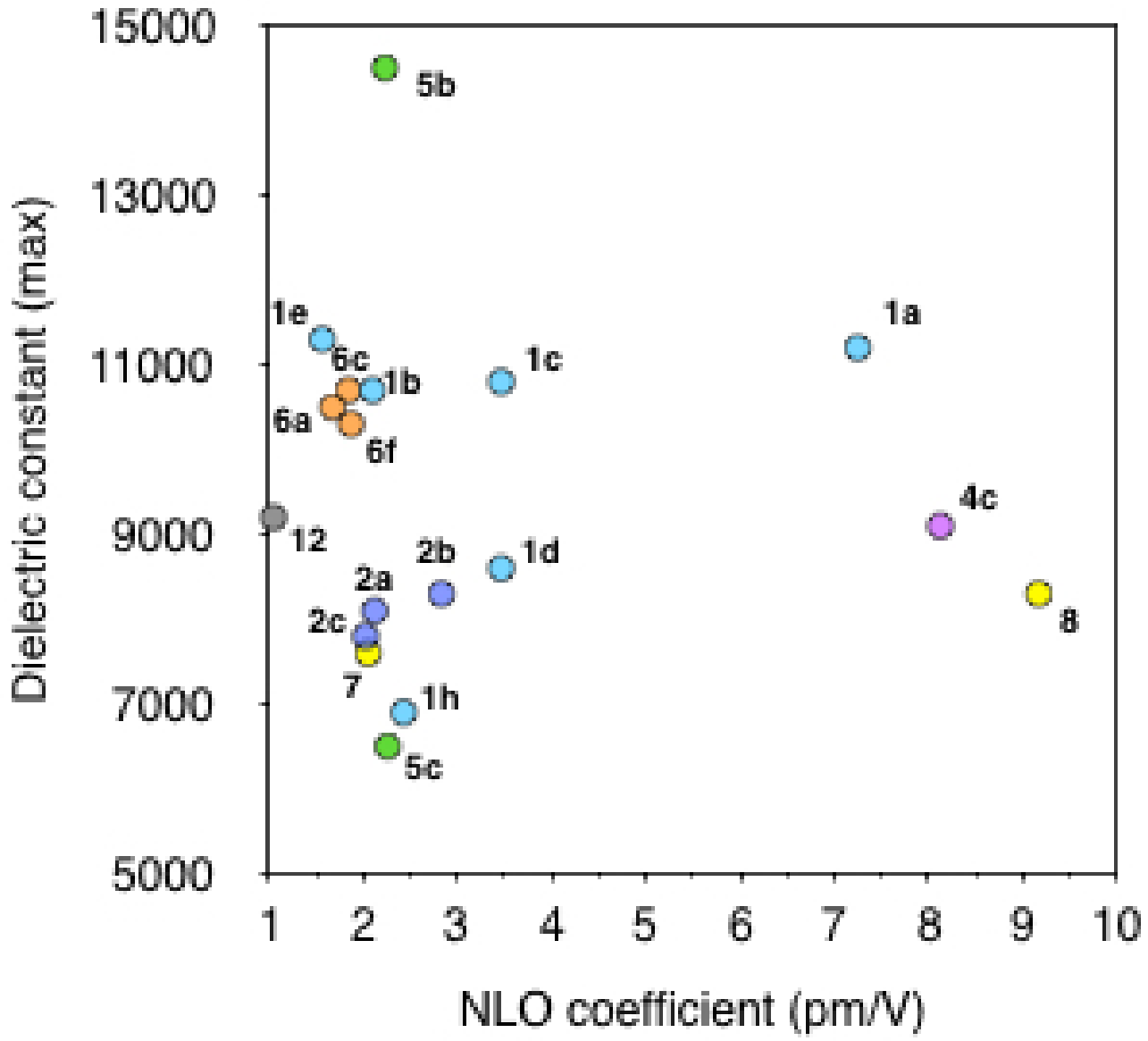

Fig. S6 Closed-up of Fig. 3p for the PN materials. The numbers indicate the names of the PN materials.

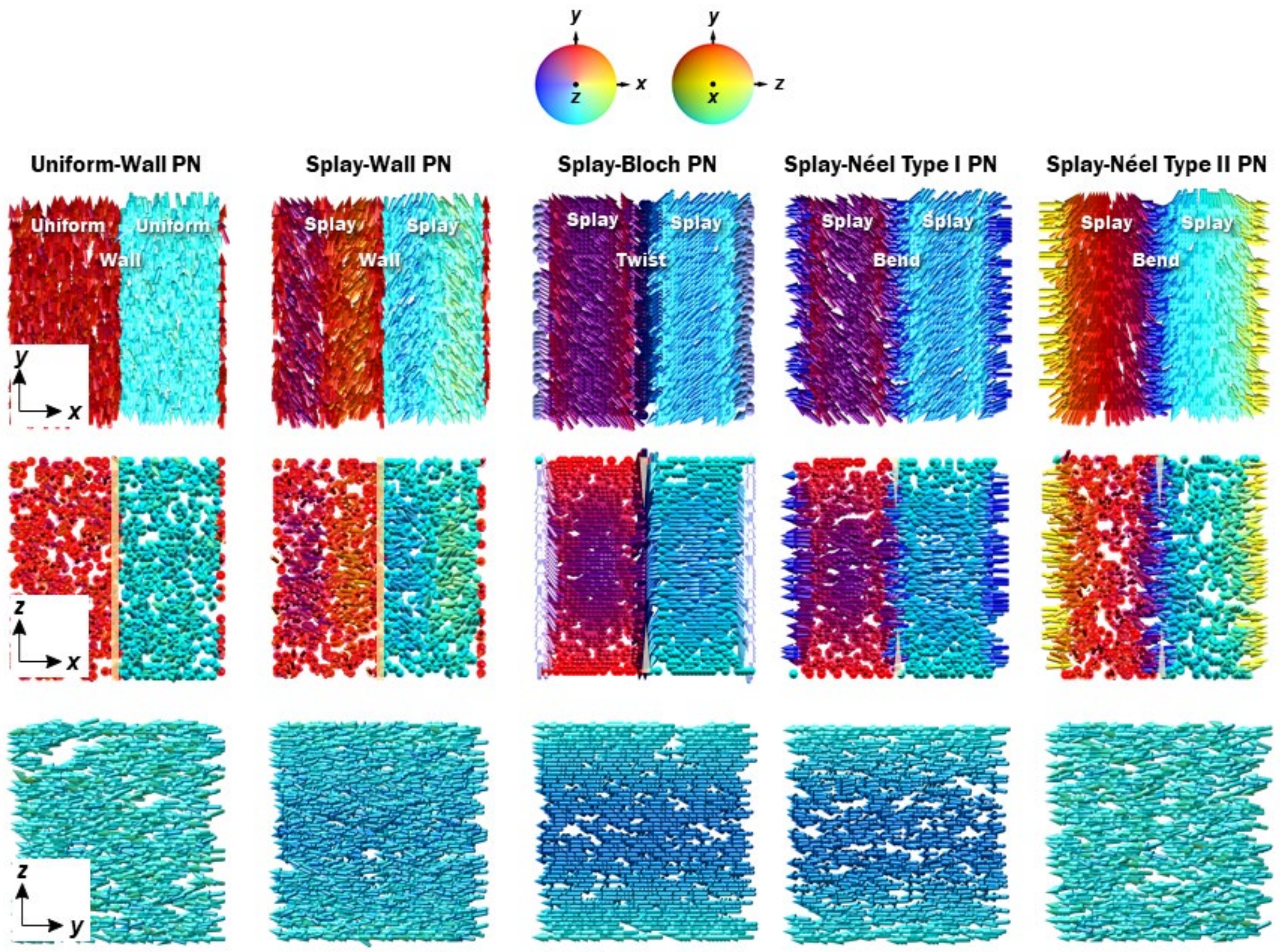

Fig. S7 Possible model structures for the PN phase.

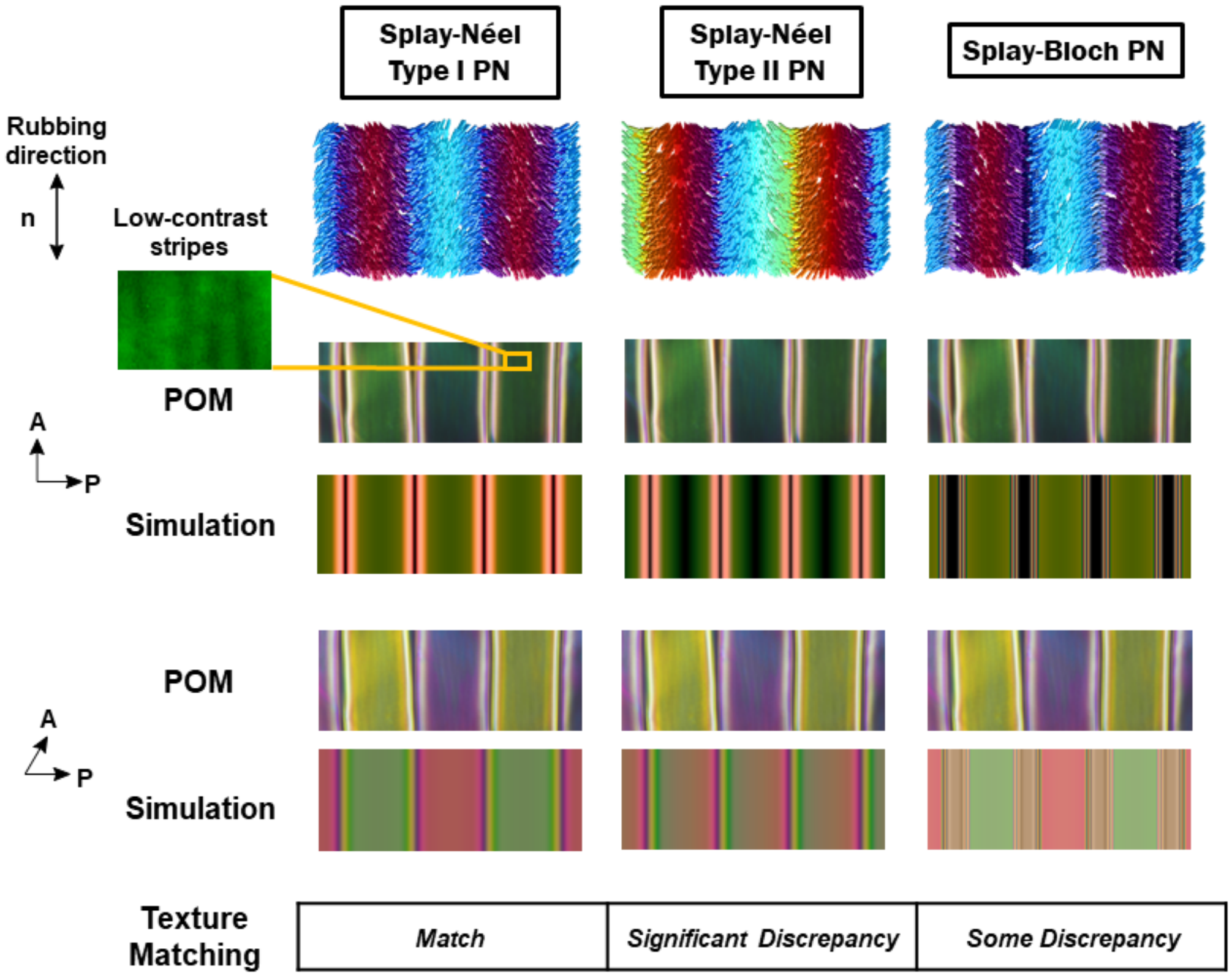

Fig. S8 Comparison of experimental and calculated POM images between different models. The stripes shown in the inset is enhanced in contrast.

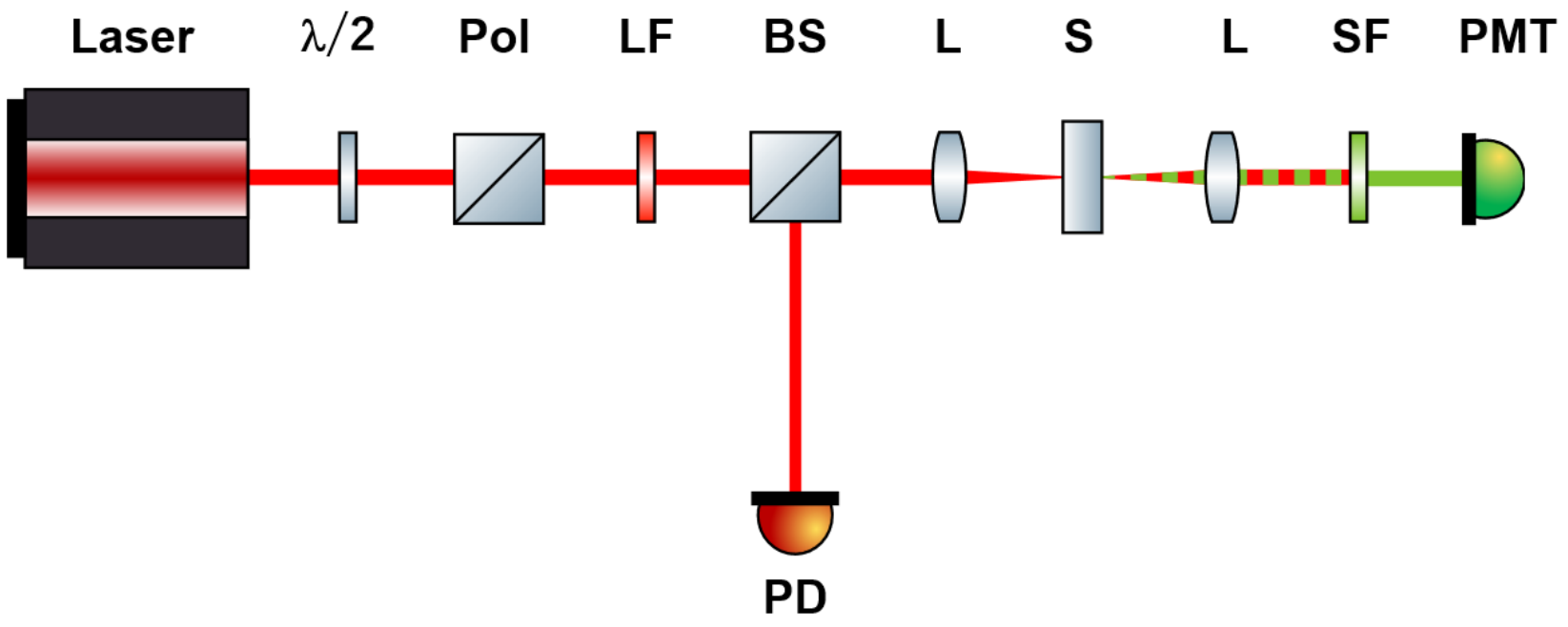

Fig. S9 The schematic of the SHG measuring system. The symbols, $\lambda/2$, Pol, LF, BS, L, S, SF, PMT and PD, denote half-wave plate, polariser, long pass filter, beam splitter, lens, sample, short pass filter or 532nm interference filter, photomultiplier tube and photodiode, respectively.

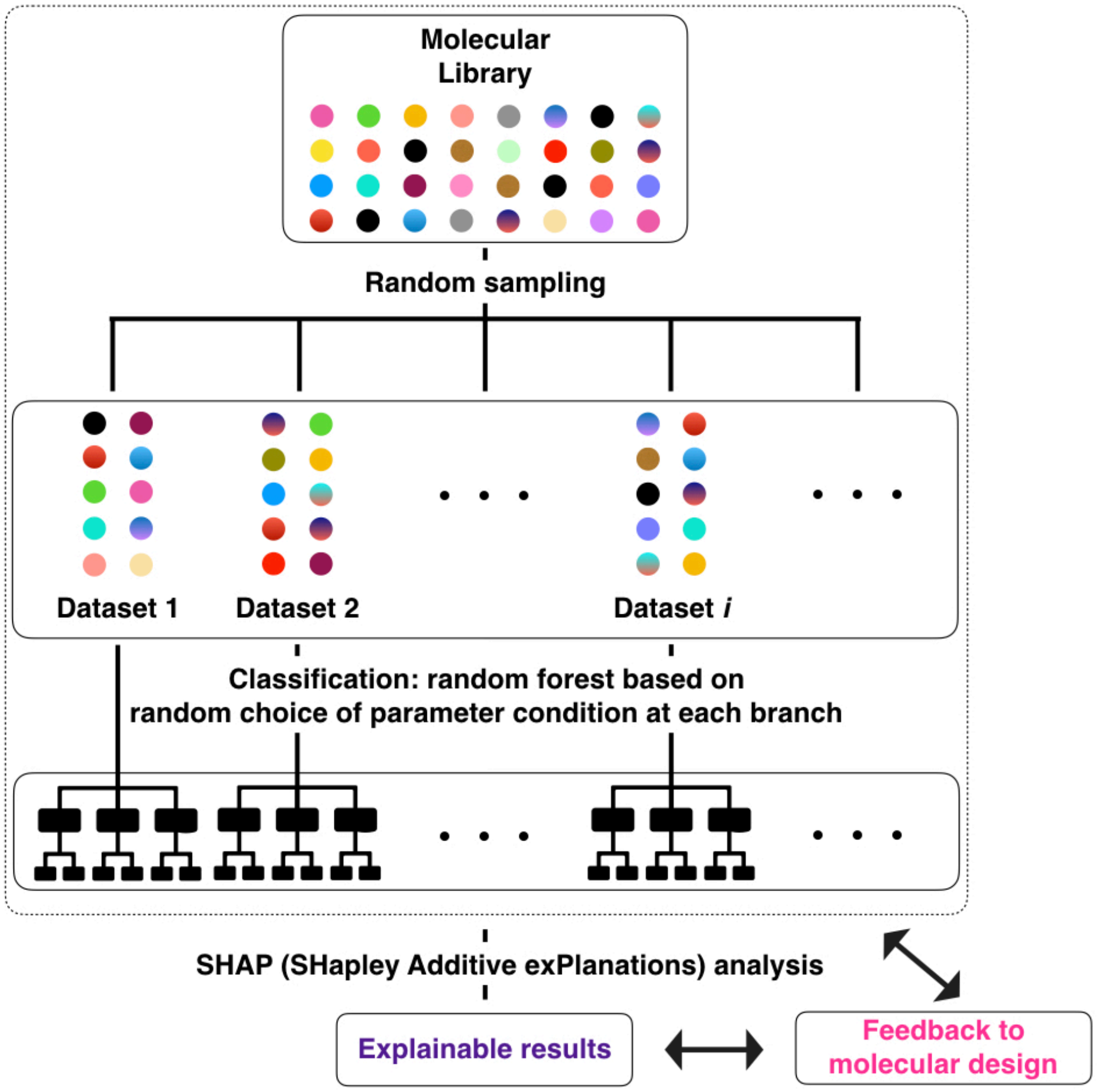

Fig. S10 Scheme of machine learning.

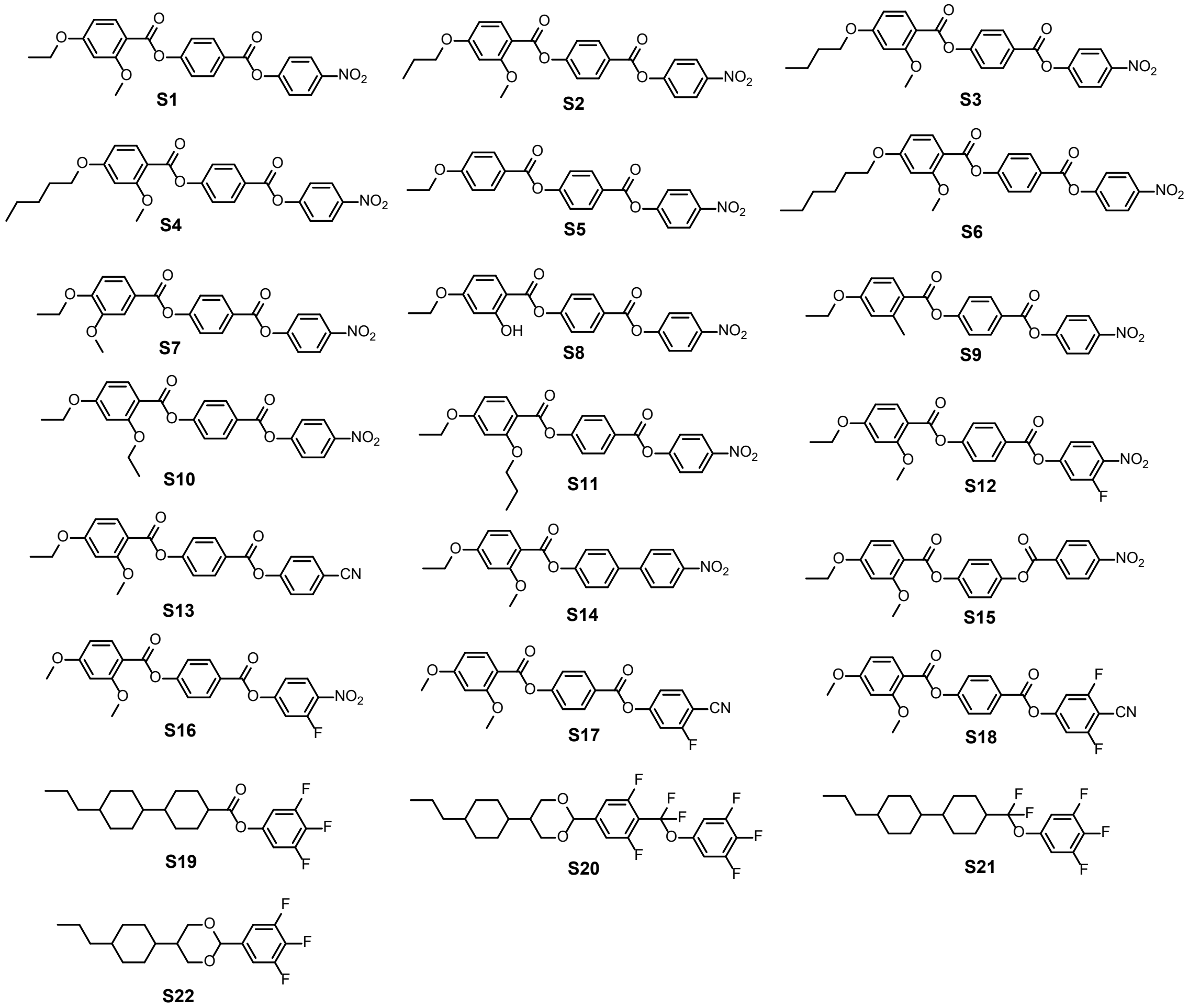

Fig. S11. Other molecular structures for machine learning.

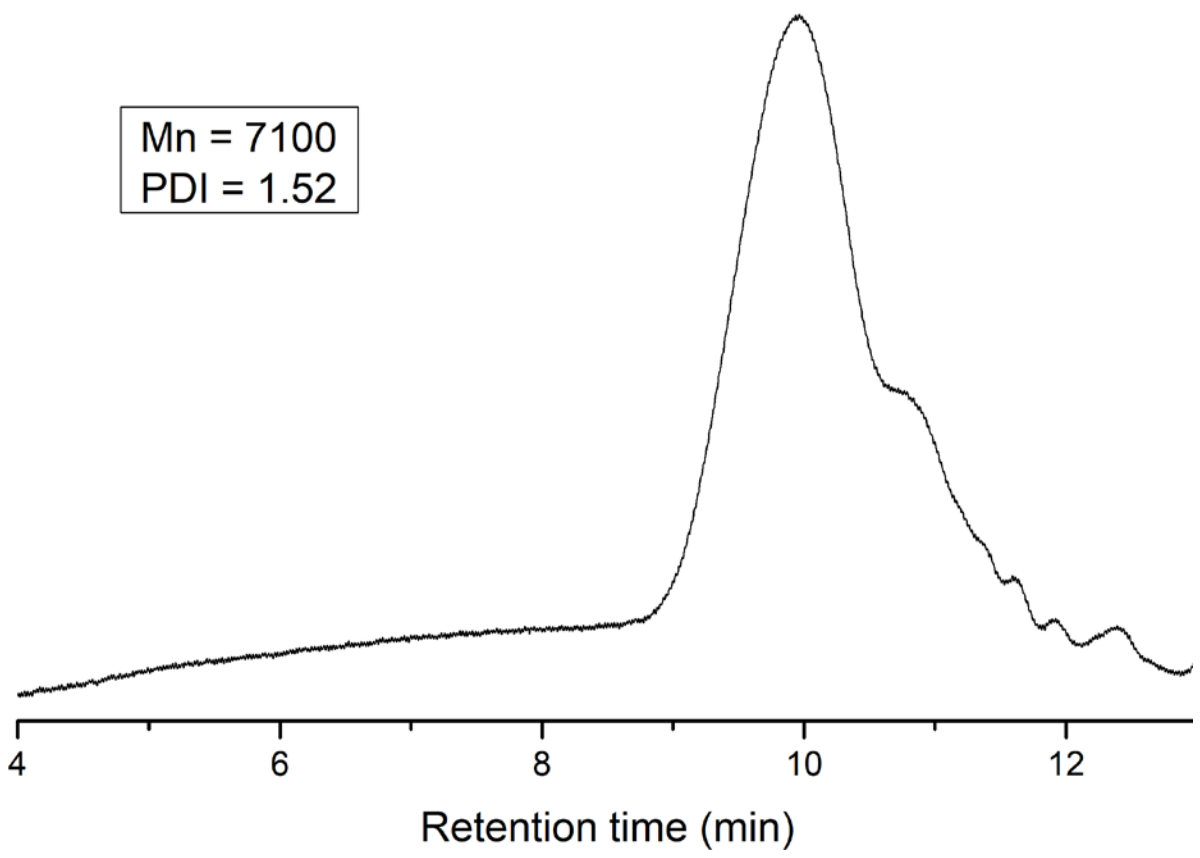

Fig. S12. The GPC result of P1.

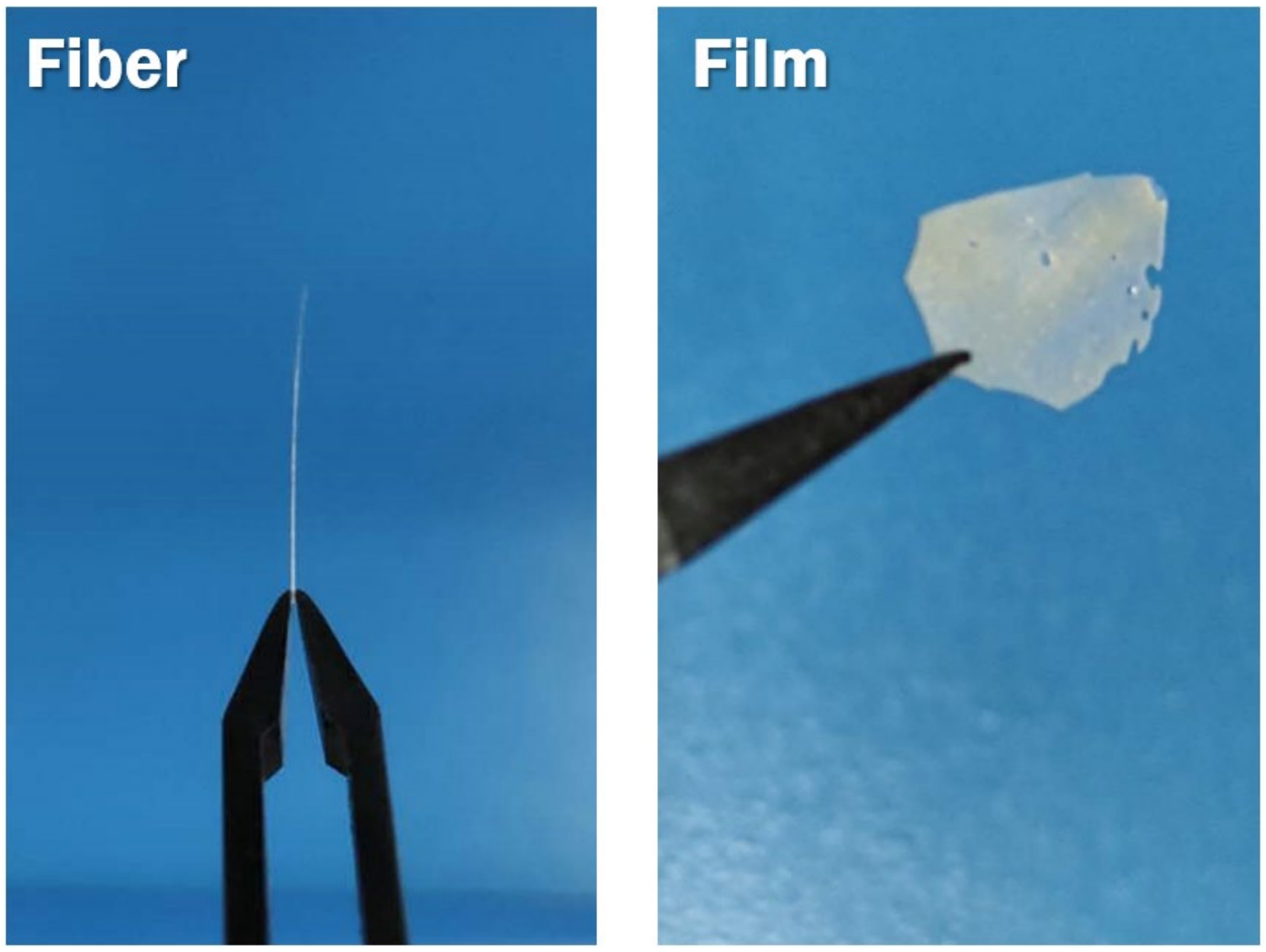

Fig. S13. Processed polar polymer fiber and polar polymer film.

Table S1. Phase sequence *vs* synthesized materials.

| Name | Phase behaviors determined by DSC (Cooling at 10 ℃/min) | Name | Phase behaviors determined by DSC (Cooling at 10 ℃/min) |
|---|---|---|---|
| 1a | Iso → 182.4℃ → N → 141.2℃ → PN → 69.1℃ → $M_x$ | **5b** | Iso → 173.5℃ → N → 83.9℃ → M2 → 68.9℃ → PN → 47℃ → Crystal |
| 1b | Iso → 96.3℃ → N → 84.8℃ → PN | **5c** | Iso →90.7℃ →N→66.1℃ →PN→ 43.6℃→ Crystal |
| 1c | Iso → 58.3℃ → N → 51.4℃ → PN | **5d** | Iso → 146.8℃ → N → 123.7℃ → PN → 96.1℃ → Crystal |
| 1d | Iso → 77.23℃ → Crystal | **6a** | Iso → 78.4℃ → Crystal |
| 1e | Iso → 62.3℃ → N → 51.4℃ → Crystal | **6b** | Iso → 97.4℃ → Crystal |
| 1f | Iso → 151.7℃ → N → 117.3℃ → Crystal | **6c** | Iso → 100.8℃ →Crystal |
| 1g | Iso → 128.2℃ → N | **6d** | Iso → 15.4℃ → N |
| 1h | Iso → 119.1℃ → N → 76.2℃ → PN | **6e** | Iso → 10.5℃ → N |
| 2a | Iso → 184.0℃ → N → 125.6℃ → PN | **6f** | Iso → 16.1℃ → N |
| 2b | Iso → 127.2℃ → N → 96.8℃ → PN → 73.2℃ → Crystal | **7** | Iso → 50.0℃ → Crystal |
| 2c | Iso → 106.3℃ → N → 105.1℃ → PN → 86.8℃ → Crystal | **8** | Iso → 110.9℃ → Crystal |
| 2d | Iso → 186.0℃ → N → 140.2℃ → Crystal | **9** | Iso → 79.4℃ → Crystal |
| 2e | Iso → 134.2℃ → N → 91.9℃ → Crystal | **10** | Iso → 61.4℃ → Crystal |
| 2f | Iso → 93.6℃ → N | **11a** | Iso → 196.3℃ → N → 122.0℃ → Crystal |
| 2g | Iso → 104.6℃ → Crystal | **11b** | Iso → 65.8℃ → N → 48.1℃ → $M_x$ |
| 2h | Iso → 97.0℃ → Crystal | **12** | Iso → 172.8℃ → Crystal |
| 3 | Iso → 161.7℃ → Crystal | **13** | Iso → 106.0℃ → Crystal |
| 4a | Iso → 209.1℃ → N → 155.5℃ → Crystal | **14a** | Iso → 67.3℃ → Crystal |
| 4b | Iso → 108.1→ Crystal | **14b** | Iso → 23.7℃ → N |
| 4c | Iso → 116.5℃ → Crystal | **15** | Iso → 134.2℃ → N → 91.9℃ → Crystal |
| 4d | Iso → 53.7℃ → N | **PLC** | Iso → 124.8℃ → N → 89.2℃ → PN → 50.7℃ → Glassy PN |
| 5a | Iso → 120.4℃ → N | | |

Table S2. Phase sequence of reported liquid crystal materials (Fig. S11) for machine learning.

| Name | Phase behaviors determined by DSC (Cooling at 10 ℃/min) | Name | Phase behaviors determined by DSC (Cooling at 10 ℃/min) |
|---|---|---|---|
| S1 | Iso → 182.1℃ → N → 85.6℃ → PN | **S12** | Iso → 149.6℃ → N → 117.1℃ → PN |
| S2 | Iso → 161.2℃ → N → 134.7℃ → Crystal | **S13** | Iso → 204.2℃ → N → 117.1℃ → Crystal |
| S3 | Iso → 160.3℃ → N → 111.4℃ → Crystal | **S14** | Iso → 165.9℃ → N → 151.6℃ → Crystal |
| S4 | Iso → 150.3℃ → N → 98.8℃ → Crystal | **S15** | Iso → 190.1℃ → N → 180.8℃ → Crystal |
| S5 | Iso → 274.8℃ → N → 155.2℃ → Crystal | **S16** | Iso → 155.2℃ → N → 139.6℃ →PN |
| S6 | Iso → 148.5℃ → N → 109.6℃ → Crystal | **S17** | Iso → 169.4℃ → N |
| S7 | Iso → 181.9℃ → N → 171.9℃ → Crystal | **S18** | Iso → 153.4℃ → N |
| S8 | Iso → 240.0℃ → N → 164.7℃ → Crystal | **S19** | Iso → 117℃ → N → 56℃ → Crystal |
| S9 | Iso → 186.2℃ → N → 138.8℃ → Crystal | **S20** | Iso → 131℃ → N → 85℃ → Crystal |
| S10 | Iso → 130.3℃ → N → 91.0℃ → PN | **S21** | Iso → 105℃ → N → 44℃ → Crystal |
| S11 | Iso → 99.2℃ → N → 77.6℃ → PN | **S22** | Iso → 74℃ → Crystal |

Table S3. Detailed data in the panel (o) of Fig. 3: comparison of the combination of the dielectric permittivity and loss tangent (tan$\delta$) for different materials.

| Type | Entry | Materials (Matrix/ Filler) | Dielectric permittivity | tan$\delta$ | Frequency (Hz) | References |
|---|---|---|---|---|---|---|
| | 1 | PS/TiO2-nanorod decorated CNTs | 37 | 16 | 1 k | (*34*) |
| | 2 | PBO / Graphene (4.0 wt%) | 16 | 5.3 | 1 k | (*35*) |
| | 3 | PCHMA / CNT (9.4 vol%) | 50 | 6.1 | 1 M | (*36*) |
| | 4 | PDMS / Functionalised graphene sheets (2.0 wt%) | 14 | 10.3 | 1 k | (*37*) |
| | 5 | PDVF / nanocrystalline Ni (27 vol%) | 210 | 209 | 1 k | (*38*) |
| | 6 | PDVF / CNT-SiC (1.78 wt%-4.82 wt%) | 894 | 115 | 0.1 k | (*39*) |
| | 7 | PDVF / SiC (1.57 wt%) | 24 | 10.5 | 0.1 k | (*40*) |
| | 8 | Cyanate ester / surface-treated MWCNTs (1.5 wt%) | ~175 | 675 | 1 k | (*41*) |
| | 9 | Cyanoethyl pullulan polymer / rGO-CNT (0.062 wt%) | 30.1 | 4.58 | 1 k | (*42*) |
| Composite | 10 | Polyimide polymer (PI) / rGO (2.7 wt%) | ~17 | 7.9 | 1 k | (*43*) |
| | 11 | Polypropylene / MWCNTs (3.0 wt%) | ~20 | 12.4 | 1 k | (*44*) |
| | 12 | PP foam / MWCNTs (1.25 vol%) | ~60 | 10 | 1 k | (*45*) |
| | 13 | P(VDF-HFP) / BT (50 vol%) | 32 | 6 | 1 k | (*46*) |
| | 14 | PVDF / BT (20 vol%) | 20 | 4.25 | 1 k | (*47*) |
| | 15 | Polyurethane / PZT (30 vol%) | 24 | 9 | 1 k | (*48*) |
| | 16 | P(VDF-CTFE) (88/12) / micro-sised CCTO (50 vol%) | 150 | 60 | 1 k | (*49*) |
| | 17 | P(VDF-TrFE-CTFE) / BT (17.5 vol%) | 70 | 46 | 1 k | (*50*) |
| | 18 | P(VDF-TrFE) (55/45) / CCTO (50 vol%) | 610 | 31 | 1 k | (*51*) |
| | 19 | P(VDF-HFP) / BT-WCNT (37.1 vol%) | 71.6 | 47 | 1 k | (*52*) |
| | 20 | PP / BT-Ag (56.8 vol%) | 160 | 5 | 1 k | (*53*) |

**Table S3** (*Continued*)

| Type | Entry | Materials (Matrix/ Filler) | Dielectric permittivity | tanδ | Frequency (Hz) | References |
|---|---|---|---|---|---|---|
| Inorganic solid | 21 | $CaCu_3Ti_4O_{12}$ (CCTO) (@25 °C) | 10286 | 0.067 | 100 k | (*54*) |
| | 22 | $Na_{1/2}La_{1/2}Cu_3Ti_4O_{12}$ (@25 °C) | 3560 | 0.074 | 100 k | (*54*) |
| | 23 | $BiCu_3Ti_3FeO_{12}$ (@25 °C) | 692 | 0.082 | 100 k | (*54*) |
| Organic solid | 24 | *N*-benzylideneaniline analogues (NBOA) | 111 | 0.1282 | 1 k | (*55*) |
| | 25 | *N*-benzylideneaniline analogues (CBOA) | 264 | 0.2049 | 1 k | (*55*) |
| | 26 | *N*-benzylideneaniline analogues (HBOA) | 495 | 0.2925 | 1 k | (*55*) |
| Hybrid solid | 27 | Phenazine-chloranilic acid (Phz-H2ca) | 3000 | < 0.05 | 1 M | (*56*) |
| | 28 | Phenazine-bromanilic acid (Phz-H2ba) | 1700 | < 0.05 | 1 M | (*56*) |
| Giant-κ fluids | 29 | **1a** | 11200 | 0.15 | 1 k | This work |
| | 30 | **1b** | 12200 | 0.17 | 1 k | This work |
| | 31 | **5b** | 14500 | 0.05 | 1 k | This work |
| | 32 | **5c** | 6500 | 0.7 | 0.1 k | This work |

Table S4. Detailed data in the panel (p) in Fig. 3: comparison of the combination of the dielectric permittivity and loss tangent (tan$\delta$) for different materials.

| Type | Entry | Materials | 2nd order NLO coefficenet dij (pm/V) | | Dielectric permittivity | References |
|---|---|---|---|---|---|---|
| Inorganic solid | 1 | $Bi_4Ge_3O_{12}$ | d14[a] | 1.28[a] | 16 | (*57, 58*) |
| | 2 | $LiKSO_4$ | d33[c] | 0.71[c] | 33 | (*57, 59*) |
| | 3 | GaS | d16[c] | 135[c] | 5.3 | (*57, 60*) |
| | 4 | GaSe | d16[c] | 972[c] | 6.1 | (*57, 60*) |
| | 5 | InSe | d16[c] | 281[c] | 7.6 | (*57, 60*) |
| | 6 | CdS | d33[d] | 25.8[d] | 10.3 | (*57, 61*) |
| | 7 | $Ba_6Ti_2Nb_8O_3$ | d33[a] | 13.2[a] | 209 | (*57, 62*) |
| | 8 | $K_3Li_2Nb_5O_{15}$ | d33[a] | 11.2[a] | 115 | (*57, 62*) |
| | 9 | $AgGaSe_2$ | d36[e] | 67.7[e] | 10.5 | (*57, 63*) |
| | 10 | $CsH_2AsO_4$ (CDA) | d36[c] | 0.22[c] | 675 | (*57, 64*) |
| | 11 | $SiO_2$ (quarz) | d11[a] | 0.335[a] | 4.58 | (*57, 65*) |
| | 12 | β-$BaB_2O_4$ (BBO) | d22[b] | 13.4[b] | 7.9 | (*57, 66*) |
| | 13 | α-$HIO_3$ | d36[a] | 5.15[a] | 12.4 | (*57, 66*) |
| | 14 | $Gd_2(MoO_4)_3$ | d31[a] | 2.49[a] | 10 | (*57, 67*) |
| | 15 | $KNbO_3$ | d33[a] | 19.6[a] | ~1000 | (*57, 68*) |
| | 16 | $NaNO_2$ (<164 °C) | d32[f] | 0.76[f] | ~600 | (*57, 69*) |
| | 17 | $BaTiO_3$ (@~25 °C) | d15[a] | 17[a] | ~2000 | (*57, 70*) |
| Organic solid | 18 | $KH_2PO_4$ (KDP) (@25 °C) | d36[a] | 0.44[a] | 46 | (*57, 71*) |
| | 19 | $KH_2AsO_4$ (KDA) (@25 °C) | d36[b] | 0.39[b] | 31 | (*57, 71*) |
| | 20 | γ-glycine (GG-CsCl) | dij | 1.97[a] | 47 | (*72*) |
| | 21 | GASH | d22[a] | 1.05[a] | 5 | (*57, 73*) |
| | 22 | $C_6H_4(NO_2)_2$ (MDB) | d32[a] | 2.7[a] | 2.28 | (*57, 74*) |
| Hybrid solid | 23 | $[NH_3(CH_2)_5NH_3]SbCl_5$ | d14[a] | 0.078[a] | 11 | (*75*) |
| | 24 | $(C_6H_{14}N)_2SbCl_5$ | d14[a] | 0.130[a] | 6 | (*76*) |
| | 25 | $(TMCM)_2$-$ZnCl_4$ | dij | 0.195[a] | 4.25 | (*77*) |
| | 26 | AZSH | dij | 0.987[a] | 9 | (*78*) |
| | 27 | LL-CTA | dij | 0.850[a] | 60 | (*79*) |

**Table S4** (*Continued*)

| | | | | | |
|---|---|---|---|---|---|
| Giant-κ fluids | **1a** | dij | 7.25[a] | 11200 | This work |
| | **1b** | dij | 2.10[a] | 10700 | This work |
| | **1c** | dij | 3.46[a] | 10800 | This work |
| | **1d** | dij | 3.46[a] | 8600 | This work |
| | **1e** | dij | 1.57[a] | 11300 | This work |
| | **1h** | dij | 2.43[a] | 6900 | This work |
| | **2a** | dij | 2.12[a] | 8100 | This work |
| | **2b** | dij | 2.83[a] | 8300 | This work |
| | **2c** | dij | 2.02[a] | 7800 | This work |
| | **4c** | dij | 8.12[a] | 9100 | This work |
| | **5b** | dij | 2.23[a] | 14500 | This work |
| | **5c** | dij | 2.26[a] | 6500 | This work |
| | **6a** | dij | 1.67[a] | 10500 | This work |
| | **6c** | dij | 1.84[a] | 10700 | This work |
| | **6f** | dij | 1.87[a] | 10300 | This work |
| | **7** | dij | 2.05[a] | 7600 | This work |
| | **8** | dij | 9.17[a] | 8300 | This work |
| | **12** | dij | 1.05[a] | 9200 | This work |

Notes: wavelength for the NLO measurement, a) 1064 nm, b) 1060 nm, c) 694.3 nm, d) 1058 nm, e) 2120 nm, f) 1153 nm. Abbreviations: ammonium zinc sulphate hydrate (AZSH), L-lysine influenced cadmium thiourea acetate (LL-CTA), $(CN_3H_6)Al(SO4)_2$- $6H_2O$ (GASH). *d*ij means the effective nonlinear optical coefficient.

References

80. B. Wul, Dielectric Constants of Some Titanates. *Nature* **156**, 480 (1945).

81. W. Jackson, W. Reddish, High Permittivity Crystalline Aggregates. *Nature* **156**, 717 (1945).

82. P. R. Coursey, K. G. Brand, Dielectric Constants of Some Titanates. *Nature* **157**, 297–298 (1946).

83. M. G. Harwood, P. Popper, D. F. Rushman, Curie Point of Barium Titanate. *Nature* **160**, 58–59 (1947).

84. B. T. Matthias, Dielectric Constant and Piezo-Electric Resonance of Barium Titanate Crystals. *Nature* **161**, 325-326 (1948).

85. M. Osada, Y. Ebina, H. Funakubo, S. Yokoyama, T. Kiguchi, K. Takada, T. Sasaki, High-*κ* Dielectric Nanofilms Fabricated from Titania Nanosheets. *Adv. Mater.* **18**, 1023–1027 (2006).

86. Y. Y. Illarionov, T. Knobloch, M. Jech, M. Lanza, D. Akinwande, M. I. Vexler, T. Mueller, M. C. Lemme, G. Fiori, F. Schwierz, T. Grasser, Insulators for 2D nanoelectronics: the gap to bridge. *Nat. Commun.* **11**, 3385 (2020).

87. T. Li, T. Tu, Y. Sun, H. Fu, J. Yu, L. Xing, Z. Wang, H. Wang, R. Jia, J. Wu, C. Tan, Y. Liang, Y. Zhang, C. Zhang, Y. Dai, C. Qiu, M. Li, R. Huang, L. Jiao, K. Lai, B. Yan, P. Gao, H. Peng, A native oxide high-*κ* gate dielectric for two-dimensional electronics. *Nat. Electron.* **3**, 473–478 (2020).

88. H. Pan, F. Li, Y. Liu, Q. Zhang, M. Wang, S. Lan, Y. Zheng, J. Ma, L. Gu, Y. Shen, P. Yu, S. Zhang, L.-Q. Chen, Y.-H. Lin, C. -W. Nan, Ultrahigh–energy density lead-free dielectric films via polymorphic nanodomain design. *Science* **365**, 578-582 (2019).

89. S. K. Kim, S. W. Lee, J. H. Han, B. Lee, S. Han, C. S. Hwang, Capacitors with an equivalent oxide thickness of <0.5 nm for nanoscale electronic semiconductor memory. *Adv. Funct. Mater.* **20**, 2989–3003 (2010).

90. D. Ji, T. Li, Y. Zou, M. Chu, K. Zhou, J. Liu, G. Tian, Z. Zhang, X. Zhang, L. Li, D. Wu, H. Dong, Q. Miao, H. Fuchs, W. Hu, Copolymer dielectrics with balanced chain-packing density and surface polarity for high-performance flexible organic electronics. *Nat. Commun.* **9**, 2339 (2018).

91. J. Robertson, High dielectric constant gate oxides for metal oxide Si transistors. *Rep. Prog. Phys.* **69**, 327–396 (2006).

92. C.-H. Lee, S.-H. Hur, Y.-C. Shin, J.-H. Choi, D.-G. Park, K. Kim, Charge-trapping device structure of SiO2/SiN/high-*k* dielectric Al2O3 for high-density flash memory. *Appl. Phys. Lett.* **86**, 152908 (2005).

93. H. Nishikawa, K. Shiroshita, H. Higuchi, Y. Okumura, Y. Haseba, S. Yamamoto, K. Sago, H. Kikuchi, A Fluid Liquid-Crystal Material with Highly Polar Order. *Adv. Mater.* **20**, 1702354 (2017).

94. R. J. Mandle, S. J. Cowling, J. W. Goodby, A nematic to nematic transformation exhibited by a rod-like liquid crystal. *Phys. Chem. Chem. Phys.* **19**, 11429–11435 (2017).

95. R. J. Mandle, S. J. Cowling, J. W. Goodby, Rational design of rod-like liquid crystals exhibiting two nematic phases. *Chemistry* **23**, 14554–14562 (2017).

96. A. Mertelj, L. Cmok, N. Sebastián, R. J. Mandle, R. R. Parker, A. C. Whitwood, J. W. Goodby, M. Čopič, Splay nematic phase. *Phys. Rev. X* **8**, 41025 (2018).

97. N. Sebastián, L. Cmok, R. J. Mandle, M. R. de la Fuente, I. D. Olenik, M. Čopič, A. Mertelj, Ferroelectric-Ferroelastic Phase Transition in a Nematic Liquid Crystal, *Phys. Rev. Lett.* **124**, 037801 (2020).

98. X. Chen, E. Korblova, D. Dong, X. Wei, R. Shao, L. Radzihovsky, M. A. Glaser, J. E. Maclennan, D. Bedrov, D. M. Walba, N. A. Clark, First-principles experimental demonstration of ferroelectricity in a thermotropic nematic liquid crystal: Polar domains and striking electro-optics. *Proc. Natl. Acad. Sci. U.S.A.* **117**, 14021-14031 (2020).

99. N. Chaturvedi, R. D. Kamien, Mechanisms to splay-bend nematic phases. *Phys. Rev. E* **100**, 022704 (2019).

100. M. Copic, A. Mertelj, Q-tensor model of twist-bend and splay nematic phases. *Phys. Rev. E* **101**, 022704 (2020).

101. M. P. Rosseto, J. V. Selinger, Theory of the splay nematic phase: Single vs. double splay. *Phys. Rev. E* **101**, 052707 (2020).

102. S. M. Stigler, Francis Galton's account of the invention of correlation, *Stat. Sci.* **4**, 73 (1989).

103. S. M. Lundber, S. -I. Lee, A unified approach to interpreting model predictions, *Proceedings of the 31st International Conference on Neural*, **30**, 4768 (2017).

104. M. Born, Über anisotrope Flüssigkeiten. Versuch einer Theorie der flüssigen Kristalle und des elektrischen Kerr-Effekts in Flüssigkeiten. Sitzungsber. *Preuss. Akad Wiss.* **30**, 614 (1916).

105. D. W. Berreman, Optics in Stratified and Anisotropic Media: 4×4-Matrix Formulation, J. Opt. Soc. Am. 62, 502-510 (1972).

106. M., Gil. Preparation of bis(benzenecarboxamides) as calcium channel blockers. PCT Int. Appl., 2007068754 (2007).

107. S. Kawashita. K. Aoyagi, H. Yamanaka, R. Hantani, S. Naruoka, A. Tanimoto, Y. Hori, Y. Toyonaga, K. Fukushima, S. Miyazaki, Y. Hantani. Symmetry-based ligand design and evaluation of small molecule inhibitors of programmed cell death-1/programmed death-ligand 1 interaction. *Bioorg. Med. Chem. Lett* **29**, 2464-2467 (2019).

108. A. K. Al-Lami. Preparation and Mesomorphic Characterization of Supramolecular Hydrogen-Bonded Dimer Liquid Crystals. *Polycycl. Aromat. Compd.* **36**, 197-212 (2016).

109. D. Ndaya, R. Bosirea, R. M. Kasi. Cholesteric–azobenzene liquid crystalline copolymers: design, structure and thermally responsive optical properties. *Polym. Chem.* **10**, 3868-3878 (2019).

110. J. Pecyna, P. Kaszynski, B. Ringstrand, M. Bremer. Investigation of high Δε derivatives of the [closo-1-CB9H10]− anion for liquid crystal display applications. *J. Mater. Chem. C* **2**, 2956-2964 (2014).

111. M.-J. Gim, D. A. Beller, D. K. Yoon. Morphogenesis of liquid crystal topological defects during the nematic-smectic A phase transition. *Nat. Commun.* **8**, 15453 (2017).

112. A. Liaw, M. Wiener. Classification and Regression by random Forest. *R News*, ISSN 1609-3631.

113. C. Wu, X. Huang, X. Wu, J. Yu, L. Xie, P. Jiang. TiO2-nanorod decorated carbon nanotubes for high-permittivity and low-dielectric-loss polystyrene composites. *Compos. Sci. Technol*. **72**, 521–527 (2012).

114. Y. Chen, Q. Zhuang, X. Liu, J. Liu, S. Lin, Z. Han. Preparation of thermostable PBO/graphene nanocomposites with high dielectric constant. *Nanotechnology* **24**, 245702 (2013).

115. K. Hayashida. Dielectric properties of polymethacrylate-grafted carbon nanotube composites. *RSC Adv.* **3**, 221–227 (2013).

116. L. J. Romasanta, M. Hernández, M. A. López-Manchado, R. Verdejo. Functionalised graphene sheets as effective high dielectric constant fillers. *Nanoscale Res. Lett.* **6**, 1–6 (2011).

117. M. Panda, V. Srinivas, A. K. Thakur. On the question of percolation threshold in polyvinylidene fluoride/nanocrystalline nickel composites. *Appl. Phys. Lett.* **92**, 12–15 (2008).

118. J. Yuan, S. Yao, W. Li, A. Sylvestre, J. Bai. Vertically aligned carbon nanotube arrays on SiC microplatelets: A high figure-of-merit strategy for achieving large dielectric constant and low loss in polymer composites. *J. Phys. Chem. C* **118**, 22975–22983 (2014).

119. A. B. Dichiara, J. Yuan, S. Yao, A. Sylvestre, L. Zimmer, J. Bai. Effective synergistic effect of Al2O3 and SiC microparticles on the growth of carbon nanotubes and their application in high dielectric permittivity polymer composites. *J. Mater. Chem. A* **2**, 7980–7987 (2014).

120. C. Han, A. Gu, G. Liang, L. Yuan. Carbon nanotubes/cyanate ester composites with low percolation threshold, high dielectric constant and outstanding thermal property. *Compos. Part A Appl. Sci. Manuf.* **41**, 1321–1328 (2010).

121. J. Y. Kim, T. Y. Kim, J. W. Suk, H. Chou, J. ‐ H. Jang, J. H. Lee, I. N. Kholmanov, D. Akinwande, R. S. Ruoff. Enhanced dielectric performance in polymer composite films with carbon nanotube-reduced graphene oxide hybrid filler. *Small* **10**, 3405–3411 (2014).

122. G. Tian, J. Song, J. Liu, S. Qi, D. Wu. Enhanced dielectric permittivity and thermal stability of graphene-polyimide nanohybrid films. *Soft Mater* **12**, 290–296 (2014).

123. C. R. Yu, D.-M. Wu, Y. Liu, H. Qiao, Z.-Z. Yu, A. Dasari, X. Du, Y.-W. Mai. Electrical and dielectric properties of polypropylene nanocomposites based on carbon nanotubes and barium titanate nanoparticles. *Compos. Sci. Technol.* **71**, 1706–1712 (2011).

124. A. Ameli, S. Wang, Y. Kazemi, C. B. Park, P. Pötschke. A facile method to increase the charge storage capability of polymer nanocomposites. *Nano Energy* **15**, 54–65 (2015).

125. P. Kim, N. M. Doss, J. P. Tillotson, P. J. Hotchkiss, M.-J. Pan, S. R. Marder, J. Li, J. P. Calame, J. W. Perry. High energy density nanocomposites based on surface-modified BaTiO3 and a ferroelectric polymer. *ACS Nano* **3**, 2581–2592 (2009).

126. Yu, K., Wang, H., Zhou, Y., Bai, Y. & Niu, Y. Enhanced dielectric properties of BaTiO3/poly(vinylidene fluoride) nanocomposites for energy storage applications. *J. Appl. Phys*. **113**, 034105 (2013).

127. K. S. Lam, Y. W. Wong, L. S. Tai, Y. M. Poon, F. G. Shin. Dielectric and pyroelectric properties of lead zirconate titanate/polyurethane composites. *J. Appl. Phys.* **96**, 3896–3899 (2004).

128. L. Zhang, X. Shan, P. Wu, Z. Y. Cheng. Dielectric characteristics of CaCu3Ti4O12/P(VDF-TrFE) nanocomposites. *Appl. Phys. A Mater. Sci. Process.* **107**, 597–602 (2012).

129. H. Tang, Y. Lin, H. A. Sodano. Synthesis of high aspect ratio batio3 nanowires for high energy density nanocomposite capacitors. *Adv. Energy Mater.* **3**, 451–456 (2013).

130. M. Arbatti, X. Shan, Z. Cheng. Ceramic-polymer composites with high dielectric constant. *Adv. Mater.* **19**, 1369–1372 (2007).

131. Y. Jin, N. Xia, R. A. Gerhardt. Enhanced dielectric properties of polymer matrix composites with BaTiO3 and MWCNT hybrid fillers using simple phase separation. *Nano Energy* **30**, 407–416 (2016).

132. S. Luo, S. Yu, R. Sun, C. P. Wong. Nano Ag-deposited BaTiO3 hybrid particles as fillers for polymeric dielectric composites: Toward high dielectric constant and suppressed Loss. *ACS Appl. Mater. Interfaces* **6**, 176–182 (2014).

133. M. A. Subramanian, A. W. Sleight. ACu3Ti4O12 and ACu3Ru4O12 perovskites: High dielectric constants and valence degeneracy. *Solid State Sci.* **4**, 347–351 (2002).

134. R. Kumari, R. Seera, A. De, R. Ranjan, T. N. Guru Row. Organic Multifunctional Materials: Second Harmonic, Ferroelectric, and dielectric properties in N-benzylideneaniline analogues. *Cryst. Growth Des.* **19**, 5934–5944 (2019).

135. S. Horiuchi, F. Ishii, R. Kumai, Y. Okimoto, H. Tachibana, N. Nagaosa, Y. Tokura. Ferroelectricity near room temperature in co-crystals of nonpolar organic molecules. *Nat. Mater.* **4**, 163–166 (2005).

136. M. J. Weber. Handbook of Optical Materials 1.9.3 Second Harmonic Generation Coefficients (CRC Press, Boca, Raton, London, New York, Washington, D.C., 2003).

137. J. Link, J. Fontanella, C. G. Andeen. Temperature variation of the dielectric properties of bismuth germanate and bismuth germanium oxide. *J. Appl. Phys.* **51**, 4352–4355 (1980).

138. M. Delfino, G. M. Loiacono, W. A. Smith. Thermal and dielectric properties of LiKSO4 and LiCsSO4. *J. Solid State Chem.* **31** 131–134 (1980).

139. A. Chevy, A. Segura, V. Muñoz. Effects of pressure and temperature on the dielectric constant of gas, gase, and inse: role of the electronic contribution. *Phys. Rev. B - Condens. Matter Mater. Phys.* **60,** 15866–15874 (1999).

140. D. Berlincourt, H. Jaffe, L. R. Shiozawa. Electroelastic properties of the sulfides, selenides, and tellurides of zinc and cadmium. *Phys. Rev.* **129**, 1009–1017 (1963).

141. R. R. Neurgaonkar, W. K. Cory. Progress in photorefractive tungsten bronze crystals. *J. Opt. Soc. Am. B* **3**, 274 (1986).

142. A. U. Sheleg, V. G. Hurtavy. The Influence of Electron Irradiation on the Dielectric Characteristics of Single Crystals of AgGaSe2. *Phys. Solid State* **61**, 1695–1698 (2019).

143. R. J. Pollina, C. W. Garland. Dielectric and ultrasonic measurements in CsH2AsOs. *Phys. Rev. B* **12**, 362–367 (1975).

144. A. De, K. V. Rao. Dielectric properties of synthetic quartz crystals. *J. Mater. Sci.* **23**, 661–664 (1988).

145. C. R. Raja, R. Gobinathan, F. D. Gnanam. Dielectric Properties of Beta Barium Borate and Potassium Pentaborate Single Crystals. *Cryst. Res. Technol.* **28**, 737–743 (1993).

146. S. W. Ko, D. A. Mourey, T. Clark, S. Trolier-Mckinstry. Synthesis, characterisation, and dielectric properties of β-Gd2(MoO4)3 thin films prepared by chemical solution deposition. *J. Sol-Gel Sci. Technol*. **54**, 269–275 (2010).

147. M. Zgonik, R. Schlesser, I. Biaggio, E. Voit, J. Tscherry, P. Günter. Materials constants of KNbO3 relevant for electro- and acousto-optics. *J. Appl. Phys.* **74**, 1287–1297 (1993).

148. K. Gesi. Electrical properties of NaNO2 single crystal in the vicinity of the ferroelectric curie temperature. *J. Phys. Soc. Japan* **22**, 979–986 (1967).

149. A. El Ghandouri, S. Sayouri, T. Lamcharfi, A. Elbasset. Structural, microstructural and dielectric properties of Ba1-xLaxTi(1-x/4)O3 prepared by sol gel method. *J. Adv. Dielectr.* **9**, 1–19 (2019).

150. K. F. Young, H. P. R. Frederikse. Compilation of the Static Dielectric Constant of Inorganic Solids. *J. Phys. Chem. Ref. Data* **2**, 313–410 (1973).

151. M. Anis, S. P. Ramteke, M. D. Shirsat, G. G. Muley, M. I. Baig. Novel report on γ-glycine crystal yielding high second harmonic generation efficiency. *Opt. Mater.* **72**, 590–595 (2017).

152. A. N. Holden, W. J. Merz, J. P. Remeika, B. T. Matthias. Properties of guanidine aluminum sulfate hexahydrate and some of its isomorphs. *Phys. Rev.* **101**, 962–966 (1956).

153. J. W. Williams, C. H. Schwrzngel. The dielectric constants of binary mixtures. VI The electric moments of certain nitro derivatives of benzene and toluene. *J. Am. Chem. Soc.* **50**, 362–368 (1928).

154. G. Q. Mei, H. Y. Zhang, W. Q. Liao. A symmetry breaking phase transition-triggered high-temperature solid-state quadratic nonlinear optical switch coupled with a switchable dielectric constant in an organic-inorganic hybrid compound. *Chem. Commun.* **52**, 11135–11138 (2016).

155. J. Zhang, S. Han, X. Liu, Z. Wu, C. Ji, Z. Sun, J. Luo. A lead-free perovskite-like hybrid with above-room-temperature switching of quadratic nonlinear optical properties. *Chem. Commun.* **54**, 5614–5617 (2018).

156. W. Q. Liao, J. X. Gao, X. N. Hua, X. G. Chen, Y. Lu. Unusual two-step sequential reversible phase transitions with coexisting switchable nonlinear optical and dielectric behaviors in [(CH3)3NCH2Cl]2[ZnCl4]. *J. Mater. Chem. C* **5**, 11873–11878 (2017).

157. S. P. Ramteke, M. I. Baig, M. Shkir, S. Kalainathan, M. D. Shirsat, G. G. Muley, M. Anisa. Novel report on SHG efficiency, Z-scan, laser damage threshold, photoluminescence, dielectric and surface microscopic studies of hybrid inorganic ammonium zinc sulphate hydrate single crystal. *Opt. Laser Technol.* **104**, 83–89 (2018).

158. I. Khan, M. Anis, U. Bhati. Influence of L-lysine on optical and dielectric traits of cadmium thiourea acetate complex crystal. *Optik.* **170**, 43–47 (2018).